\documentclass[12pt,twoside]{article}

\pdfoutput=1

\usepackage[usenames]{color}
\usepackage{lscape}
\usepackage{amssymb}
\usepackage{rotating}
\usepackage[T1]{fontenc}
\usepackage{graphicx}
\usepackage{fancyhdr}
\usepackage{amsmath}
\usepackage{ae}
\usepackage{aecompl}
\usepackage{hyperref}

\usepackage[tight]{subfigure}

\usepackage{xcomment}

\def\sgn{{\rm\,sgn}}
\def\nn{\nonumber \\}

\newcommand{\be}{\begin{equation}}
\newcommand{\ee}{\end{equation}}
\newcommand{\ba}{\begin{eqnarray}}
\newcommand{\ea}{\end{eqnarray}}

\newcommand{\gev}{\, {\rm GeV}}

\newcommand{\lsim}
{\;\raisebox{-.3em}{$\stackrel{\displaystyle <}{\sim}$}\;}
\newcommand{\gsim}
{\;\raisebox{-.3em}{$\stackrel{\displaystyle >}{\sim}$}\;}

\newlength{\dinwidth}
\newlength{\dinmargin}
\begin{document}

\thispagestyle{empty}
%{\color{red}version: \today}
\begin{flushright}
CERN-PH-TH-2015-141,
KCL-PH-TH/2015-27,
LCTS/2015-19
\end{flushright}

\vspace*{15mm}

\centerline{\Large\bf Detecting underabundant neutralinos} 
\vspace*{15mm}

\centerline{M. Badziak${}^a$, A. Delgado${}^{b,c}$,  M. Olechowski${}^a$,  S.  Pokorski${}^a$ and K. Sakurai${}^d$}
\vspace*{5mm}

\centerline{${}^a$\em Institute of Theoretical Physics,
Faculty of Physics, University of Warsaw} 
\centerline{\em ul.~Pasteura 5, PL--02--093 Warsaw, Poland} 
\centerline{${}^b$\em Department of Physics, 225 Nieuwland Science Hall,}
\centerline{\em University of Notre Dame,
Notre Dame, IN 46556, USA}
\centerline{${}^c$\em
Theory Division, Physics Department CERN,
CH-1211 Geneva 23, Switzerland} 

\centerline{${}^d$\em Department of Physics, King's College London, London WC2R 2LS, UK}

\vskip 1cm

\centerline{\bf Abstract}
The electroweak sector may play a crucial role in discovering supersymmetry. We systematically investigate the patterns of the MSSM-like electroweakinos,  when the neutralino relic abundance
$\Omega_\chi h^2\leq 0.12$, that  is, also admitting for multi-component Dark Matter, in a broad range of the parameter space. We find that for a very large range of parameters the Direct Detection  experiments are/will be sensitive to   underabundant neutralinos, in spite of  the strong rescaling
of the flux factor. The second general conclusion is that the bound  
$\Omega_\chi h^2\leq 0.12$
together with the LUX (XENON1T) limits for the neutralino spin independent scattering cross sections
constrain the electroweakino spectrum so that the mass differences between
the NLSP and the LSP are smaller  than 40 (10) GeV, respectively, with important implications for the collider searches. 
The future Direct Detection experiments and the high luminosity LHC run will probe almost the entire range of the LSP and NLSP mass spectrum that is consistent with
the bound $\Omega_\chi h^2\leq 0.12$.

\vskip 3mm

%%%%%%%%%%%%%%%%%%%%%%%%%%%%%%%%%%%%%%%%%%%%%%%%%%%%%%%%%%%%%%%%%%%%%%
%%%%%%%%%%%%%%%%%%%%%%%%%%%%%%%%%%%%%%%%%%%%%%%%%%%%%%%%%%%%%%%%%%%%%%
\newpage
%%%%%%%%%%%%%%%%%%%%%%%%%%%%%%%%%%%%%%%%%%%%%%%%%%%%%%%%%%%%%%%%%%%%%%
%%%%%%%%%%%%%%%%%%%%%%%%%%%%%%%%%%%%%%%%%%%%%%%%%%%%%%%%%%%%%%%%%%%%%%

\section{Introduction}

Supersymmetry remains to be an attractive extensions of the SM even if merely for being one of the few UV complete frameworks, with no quadratic sensitivity to large mass scales.
It may well  be that the supersymmetric electroweak sector will play the leading role in discovering supersymmetry.  Indeed, experimental searches for coloured superpartners, motivated by the issue of naturalness, have so far given  null results. It is conceivable that  supersymmetric models remain natural even
with new coloured degrees of freedom heavier than few TeV (e.g. because the soft masses are correlated 
\cite{ft_correlations},  because the Higgs mass is further protected 
\cite{doubleprotection} or  it is protected by colourless degrees of freedom \cite{colorless}) or  it could be that supersymmetry plays no role in stabilizing the weak scale (like in split supersymmetry \cite{splitSUSY}). 

However, in models with a stable neutralino and if one
accepts the thermal history of the universe, the higgsino and gaugino masses 
are bounded from above by the thermal overabundance of the LSP. 
\footnote{Light electroweakinos are additionally motivated by the supersymmetric solution to the muon $g-2$ anomaly, see e.g. refs.~\cite{g2_Shafi,g2_upper} for a recent works on this topic.
Throughout this paper we assume the stability of the lightest neutralino and the standard scenarios of thermal history of the Universe. For example we do not consider the scenario of late entropy production, discussed e.g. in \cite{Gelmini:2006pw}.
}
So far, the interest in  the supersymmetric electroweak sector has been
often linked to the fact that a stable neutralino can account for the observed Dark Matter (DM) in the universe.   It is, however,   conceivable that  DM  has a multi-component structure. There is no reason why in the ``dark'' sector there would be only one stable particle.  It is
possible that the neutralino component in the observed DM is small or even very small and the bulk of DM has a different origin.  
The electroweak sector is then an interesting signature of supersymmetry in its own sake  and there should be investigated  the prospects for its discovery for any $\Omega_{\chi}\leq\Omega_{\rm DM}$.

Partial results of such investigations do exist in the literature, particularly for collider signatures and for the pure higgsino and wino limits 
%\cite{welltempered, Hall_blindspots, Baer:2013vpa, rolbiecki, cohen, taohan, SchwallerZurita, baer, Han:2013usa, Han:2014kaa, Neutralino_Wang, Raby, nagata, Bramante_neutralino, Martin_EWino_LHC, ChaNeu100_Sakurai, Gori100, NeutralinoSurface100, diCortona, Higgsino_Munir, NeutralinoVBF, chala, Barducci}.  
\cite{welltempered}-\cite{Barducci}.\footnote{
    More general studies based on $S$-matrix unitarity and the dynamics of thermal freeze-out 
    can be found e.g. in \cite{Blum:2014dca}.
}
In this paper we readdress this question, with the aim  of an overall, systematic
analysis in a large parameter space of the electroweak sector  and with the focus on the interplay of the direct detection (DD) and collider experiments. Our global analysis leads to several new general conclusions.  An additional theoretical motivation for such a global  analysis is  the variety of possible mechanisms for mediation of supersymmetry breaking, with different patterns of gaugino masses.

There are at least two different ways to discover those underabundant neutralinos, in DD experiments and in colliders.
%\footnote{The indirect detection may be more difficult if neutralinos are only a small component of dark matter.} For the former the relevant quantities to calculate is their relic abundance $\Omega_\chi$ and their spin independent nuclear cross sections $\sigma_\chi^{\rm SI}$.\footnote{Spin dependent bounds are not competitive.} 
The latter are at present excluded by LUX \cite{LUX} for 
\begin{equation}
\frac{\sigma_{\chi}^{\rm SI}}{\sigma_{\rm LUX}^{\rm SI}}
\frac{\Omega_{\chi}}{\Omega_{\rm DM}}>1 \,,
\label{bounds}
\end{equation}
where $\sigma_{\rm LUX}^{\rm SI}$ is the upper bound on the cross section obtained with  the local DM density $\rho_0=0.3$ GeV/cm$^2$, corresponding to the observed
dark matter abundance $\Omega_{\rm DM}$ \cite{Planck}. For the future discovery potential in the direct detection, the $\sigma_{\rm LUX}^{\rm SI}$ has to be
replaced by
the corresponding new experimental limits.
For the colliders one needs to know the spectrum of the electroweakinos which will then dictate the different signatures. 
Indirect DM detection experiments may also be sensitive to our underabundant neutralino scenario, although,
as we will see later,
the current limit is weaker than the direct detection limit most of the cases.  

Motivated by the absence, so far, of any signal of the scalar superpartners  and heavy Higgs bosons at the LHC, we will assume that they are sufficiently decoupled 
to be neglected in the neutralino annihilation and scattering processes. Most of the results presented here are for heavy degenerate slepton and squark masses, as for instance in the split supersymmetry models. There, the gluinos are light and the renormalization effects by strong interactions are weak.
However, we have checked that our conclusions remain unchanged even when sleptons are just about 20$\%$ heavier than the LSP and we show some results for that case, too. The  relevant supersymmetric parameter space consists then of the soft bino ($M_1$) and  wino ($M_2$) 
masses, the higgsino  mass ($\mu$) and the ratio of the vacuum expectation values of the two Higss fields $\frac{\langle H_u \rangle}{\langle H_d \rangle}$ ($\tan\beta$). In our scan the above mass parameters are bounded from below by 100 GeV, so we do not consider the Higgs and $Z$ funnel effects. 
We also do not consider the funnel effects of the heavy CP-even ($H$) and odd ($A$) Higgses,
because we assume they are as heavy as sfermions. %for simplicity, by decoupling them explicitly.
Our analysis applies to the MSSM and to all models where the admixture of additional states to
the LSP and NLSP mass eigenstates is small like  in certain versions of the NMSSM \cite{ellwanger} or folded SUSY \cite{colorless} where one expects the higgsinos to be light. In general one can view our analysis as that of a simplified model with a singlet (the bino), a pair of doublets (the higgsino) and a triplet (the wino) of $SU(2)$ with couplings close to the gauge couplings.

Our approach will be to first study certain limits of the LSP composition:
 
 \begin{itemize}
 \item bino-higgsino with wino decoupled; this spectrum is motivated by minimal supergravity \cite{msugra} or gauge mediation \cite{gaugemed}.

 \item wino-higgsino with bino decoupled; 
 this spectrum may appear in theories   with anomaly mediation and light higgsinos 
 \cite{anomalymed, ArkaniHamed:2004fb, Bagnaschi:2014rsa}.

 \item bino-wino with a heavy but not totally decoupled  higgsino,
  to mediate some bino-wino mixing; this situation may happen in an anomaly mediated scenario with the higgs superfields acting  as messengers  of supersymmetry breaking \cite{Ibe:2012hu, Bagnaschi:2014rsa}.

\end{itemize}

Those limiting cases are a good reference point for  subsequently investigating  the general case  in which higgsinos and both electroweak (EW) gauginos simultaneously
contribute to the LSP composition (that might possibly occur in more sophisticated SUSY breaking schemes like e.g. mirage mediation \cite{mirage} or mini-split supersymmetry \cite{Ibe:2012hu} ). 

The main conclusions of this paper are the following: first of all, in spite of the small relic abundance and the rescaling in the flux factor, the present and future DD experiments are/will be sensitive to
neutralinos with $\Omega_{\chi} h^2$  even two to three orders of magnitude below the
observed value $\Omega_{\rm DM} h^2 = 0.12$.\footnote{
{
    This has been already pointed out for higgsino-like LSPs 
    with $m_{\tilde \chi_1^0} \lsim 350$ GeV \cite{Baer:2013vpa}.
    }
} 
Thus,  a discovery of  a signal in DD experiments would not necessarily mean the discovery of the main DM component  but might be a strong hint for supersymmetry. Similarly, the present and future
limits on the spin independent  scattering cross section put strong constraints  on the electroweakino
parameter space with underabundant neutralinos.

Secondly, the results of the DD experiments have strong impact on the search strategies in colliders
for electroweakinos in the bulk of the supersymmetric parameter space, if sfermions are heavy. This is because  then  the smaller the $\sigma_{\chi}^{\rm SI}$  and/or the $\Omega_\chi$ are/is, the smaller the mass splitting between the LSP and NLSP becomes, which is crucial for collider searches.  Already the current LUX limit constrains
this splitting to be less than 40 GeV.  With the future direct detection experiments, such as XENON1T \cite{XENON1T} and LZ \cite{LZ}, the bound on the LSP-NLSP
mass splitting can be brought below 10 GeV, as a generic result for a MSSM-like spectrum with heavy sfermions.  Qualitatively, that result can be understood as
a consequence of a smaller bino contribution  to the LSP, strongly constrained from above by the limits from  the DD experiments, in addition to  the
constraint from the upper bound on $\Omega_{\chi} h^2$. Of course, if a signal  was discovered in the DD experiments, similar remarks would apply to the searches for
its direct confirmation in colliders.

Finally, focusing on the interplay of  the DD and collider experiments, we identify the parameter regions where both may have comparable discovery potential and supplement each other and those  where they are complementary to each other. 
There are regions of the parameter space in which the LSP escapes direct detection and  we survey the
prospects of the 13/14 TeV LHC to probe the SUSY EWkino states in these regions.    
In many cases existing search techniques for quasi-degenerate states at the 13/14 TeV LHC are able to close the gaps in the parameter
space that are inaccessible to future direct detection experiments. However, for some regions of the  parameters the spin independent scattering cross
section is below the neutrino background \cite{Cushman:2013zza} and parts of those regions are also inaccessible to the 13/14 TeV LHC experiments using present search techniques. We classify these regions  to encourage search for new techniques to probe them at the LHC and future colliders.

%%%%%%%%%%%%%%%%%%%%%%%%%%%%%%%%%%%%%%%%%%%%%%%%%%%%%%%%%%%%%%%%%%%%%%
%%%%%%%%%%%%%%%%%%%%%%%%%%%%%%%%%%%%%%%%%%%%%%%%%%%%%%%%%%%%%%%%%%%%%%
\section{Bino-Higgsino LSP}
\label{sec:Bino-Higgsino-DM}
%%%%%%%%%%%%%%%%%%%%%%%%%%%%%%%%%%%%%%%%%%%%%%%%%%%%%%%%%%%%%%%%%%%%%%
%%%%%%%%%%%%%%%%%%%%%%%%%%%%%%%%%%%%%%%%%%%%%%%%%%%%%%%%%%%%%%%%%%%%%%

The electroweakino sector with decoupled wino component in the LSP and the prospects for its experimental discovery have already been studied in a number of
papers \cite{rolbiecki, Baer:2013vpa, SchwallerZurita,baer,Neutralino_Wang,nagata,diCortona,Higgsino_Munir,Barducci,Calibbi1}. In this paper we reanalyze this case for completeness of our presentation
and also for stressing several  general conclusions.

With heavy sfermions, pure bino has no annihilation channels and  is excluded  by the bound
$\Omega_{\chi} h^2\leq 0.12$ \cite{Planck}.  The annihilation cross section for  pure higgsino  ${\tilde h}$ is determined by the vertex 
$\tilde h^0 \tilde h^\pm W^\mp$.
The dominant annihilation channel is into  the gauge bosons  and co-annihilation with almost mass degenerate  remaining higgsino states is very important.
The approximate formula for the relic abundance reads \cite{welltempered}:

\begin{equation}
\label{Omega_Bino}
\Omega_{\tilde h} h^2=0.10\left(\frac{\mu}{\mbox{1 TeV}}\right)^2
\,.
\end{equation}

\begin{figure}
	\centering \vspace{-0.0cm}
		\includegraphics[width=0.48\textwidth]{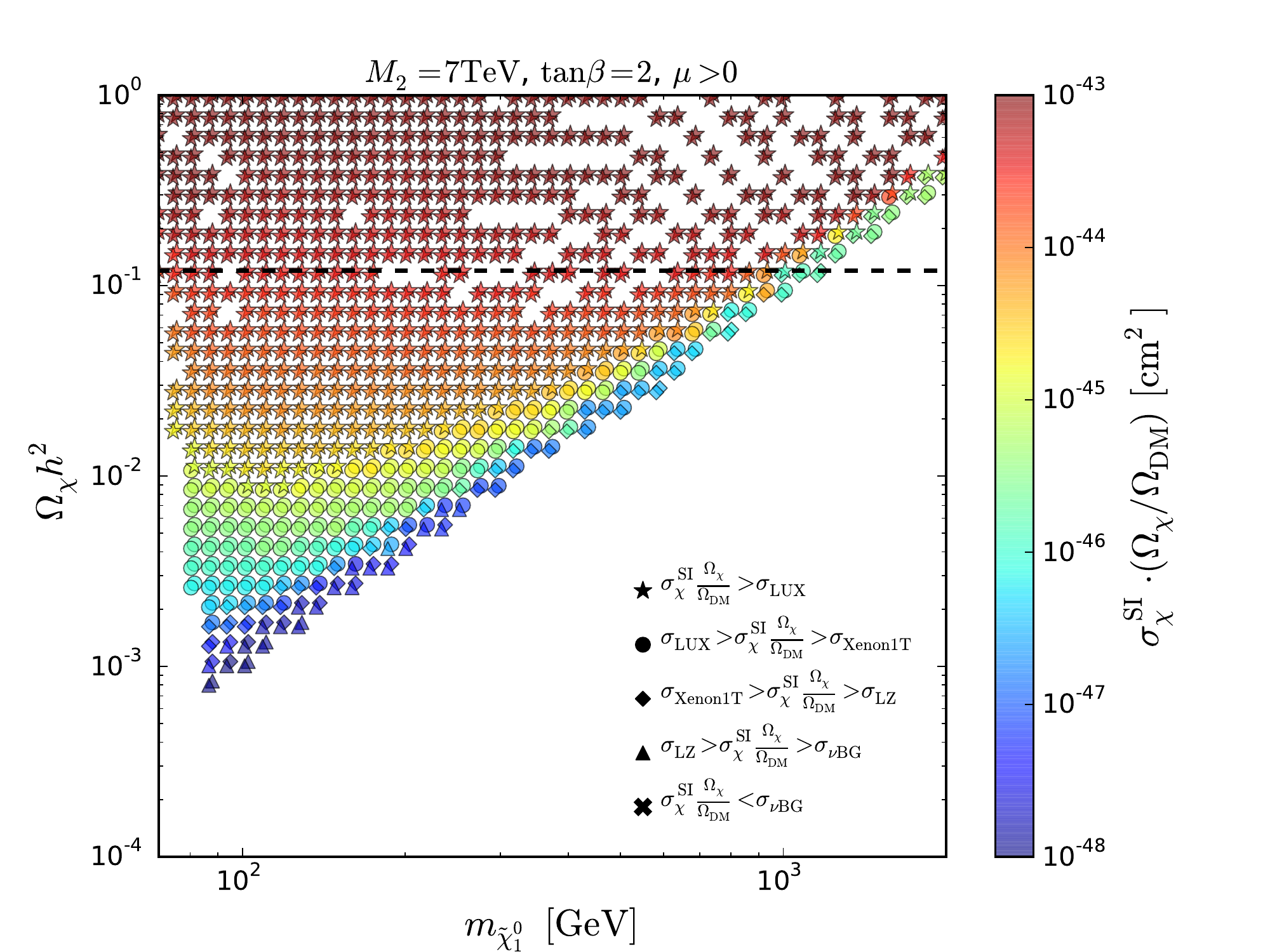}
		\includegraphics[width=0.48\textwidth]{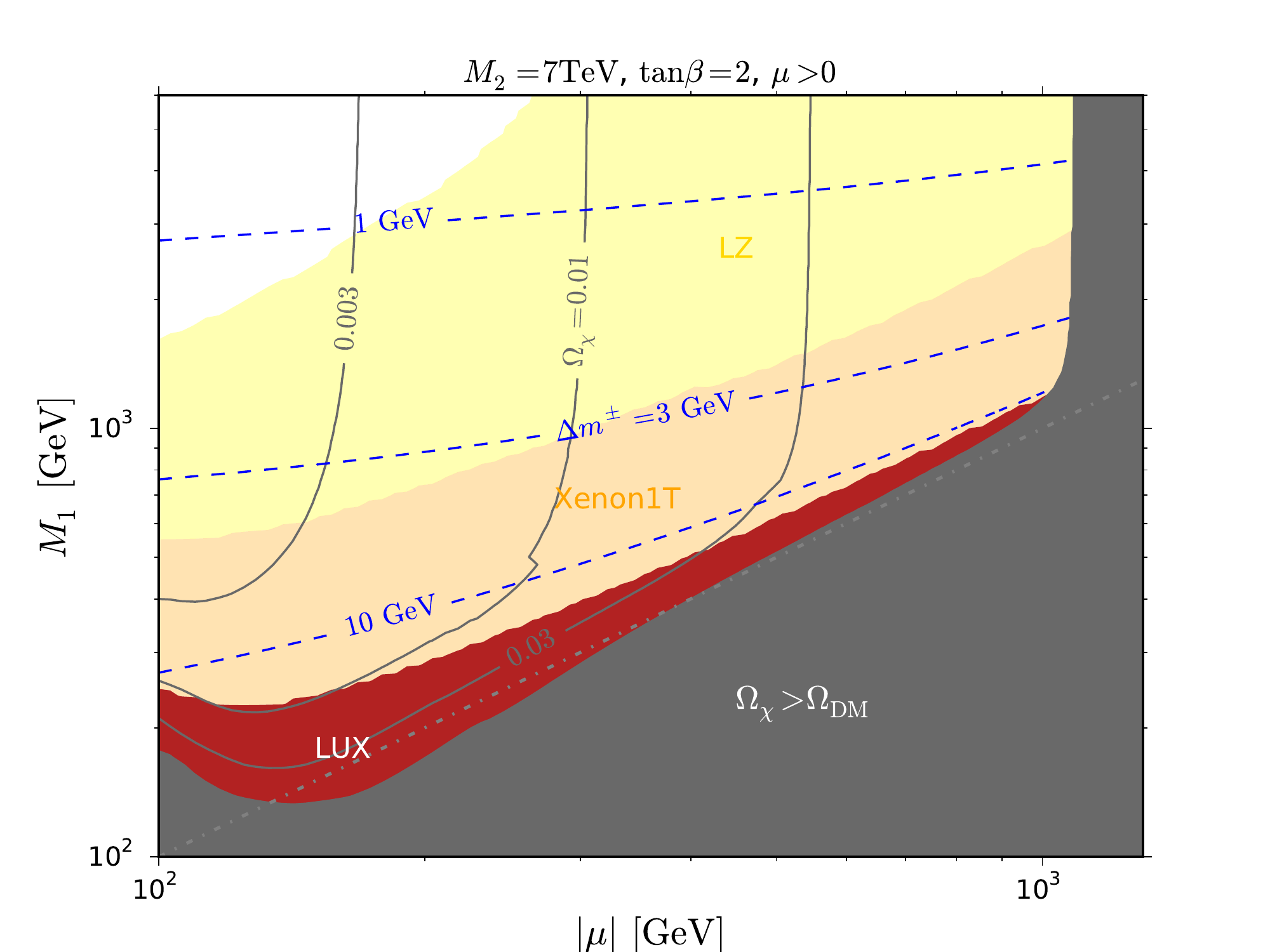}
		\includegraphics[width=0.48\textwidth]{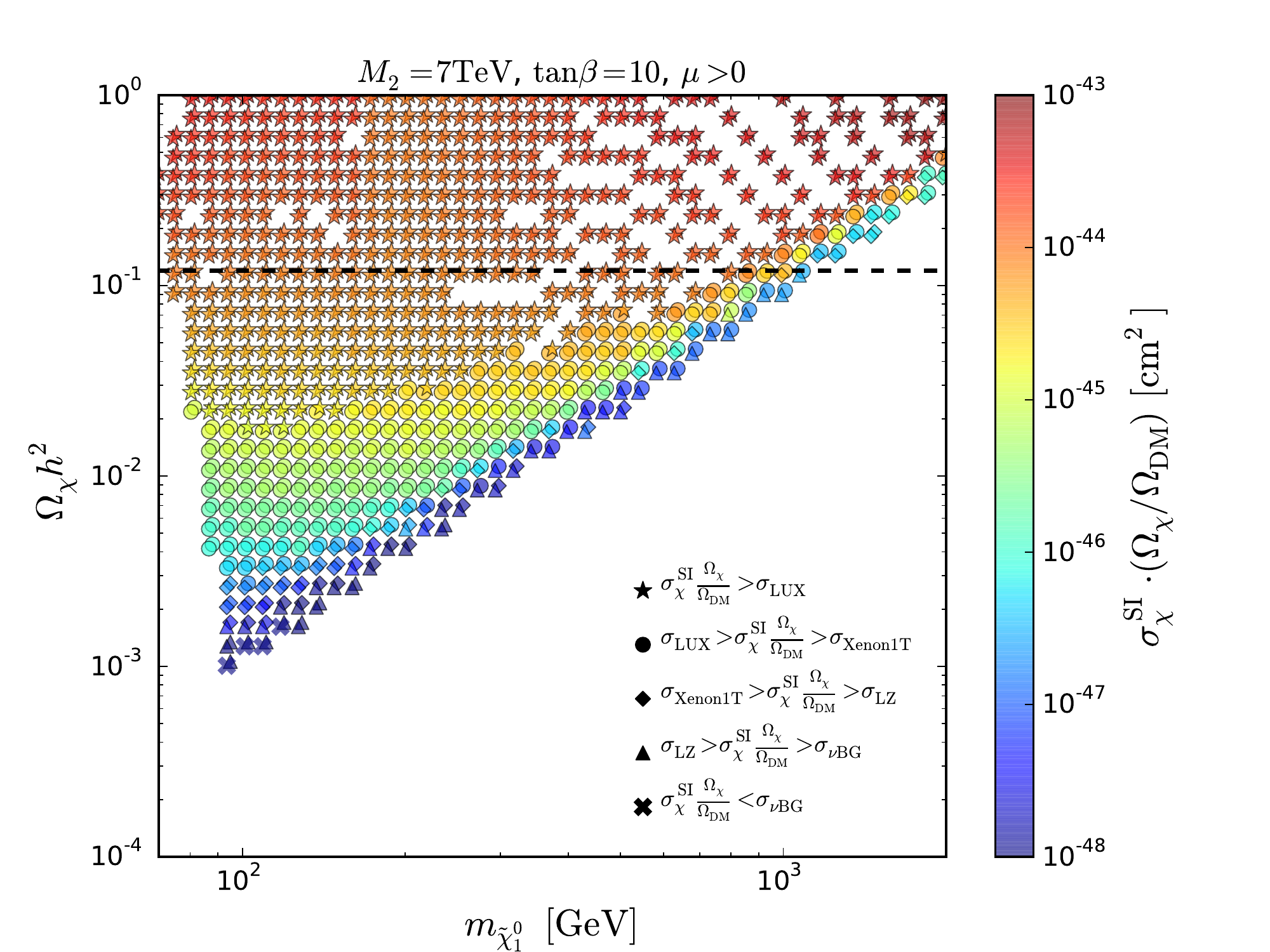}
		\includegraphics[width=0.48\textwidth]{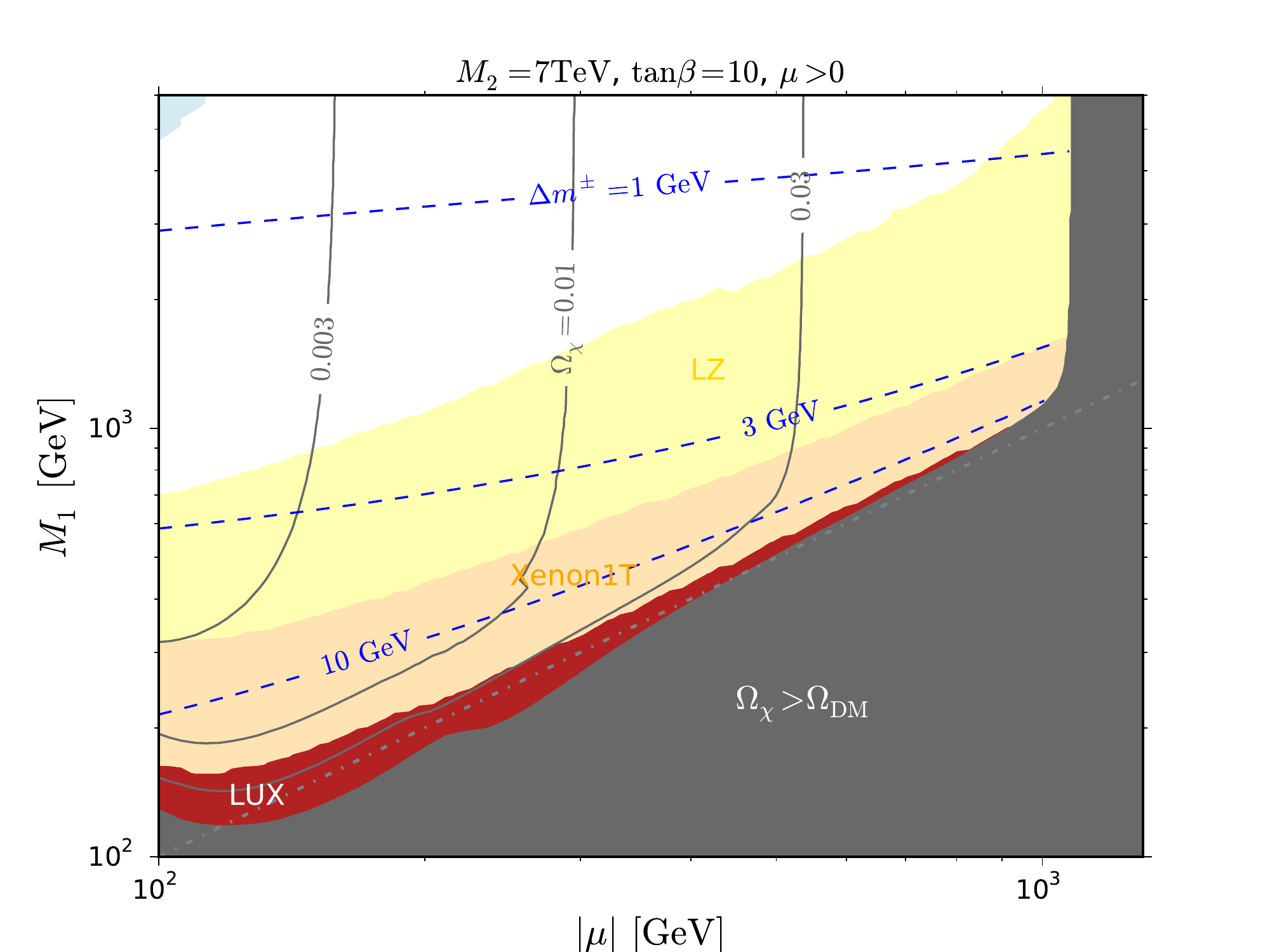}
	\includegraphics[width=0.48\textwidth]{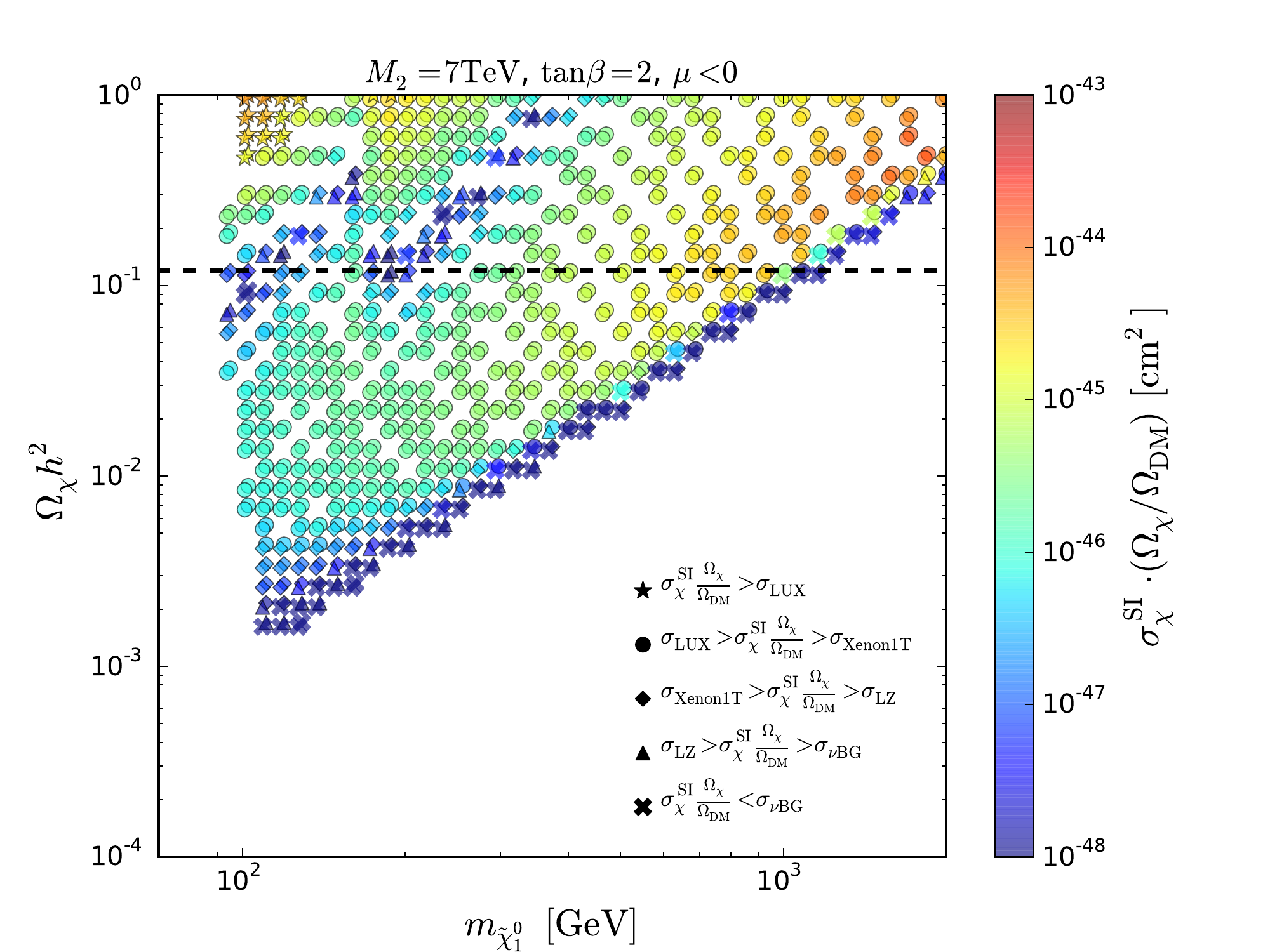}
	\includegraphics[width=0.48\textwidth]{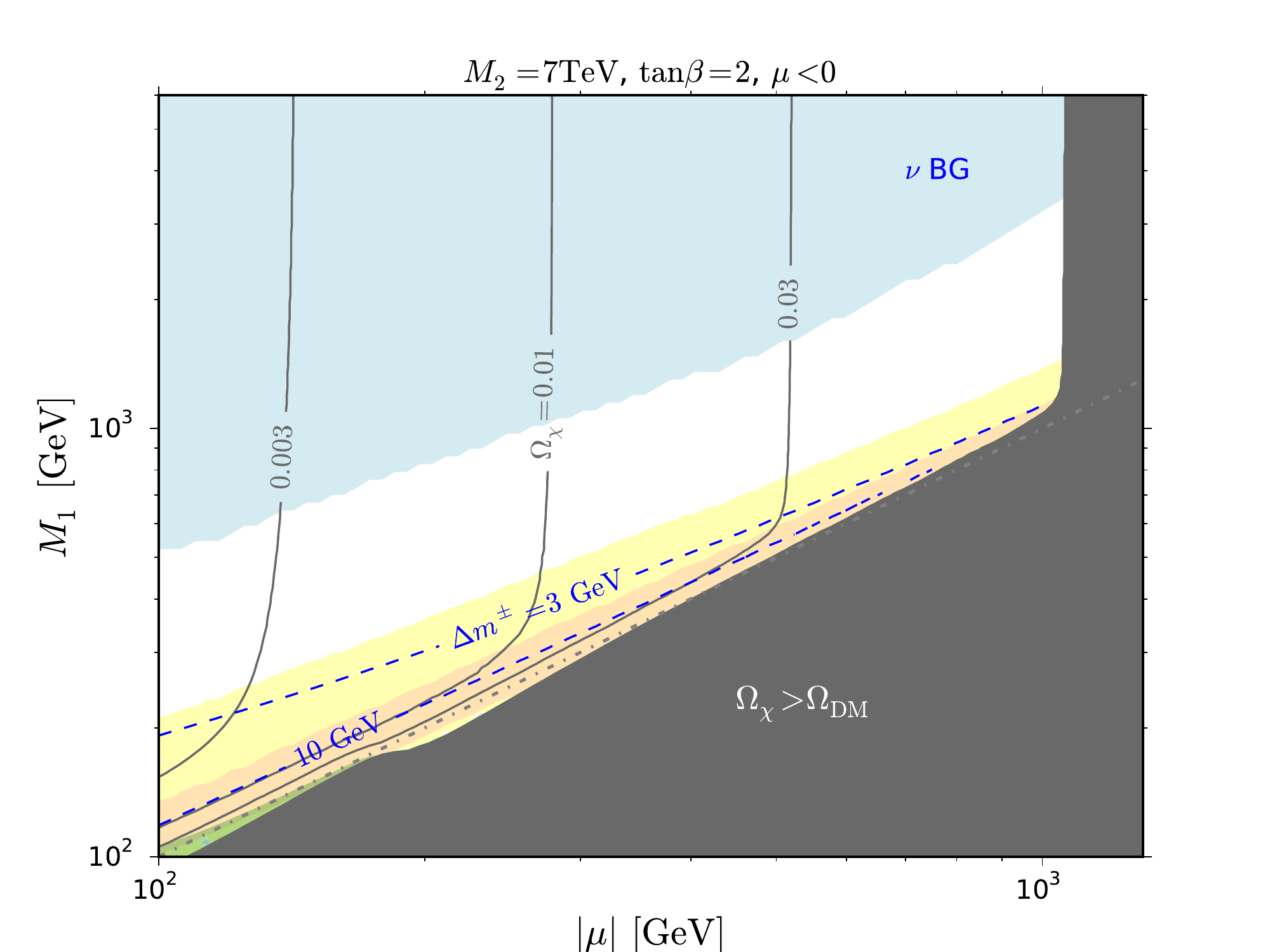}
\caption{Some results of the scan over $M_1$ and $\mu$ with winos decoupled ($M_2=7$  TeV).
In the plots on the right, the white region is not accessible in the XENON1T and LZ experiments but
the cross sections are above the neutrino background, and the light blue region is below that background. The red, orange and yellow regions are: excluded by LUX, accessible by XENON1T and accessible by LZ experiments, respectively.  The dashed-dotted lines mark $M_1=|\mu|$. All  other details of the plots are explained in the figures.
The small green region at $M_1 \sim |\mu| < 200$ GeV in the lower right plot is the excluded region by 
the ICECUBE experiment \cite{Aartsen:2012kia}. 
\label{fig:mu-m1_dec_1}}
\end{figure}

Thermally produced pure higgsinos overclose the universe  
for  $|\mu|\gtrsim 1$~TeV. 
In the left column of Fig.~\ref{fig:mu-m1_dec_1} we exemplify the  results of the scan over $\mu$
in the range $|\mu|=$ (100 GeV, 2 TeV) and $M_1$ in the range (100 GeV,  6 TeV),\footnote{
Since only relative signs of $M_1, M_2$ and $\mu$ are important, we choose $M_1>0$ as our convention. 
The lower bound of the scan over $|\mu|$ corresponds approximately to the lower experimental limit for the chargino mass.
For $M_1<100$ GeV, the LSP is bino-dominated and the bound $\Omega_{\chi} h^2\leq 0.12$ can be satisfied only for the LSP mass close to the half of the $Z$ or
$h$ mass, where the resonant annihilation takes place. Such scenario was studied in detail in Refs.\cite{Calibbi1,Calibbi2}.}
with $|M_2|=m_{\tilde f}=m_A=7$ TeV,\footnote{
    As we will see in Sec.~\ref{sec:concl}, our result is not sensitive to the precise value of the sfermion mass (7 TeV) in our scan.  Although we do not impose the Higgs mass constraint in our sample, one can tune the stop mass 
    to fit the lightest CP-even Higgs mass to the observed value without changing the result. 
} for $\Omega_\chi h^2$ 
versus the neutralino mass.
Throughout this paper we calculate the SUSY spectrum using {\tt SOFTSUSY} \cite{softsusy} and compute the $\Omega_\chi$ and the $\sigma_{\chi}^{\rm SI}$
using {\tt micrOMEGAs} \cite{Micromega}.    
We set the values of the  soft SUSY breaking parameters and the $\mu$ parameter at 7\,TeV.
In the plots we divide the $(m_{ \tilde \chi_1^0}, \Omega_{\chi})$ plane into ($45 \times 40$) bins and keep the maximum and minimum of the
scaled neutralino-nucleon cross section, $\sigma_\chi^{\rm SI} \cdot \Omega_{\chi} / \Omega_{\rm DM}$,
if there are more than one entries in the bin. 
For each bin we plot two points slightly displaced indicating these maximum and minimum values 
with the colour-coding that represents the value of 
$\sigma_\chi^{\rm SI} \cdot \Omega_\chi / \Omega_{\rm DM}$.
We use five different markers to classify the constraint/sensitivity on the cross section versus neutralino mass plane.
The scaled cross section $\sigma_\chi^{\rm SI} \cdot \Omega_\chi / \Omega_{\rm DM}$ is: 
larger than the LUX limit and excluded for stars,
allowed by LUX but within the reach of XENON1T for circles,
beyond the reach of XENON1T but within the reach of LZ for diamonds,
beyond the reach of LZ but above the neutrino background floor for triangles
and below the neutrino floor for crosses.\footnote{
        If the neutralino-nucleon scattering cross section becomes smaller than some threshold, so-called the neutrino floor,
        the signal is swamped by the overwhelming neutrino background and the traditional technique for the direct DM detection does not work efficiently.   
        However, this parameter region may be searched for by more creative technique, discussed e.g. in \cite{Grothaus:2014hja}.
}
We use the expected sensitivity of XENON1T corresponding to a three year run
from 2015 and
that of LZ also corresponding to a 3-year run with a projected start date of 2017.
The projected sensitivities are taken from \cite{Cushman:2013zza}.

In the scanned parameter range, there is  only a small admixture of
the wino component (although it is not totally decoupled). As it can be seen in those plots, the
minimal value of $\Omega_{\chi} h^2\approx 10^{-3}$ corresponds to $|\mu| =100 \gev$.
For a given value of
the neutralino mass, approximately equal $|\mu |$ in the whole range of $ \Omega_{\chi} h^2\leq 0.12$, 
the approximately pure higgsino  case, $M_1\gg|\mu|$,  corresponds to  the smallest values of $\Omega_\chi$, that is to the boundary of the scan.
The neutralino 
relic abundance increases with the increasing bino component, that is for a decreasing value of $M_1$, eventually reaching the observed value  $\Omega_{\rm DM} h^2 =0.12$.
There, $|\mu| \approx M_1$, which is called the well tempered bino/higgsino region \cite{welltempered}.

It is instructive to present the results of the scan in the ($|\mu|, M_1$) plane as in the right column of Fig.~\ref{fig:mu-m1_dec_1}. First  of all, we indeed see that
the bound $\Omega_{\rm DM} h^2=0.12$ corresponds to $M_1\approx|\mu|$  with $M_1<|\mu|$ and $M_1>|\mu|$ for low and close to 1 TeV values of
$|\mu|\approx m_{\tilde \chi_1^0}$, respectively. This is easily understandable as the effect of the $\mu$ dependence of the pure higgsino relic abundance,
eq.~\eqref{Omega_Bino}. Secondly, it is interesting to compare the contours of constant $\Omega_{\chi} h^2$ in the plots in the right column of Fig.~\ref{fig:mu-m1_dec_1}  with the results in the plots in the left column, remembering that $m_{\tilde \chi_1^0} \approx|\mu|$. A given value
of $\Omega_{\chi} h^2$ can be obtained only for a certain range of $m_{\tilde \chi_1^0} \approx|\mu|$  and this is clearly seen in both sets of plots. Furthermore, the bottom-up increase of $\Omega_{\chi} h^2$ with the
increasing bino component corresponds to the top down change of $M_1$ for fixed $\mu$.

Also in Fig.~\ref{fig:mu-m1_dec_1} 
there is given information about the magnitude of
the spin independent scattering cross section to the nucleons and the present exclusion regions and the regions accessible  in the future,  in the direct detection experiments.  Since the $h\chi\chi$   vertex vanishes for pure states, the tree level
$\sigma_\chi^{\rm SI}$
vanishes for pure higgsino. 
The loop  effects included in the {\tt micrOMEGAs} code and the residual mixing effects give non-zero but very small cross sections along the boundary of
the scan in the ($m_{{\tilde \chi}^0_1},\,\Omega_\chi h^2$) plane. 
For a given neutralino mass, the larger the value of $\Omega_\chi$ is, the larger the event rate in the spin independent scattering cross section becomes because of the flux factor 
$\frac{\Omega_{\chi}}{\Omega_{\rm DM}}$  but also because of the larger bino-higgsino mixing angle.
One can see the difference in the magnitude of the spin independent scattering cross sections  between small and large values of $\tan\beta$  and between the two
signs of $\mu$, especially for small $\tan\beta$.  They can be 
understood in terms of the formulae for the mass eigenstates and the bino-higgsino mixing angle, collected for the sake of easy reference in the Appendix.
Approximately,  the  $\sigma_\chi^{\rm SI}$   reads \cite{Hall_blindspots}:
\begin{equation}
\sigma_\chi^{\rm SI}=8\times 10^{-45} {\rm cm}^2 \left(\frac{c_{h\chi\chi}}{0.1}\right)^2
\label{eq:sigma}
\end{equation}
where the coupling  $c_{h\chi\chi}$ is defined as 
${\cal L} \sim\frac{1}{2}c_{h\chi\chi}(\chi\chi+\chi^{\dagger}\chi^{\dagger}) h $  
and in the limit $M_Z\sin\theta_W \ll ||M_1|-|\mu||$ reads:
\begin{equation}
\label{eq:c_mu>M1}
 c_{h\chi\chi} \approx \frac{g_1}{2} \sin\theta_W M_Z \frac{M_1 + \mu \sin\left(2\beta\right)}{\mu^2-M_1^2}
\end{equation}
for $|\mu|>|M_1|$ (this is the region almost totally excluded by the bound $\Omega_\chi h^2\leq 0.12$) and
\begin{equation}
\label{eq:c_mu<M1}
 c_{h\chi\chi} \approx \frac{g_1}{4} \sin\theta_W M_Z  \frac{1 + \sgn\left(M_1\mu\right) \sin\left(2\beta\right) }{M_1 - \sgn(M_1\mu)\mu}
\end{equation}
for $|\mu|<|M_1|$ 
(see eqs.~(\ref{mu-c_noM2}) and (\ref{M1-c}) as well as the 
corresponding comments in the Appendix). 
Moreover,  in the limit $|M_1|+|\mu| \gg M_Z\sin\theta_W \gg ||M_1|-|\mu||$, that is along the diagonal $|M_1|\approx |\mu|$,  one has:
\begin{equation}
\label{eq:diagonal}
 c_{h\chi\chi} \approx \frac{\sqrt{2}\,g_1}{4}\,\sgn(M_1)
\left[
\sqrt{1+\sgn(M_1\mu)\sin(2\beta)}
-\frac{1}{\sqrt{2}}\,\frac{M_Z \sin\theta_W}{|M_1|+|\mu|} (1-\sgn(M_1\mu)\sin(2\beta)) \right]
\end{equation}
(see eq.~(\ref{M1=mu-c}) and the following discussion in the Appendix).

Those formulae explain the behaviour of $\sigma_\chi^{\rm SI}$, its $\tan\beta$ dependence and  the dependence on the sign of $\mu$ for small $\tan\beta$.   
An important  effect is the existence of the so-called blind spots 
for the $h\chi\chi$ coupling,\footnote{
    Another type of blind spots involving 
    cancellation between the light and heavy Higgs exchanges has also been discussed e.g. in \cite{Baer:2006te,Feng:2010ef,Huang:2014xua}.
} 
that is the parameter regions where it is very small,  for small $\tan\beta$  and $\mu M_1<0$, as it is evident from the above equations \cite{Hall_blindspots}.

The main conclusion following from Fig.~\ref{fig:mu-m1_dec_1}  is that, in spite of the rescaling by the flux factor 
$\frac{\Omega_{\chi}}{\Omega_{\rm DM}}$, the whole range of $\Omega_\chi h^2$ values between $10^{-3}$ and 0.12 is reachable in the direct detection experiments. A large part of the ($\mu,\,M_1$) plane is either already ruled out by the present DD  experiments or will be tested by the XENON1T and LZ experiments or at least corresponds to the  spin independent cross sections above the neutrino background.  The  only exception is the large blind spot region for small $\tan\beta$  and $\mu<0$, not accessible in the DD experiments. We also  note that \eqref{eq:diagonal} gives a blind spot near the diagonal $|M_1|\approx|\mu|$ for small $\tan\beta$ and $\mu<0$. 

The results shown in Fig.~\ref{fig:mu-m1_dec_1}  also confirm the known results
\cite{welltempered, Hall_blindspots},  that the bino-higgsino LSP can still account for the  observed DM. For large $\tan\beta$ with
positive (negative) $\mu$
that would be a well-tempered bino-higgsino in a mass  range from 800 (700) up to 1000 GeV corresponding to a pure higgsino.   For small $\tan\beta$
and negative $\mu$ the whole range of masses between 100 and 1000 GeV is still allowed. The latter follows from the fact that for small $\tan\beta$
the well-tempered neutralino is close to the blind spot for the $h\chi\chi$ coupling. For small $\tan\beta$ and positive
$\mu$ the $h\chi\chi$ coupling is enhanced by the $\mu\sin(2\beta)$ term in \eqref{eq:c_mu<M1} so only masses very close to 1 TeV, corresponding essentially to pure higgsino LSP, are allowed.

From eq.~\eqref{eq:c_mu<M1}, one can see that  for $M_1\gg|\mu|$, $\sigma_\chi^{\rm SI}\approx\frac{1}{M^2_1}$.
The exclusion/accessibility regions  depend  on the experimental limits which are  in addition a function of the LSP mass (weaker bounds for heavier LSP).  The
interplay of both factors explains the  pattern observed in  the right column of Fig.~\ref{fig:mu-m1_dec_1}.   

We also check the indirect DM detection experiments in our parameter scan.
The indirect detection bound is usually not as strong as that for direct detections.   
However, it can effectively constrain the spin-dependent scattering cross section.
Since the structures of spin-dependent and spin-independent cross sections are not correlated   to each other, the spin-dependent bound may become superior in the blind spot region of the spin-independent cross section.

One of the most stringent bounds on the spin-dependent cross section comes from the ICECUBE experiment.
ICECUBE looks for the neutrinos coming from the annihilation of the neutralinos captured by the sun.
When the capture and annihilation reach the equilibrium, the total annihilation of the neutralinos is fixed
by the capture rate, which is proportional mainly to the spin-dependent cross section 
and the local neutralino density on the sun's trajectory. 
Therefore, if the neutralino annihilation produces neutrinos (e.g. $\tilde \chi_1^0 \tilde \chi_1^0 \to W^+W^-$), 
the limit on the high energy neutrino flux from the sun
can be translated to the limit on the spin-dependent cross section times the local neutralino density 
\cite{Jungman:1995df}.\footnote{
    We thank Marc Kamionkowski for discussion on this point.
}        

We have checked the spin-dependent cross section at each parameter point of our 
scan and confronted it with the ICECUBE spin-dependent bound  
shown in Fig.2 of ref.~\cite{Aartsen:2012kia} (obtained
assuming\footnote{
We have checked that this is indeed the dominant annihilation 
process in the majority of the parameter region of our scan.}
$\tilde \chi_1^0 \tilde \chi_1^0 \to W^+W^-$).
In ref.~\cite{Aartsen:2012kia}
the limits on the capture rate are converted into the limits on  the spin 
dependent cross section assuming the standard local dark matter density of 
0.3 GeV/${\rm cm}^3$. For smaller DM density the bounds for spin dependent 
cross section  given in that Fig.2 of ref.~\cite{Aartsen:2012kia}  have to 
be rescaled by $\Omega_\chi/\Omega_{DM}$ and are, of course, weaker. 
Another important point for understanding our results is the 
dependence of the SD scattering cross section on the LSP composition.
The SD cross section is dominated by the $Z$ boson exchange and the 
$Z\chi\chi$ coupling is proportional to $(N_{13}^2-N_{14}^2)$.
This coupling vanishes for any of the pure neutralino states 
(there is no direct $Z$-wino-wino or $Z$-bino-bino coupling while 
in the case of a pure higgsino this follows from the equality 
$N_{13}^2=N_{14}^2=\frac12$). 
For the higgsino-dominated LSP its coupling to $Z$, given by 
eq.~(\ref{mu-cZ}), is non-zero and suppressed by $M_1$ and $M_2$. 
Thus, the smaller $M_1$ and/or $M_2$ (i.e.\ the larger the wino and/or 
bino component), the larger the $Z\chi\chi$ coupling and the SD 
cross section, up to the point where the higgsino component becomes very 
small (in this limit the $Z\chi\chi$ coupling is given by 
eqs.~(\ref{M1M2-cZB}) and (\ref{M1M2-cZW})).
Thus, the $Z\chi\chi$ coupling becomes sizable for bino-higgsino or  
wino-higgsino mixed LSP (see also eq.~(\ref{M1=mu-cZ})) but in the second 
case the neutralino relic abundance is smaller and after the rescaling 
the bounds on the SD cross section are weaker.  
This is why in our plots the ICECUBE bounds are relevant only 
for the bino-higgsino mixed LSP. 
We found that the ICECUBE spin-dependent constraint is less stringent 
than the LUX spin-independent constraint in most of the parameter region.
We only found small excluded regions by ICECUBE around the 
$M_1 \sim |\mu| < 200$~GeV in the ($M_1, \mu$) plane with 
$\tan\beta = 2$, $\mu < 0$. This can be understood from eq.~(\ref{M1=mu-cZ}). 
The leading term of the $Z\chi\chi$ coupling is suppressed by 
$|M_1|+|\mu|$ and enhanced when $M_1\mu<0$, especially for small 
values of $\tan\beta$. Those regions happen to be the ones
where the blind spot cancellation in the spin-independent cross 
section is operative. They can be seen as the green regions 
in the lower right plot of Fig.~\ref{fig:mu-m1_dec_1} and the upper 
left plot of Fig.~\ref{fig:mu-m1_5}.  We have checked that they are the only regions 
where the ICECUBE constraint is more stringent than the LUX one 
throughout our analysis presented in this paper.

\begin{figure}
	\centering \vspace{-0.0cm}
		\includegraphics[width=0.48\textwidth]{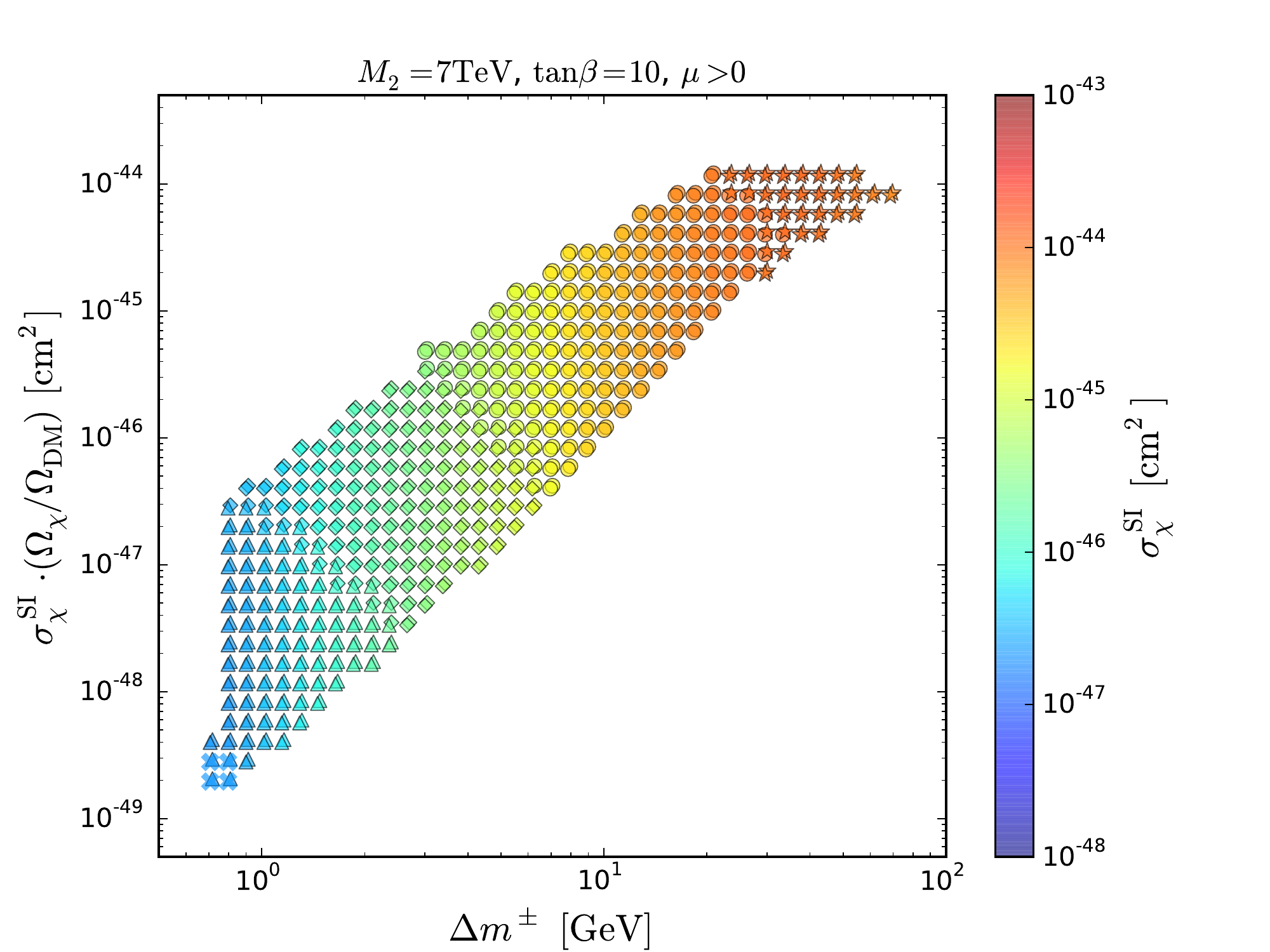}	
\caption{
The spin independent scattering cross section renormalized  by the proper flux factor  as a function of the NLSP-LSP mass difference. The  meaning of different marks is the same as in Fig.~\ref{fig:mu-m1_dec_1}
\label{fig:mu-m1_dec_4}}
\end{figure}

\begin{figure}[h]
	\centering \vspace{-0.0cm}
		\includegraphics[width=0.48\textwidth]{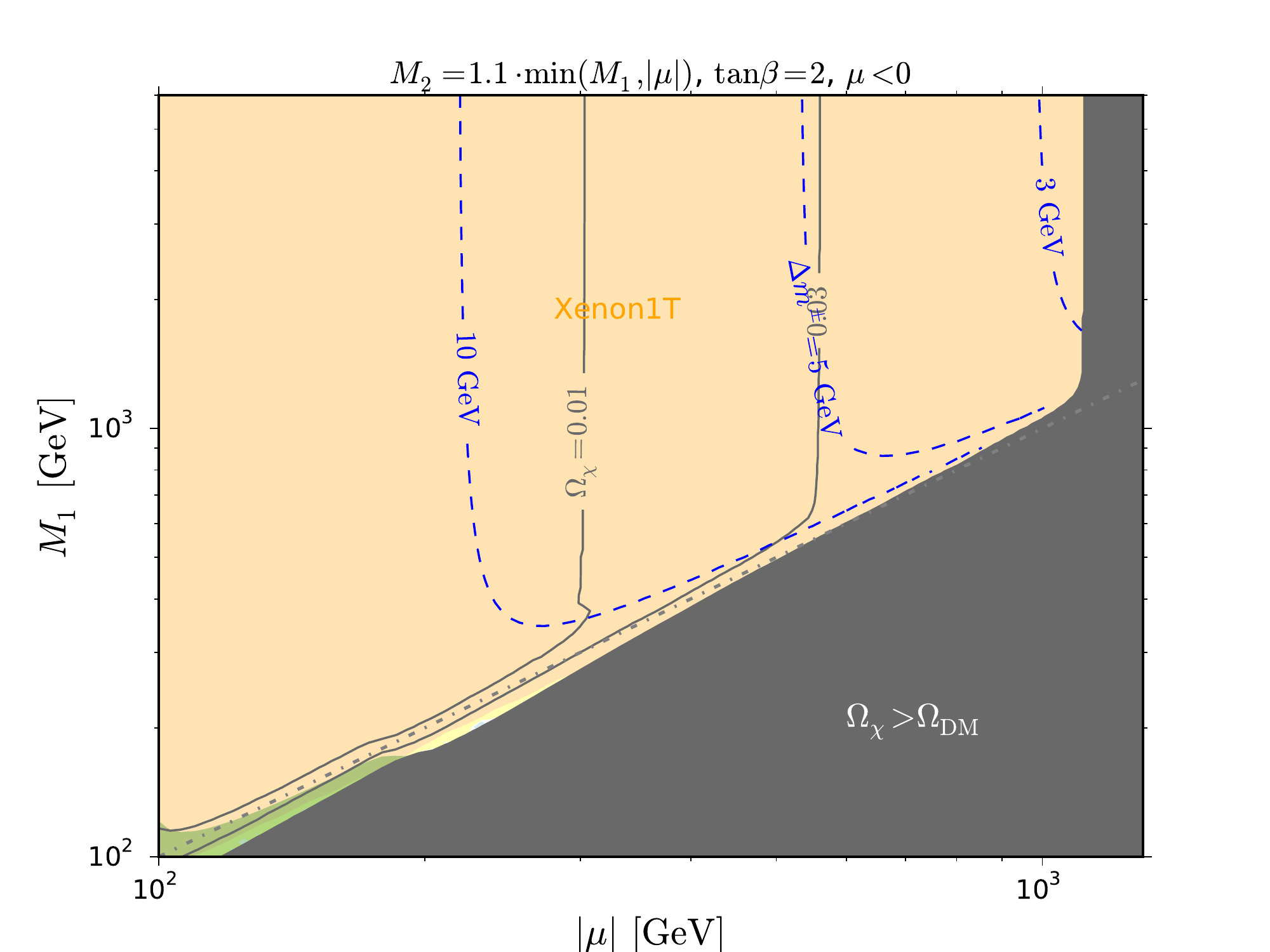}
		\includegraphics[width=0.48\textwidth]{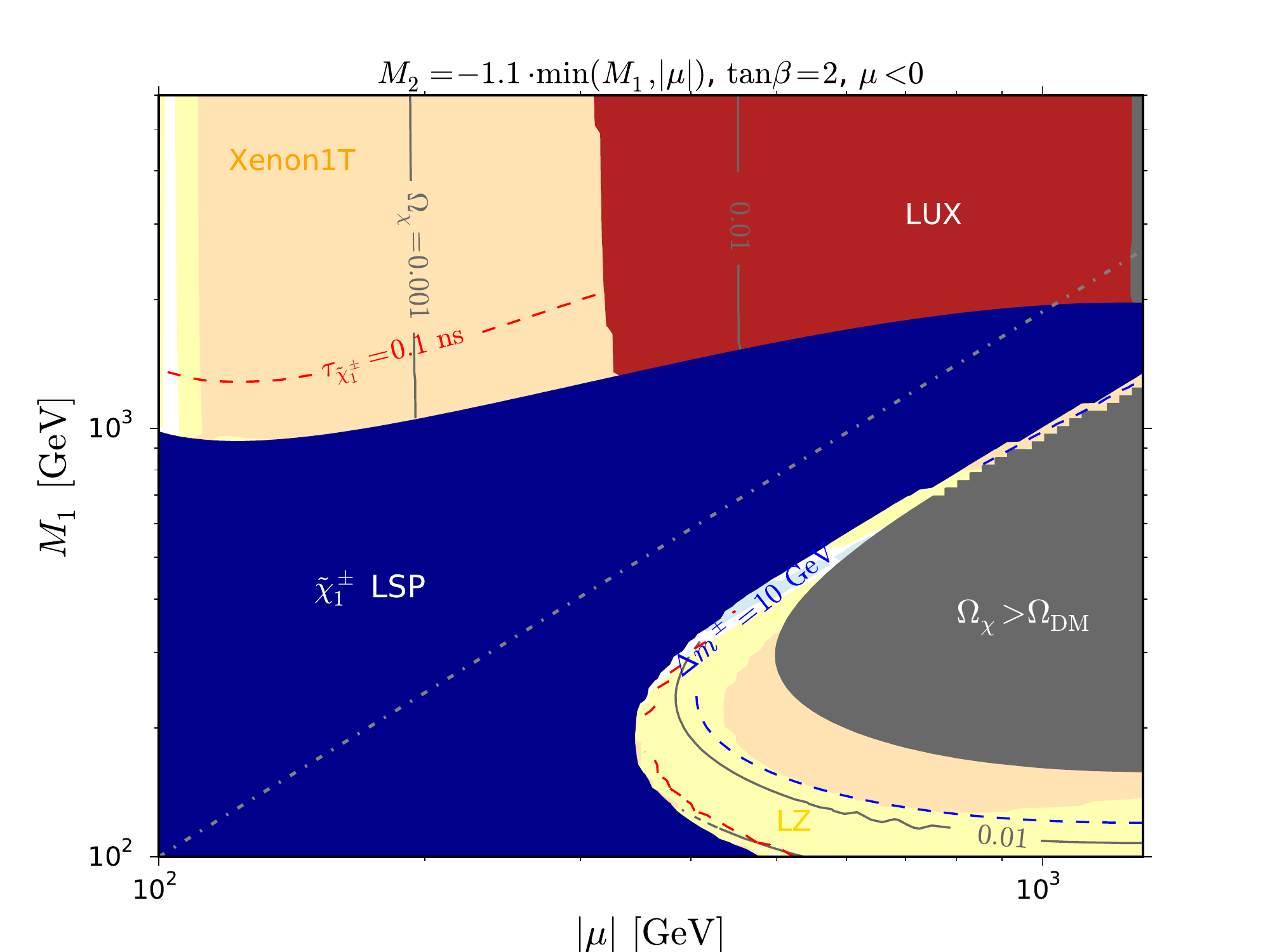}
               \includegraphics[width=0.48\textwidth]{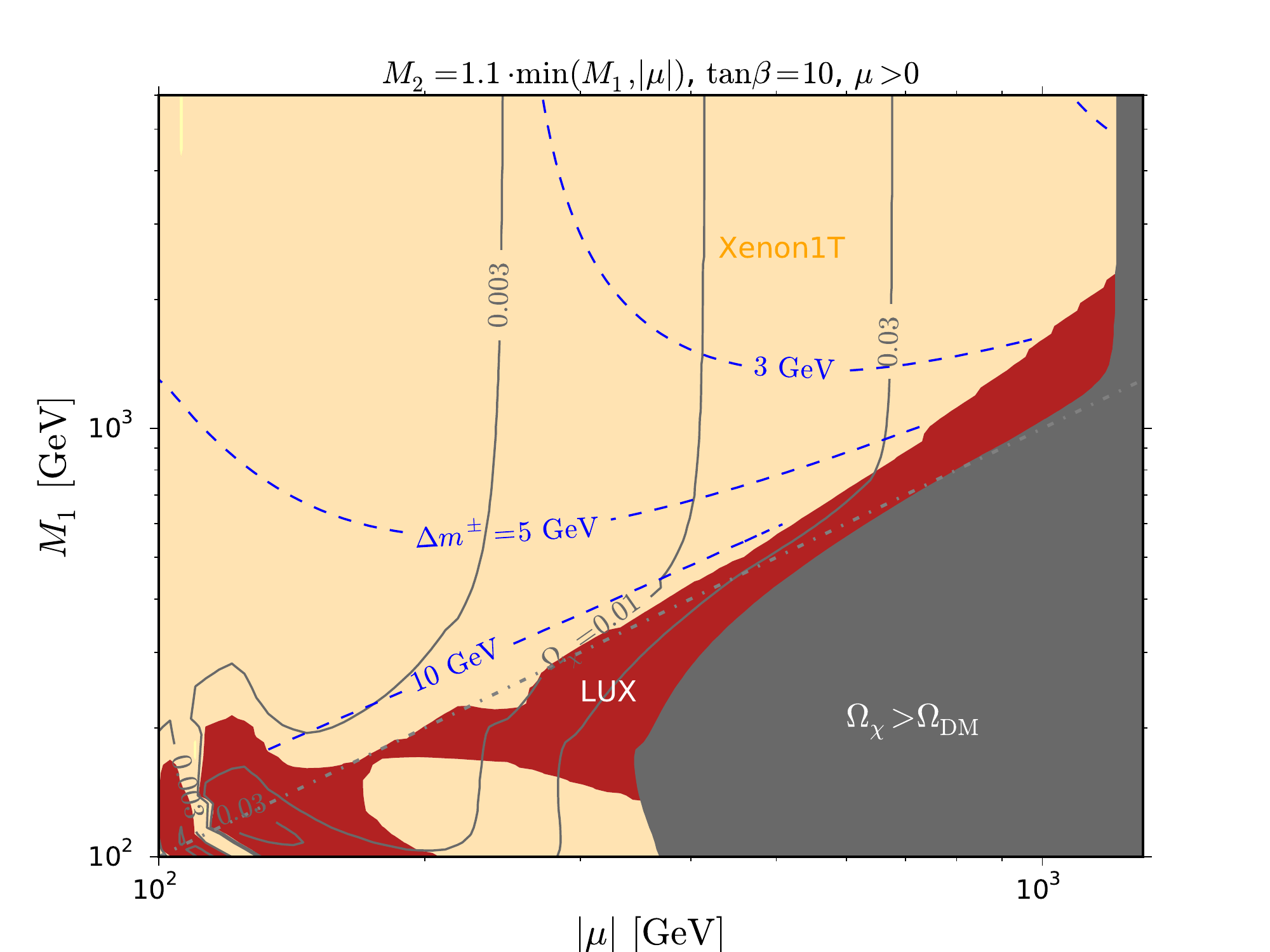}
                \includegraphics[width=0.48\textwidth]{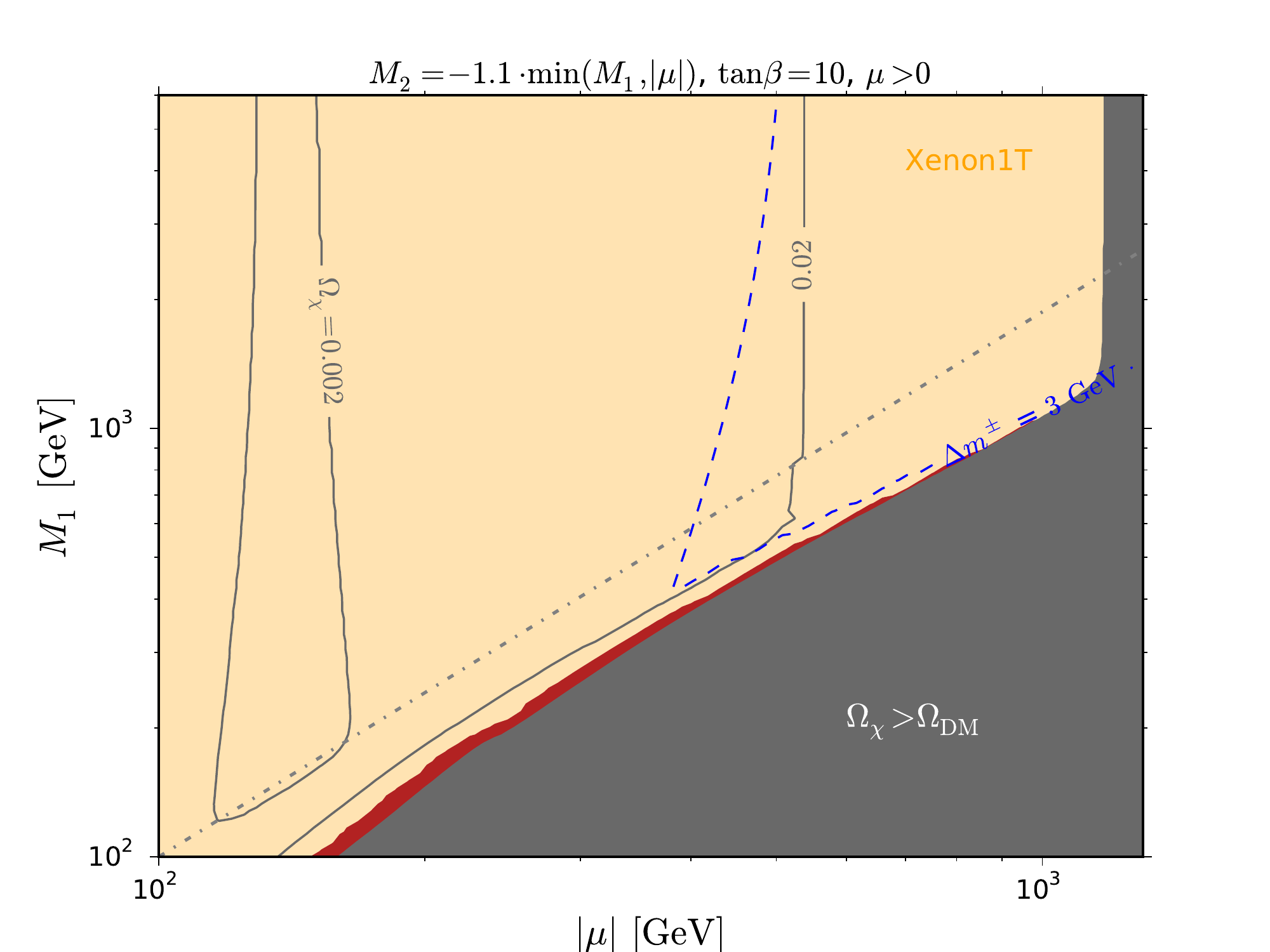}
\caption{
Same as the plots on the right of Fig.~\ref{fig:mu-m1_dec_1} but for non-decoupled winos.  The values of $M_2$ are marked on the plots. The NLSP life-time is
calculated following \cite{ChenDreesGunion} with the help of {\tt SUSYHIT} program \cite{Susyhit}. In the dark blue region the lighter chargino is the LSP.
The small green region at $M_1 \sim |\mu| < 200$ GeV in the upper left plot is the excluded region by 
the ICECUBE experiment \cite{Aartsen:2012kia}. 
\label{fig:mu-m1_5}}
\end{figure}

For collider searches, an important information is the mass difference between the NLSP  (usually  chargino) and the LSP as well as between $\chi^0_2$ and the
LSP, and also the life time of the NLSP.  
In Fig.~\ref{fig:mu-m1_dec_1}  we show the contours of $\Delta m^{\pm} \equiv m_{\tilde \chi_1^\pm}- m_{\tilde \chi_1^0}$ in the scanned region.
In Fig.~\ref{fig:mu-m1_dec_4} we plot the points with $\Omega_{\chi} h^2 < 0.12$ from our scans in the ($\Delta m^{\pm}$, $\sigma_\chi^{\rm SI} \frac{\Omega_{\chi}}{\Omega_{\rm DM}}$) plane for $\tan\beta = 10$ and $\mu > 0$.  
The colour-coding indicates $\sigma_\chi^{\rm SI}$ and we use the same markers as used in the left column of Fig.~\ref{fig:mu-m1_dec_1} to classify 
the DD constraint/sensitivity. 
One can see that after imposing the LUX bound the mass differences $\Delta m^{\pm}$ are in the range (1-30) GeV and the expected sensitivity of the XENON1T
 will bring them down to (1-10) GeV.
This is the general  result for most of the parameter space, except for small $\tan\beta$ and $\mu<0$ region where  the mass difference will remain to be up to 40 GeV even after the XENON1T and LZ  results.


The results can
be qualitatively
understood from eq.~\eqref{mu-mC1N1}  in  the Appendix. 
We see that smaller the spin independent cross section is,  smaller  the mass difference between the NLSP and LSP is.    For $M_1\gg|\mu|$, the former decreases like $\frac{1}{M_1^2}$  and the latter as $\frac{1}{M_1}$.  We note,  however,   that the  decoupling
of the  wino and bino components in the formulae for the gaugino masses and mass differences is slow, only linear in $M_2$ and/or $M_1$. 
Even for $M_1=M_2=$7 TeV one is still far from the pure higgsino limit.
For pure  higgsino states, the tree level mass difference between the lighter chargino and the LSP neutralino is zero. The loop
effects give the mass difference around 350 MeV \cite{Wells_higgsinosplitting} \footnote{For the higgsino mass around 100 GeV the chargino-LSP mass splitting
is somewhat smaller, around 250 MeV, due to corrections of order ${\mathcal{O}(m_Z^2/m_{LSP}^2)}$. } whereas in Fig.~\ref{fig:mu-m1_dec_1} it is at least
twice that large. In the parameter range of Fig.~\ref{fig:mu-m1_dec_1}, the chargino life time $c\tau$ is shorter than $10^{-1}$ cm, where the upper
bound corresponds
to the smallest mass difference. The mass difference between $\tilde \chi^0_2$ and the LSP is about twice as large as between $\tilde \chi_1^+$  and the LSP (see
eq.~\eqref{mu-mN2N1}). The implications of all those facts for collider searches are discussed in the next subsection.

With the discussed above patterns for large $|M_2|$ (small wino component) as a reference frame, it is interesting to consider how those patterns change when the
value of $|M_2|$ decreases and approaches the values of $|\mu|$ and/or $M_1$. 
In Fig.~\ref{fig:mu-m1_5} we show some generic examples of the results of the scan over $M_1$ and $\mu$, with $M_2=\pm 1.1  \min(M_1,|\mu|)$.  The plots for both signs of $M_2$  and $\mu$ flipped are similar to those shown in Fig.~\ref{fig:mu-m1_5}. The blue region  in the plot for $\tan\beta=2, \mu<0$
is excluded because  the lighter chargino is the LSP and after the change of signs of $M_2$ and $\mu$ similar region is excluded by the LUX bound. 
The main effect is that  
in the $M_1<|\mu|$ region the 10\% degeneracy $M_2=\pm 1.1  M_1$ introduces some wino component in the LSP.
The large wino annihilation cross section ensures  that even  with   a  small admixture  of winos,
the previously excluded regions (see Fig.~\ref{fig:mu-m1_dec_1}) are now partially allowed.  
This  effect is weaker for  opposite  signs of $M_2$  and $\mu$ since the admixture of wino is then  smaller.

Another very important effect of a larger wino component in the LSP are  generically larger  values
of the spin independent scattering cross sections  (the blind spots disappear),  so that   neutralinos
with the relic abundance in the whole range of $\Omega_{\chi} h^2$ between $10^{-3}$ and 0.12 are mostly within the reach of XENON1T.
Finally,  the $\Delta m^{\pm}$ remains  below  ${\mathcal O}(10)$ GeV  and  $\Delta m^0 \equiv m_{\tilde \chi_2^0}-m_{\tilde \chi_1^0}\approx 2\Delta m^{\pm}$ 
(for the higgsino-dominated LSP this can be easily seen 
from eqs.~(\ref{mu-mN2N1}) and (\ref{mu-mC1N1}) in the Appendix).

%%%%%%%%%%%%%%%%%%%%%%%%%%%%%%%%%%%%%%%%%%%%%%%%%%%%%%%%%%%%%%%%%%%%%%
%%%%%%%%%%%%%%%%%%%%%%%%%%%%%%%%%%%%%%%%%%%%%%%%%%%%%%%%%%%%%%%%%%%%%%
\subsection{Collider}
%%%%%%%%%%%%%%%%%%%%%%%%%%%%%%%%%%%%%%%%%%%%%%%%%%%%%%%%%%%%%%%%%%%%%%
%%%%%%%%%%%%%%%%%%%%%%%%%%%%%%%%%%%%%%%%%%%%%%%%%%%%%%%%%%%%%%%%%%%%%%

We have shown that in the bino-higgsino scenario the region where $|\mu| \gg M_1$ is excluded by the 
overproduction of the thermal DM or the direct DM detection limit given by the LUX experiment.
The conventional technique for the collider search is to focus on the production of heavier states and look for the energetic particles and missing energy
originated from the decays of the produced particles.
This strategy however does not work very efficiently in the phenomenologically allowed region, $M_1 > |\mu|$ and $|\mu| \gsim M_1$.
In the former case the production cross section of the heavier state namely the bino-like state is too small if squarks are decoupled. 
In the latter case the mass difference between the higgsino-like states and the bino-like LSP is small and
the sensitivity of the conventional searches is degraded because of the softness of the decay products of the higgsino-like states, although some alternative channels based on radiative photon decays may work \cite{Bramante_neutralino}.
The same is also true for the $M_1 > |\mu|$ case for the events with production of higgsino-like states, since higgsino-like states are almost mass degenerate and
the decays among these states only produce very soft particles. 

\begin{figure}
	\centering \vspace{-0.0cm}
		\includegraphics[width=0.8\textwidth]{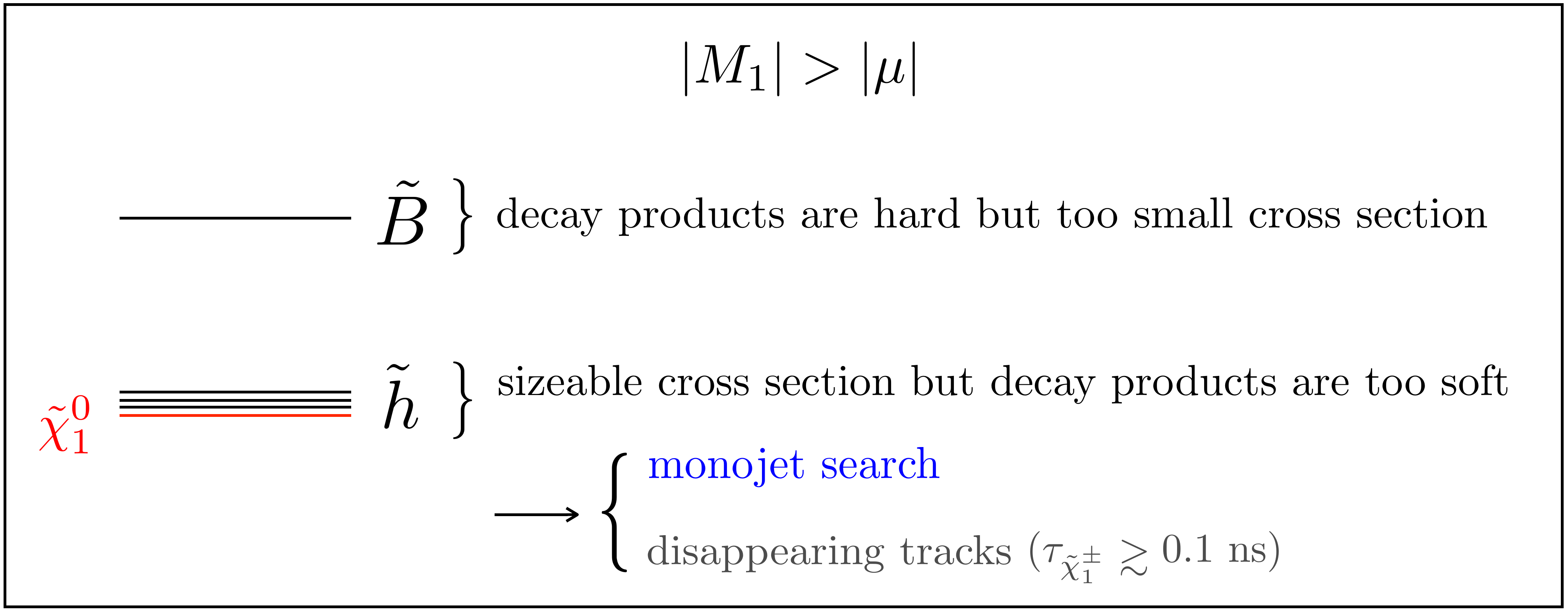}
\caption{ Strategy for collider search. 
\label{fig:mu-m1_schematic}}
\end{figure}
\begin{figure}
	\centering \vspace{-0.0cm}
		\includegraphics[width=0.48\textwidth]{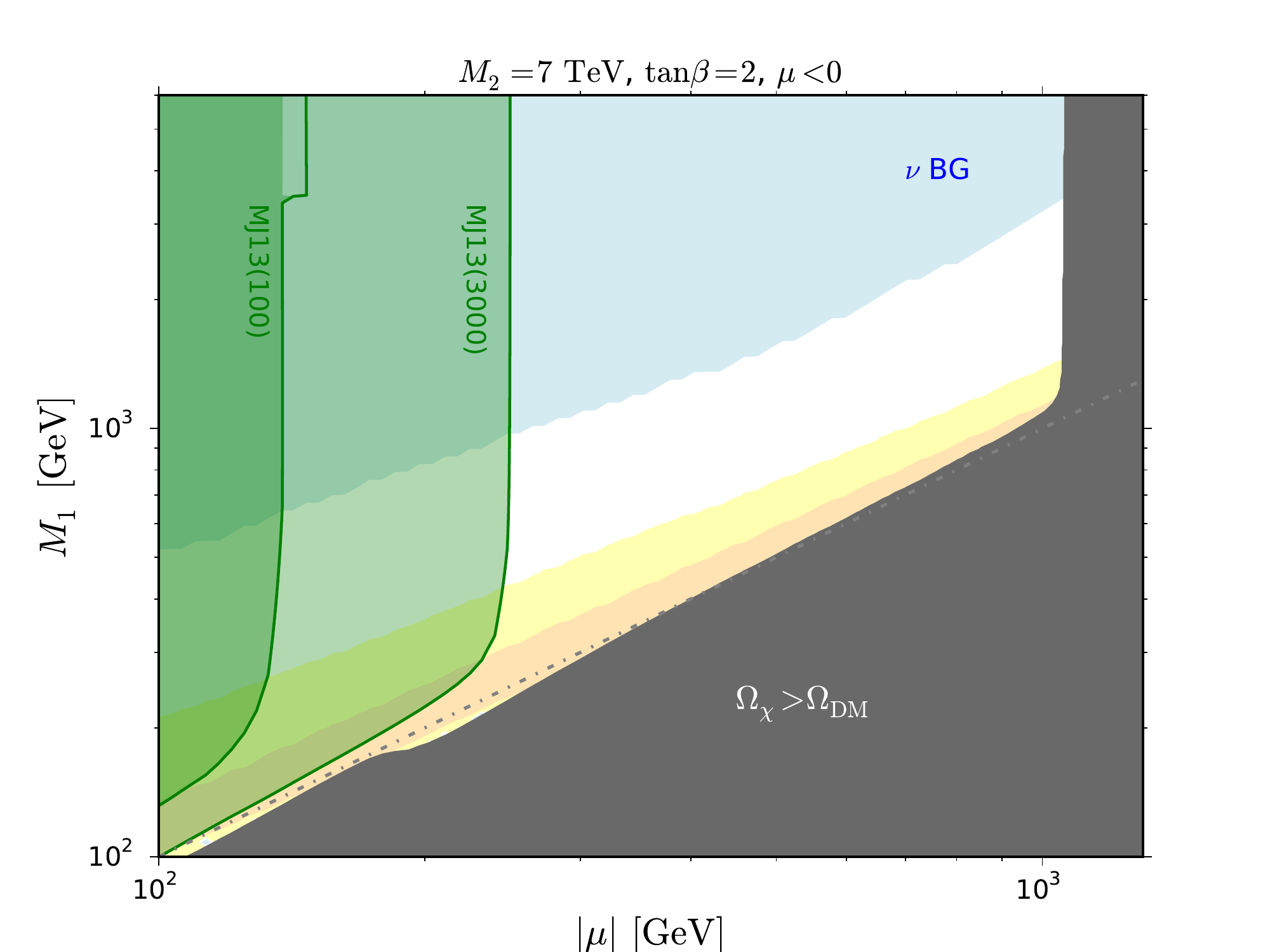}
                 \includegraphics[width=0.48\textwidth]{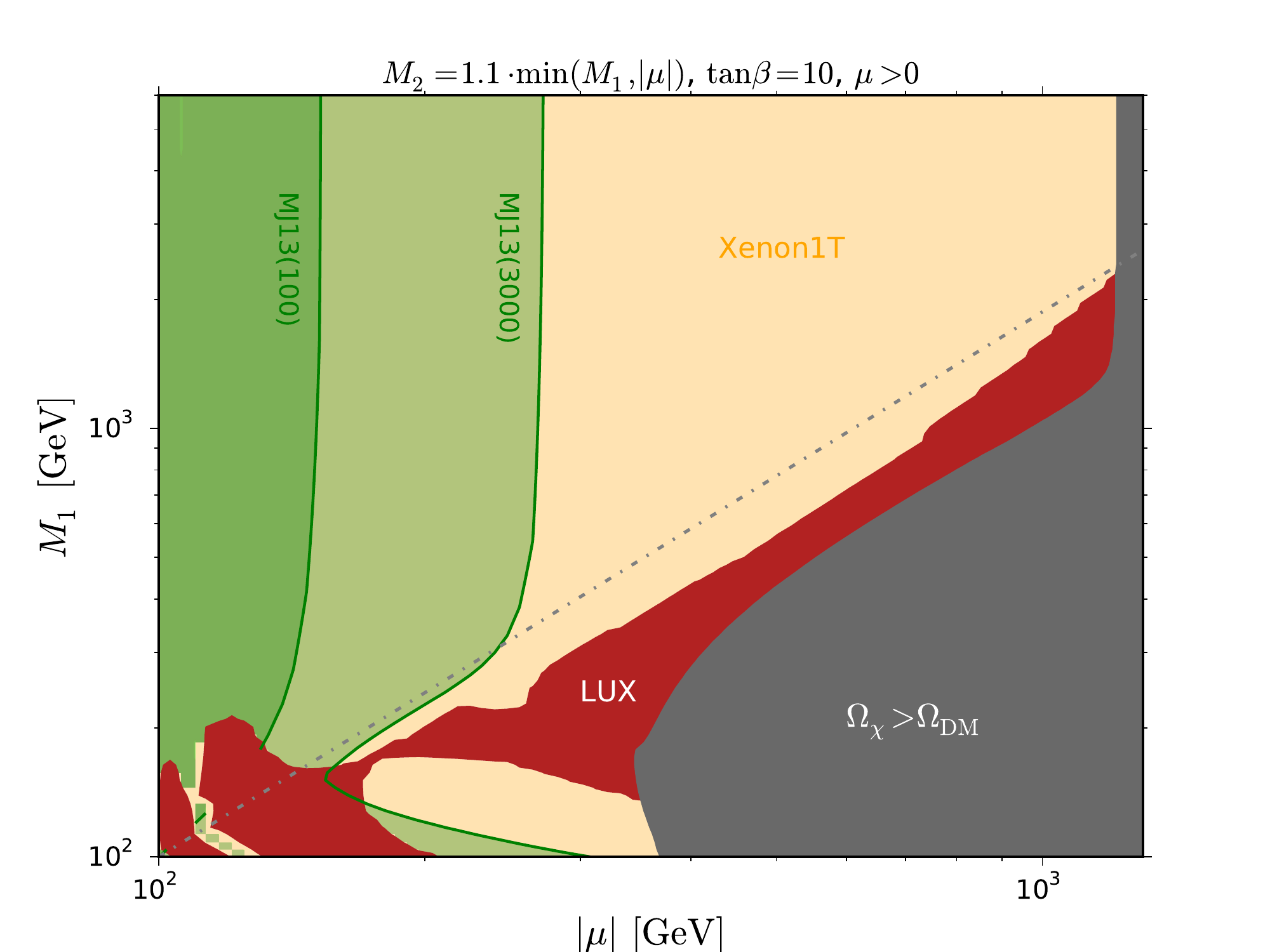}
\caption{The estimated LHC discovery limits for decoupled winos 
\label{fig:mu-m1_dec_6}}
\end{figure}

The production of such quasi mass degenerate particles can nevertheless be detected if the mass difference between 
the charged state and the LSP is small enough so that the charged particle has a collider scale lifetime ($\gsim 0.1$ ns).
Such long-lived charginos may leave the distinctive disappearing track signature in the trackers.
We however found that higgsino-like states are almost always short lived in our scan except for the vicinity of the chargino LSP region 
shown in the up-right plot of Fig.~\ref{fig:mu-m1_5}.

Another way of detecting quasi degenerate particles is to look for events where these particles are produced associated with hard 
initial state radiation.
These events are sensitive to the monojet search which requires large missing energy recoiling against one or two energetic jets 
(See Fig.~\ref{fig:mu-m1_schematic}.).
In refs.~\cite{SchwallerZurita, baer, Han:2013usa, Barducci} the sensitivity of the monojet search to the higgsino-like LSP scenario has been studied.
The results found in these literature vary because their simulation set up and selection cuts are different. 
The authors of \cite{Barducci} found the most optimistic results, 
albeit with very aggressive cuts.
We use their result because they presented the expected sensitivities at the 13 TeV LHC 
as functions of $m_{\tilde \chi_1^0}$ and $\Delta m^\pm$,
which enables us to translate their limit on our parameter plane.
We project their 2-$\sigma$ regions into our ($|\mu|, M_1$) parameter plane in Fig.~\ref{fig:mu-m1_dec_6}, where
the darker (lighter) green region corresponds to the 100 (3000) fb$^{-1}$ integrated luminosity. 
The monojet search is sensitive only for the $\Delta m^\pm < 20 - 30$ GeV.  
Therefore the sensitivity dies off in the $|\mu| \simeq M_1$ region because $\Delta m^\pm$ becomes too large.
The LHC sensitivity (green) regions are similar for other combinations of the relative signs of the parameters
and all values of $\tan\beta$. They depend very weakly on the value on $M_2$.
We can see some complementarity between the DM direct detection experiments and the collider experiments, especially for the $M_2$ decoupled case  in Fig.~\ref{fig:mu-m1_dec_6}. For negative
values of $\mu$, there remain, unfortunately, some regions  of the parameters not accesible neither at the LHC nor in the DD experiments.

%%%%%%%%%%%%%%%%%%%%%%%%%%%%%%%%%%%%%%%%%%%%%%%%%%%%%%%%%%%%%%%%%%%%%%
%%%%%%%%%%%%%%%%%%%%%%%%%%%%%%%%%%%%%%%%%%%%%%%%%%%%%%%%%%%%%%%%%%%%%%

\section{Wino-Higgsino}

%%%%%%%%%%%%%%%%%%%%%%%%%%%%%%%%%%%%%%%%%%%%%%%%%%%%%%%%%%%%%%%%%%%%%%
%%%%%%%%%%%%%%%%%%%%%%%%%%%%%%%%%%%%%%%%%%%%%%%%%%%%%%%%%%%%%%%%%%%%%%

The above discussion can be easily extended to neutralinos that are mainly a mixture of higgsino and wino (we take $M_1=7$ TeV), which are less explored in the
literature (see, however, \cite{Raby,cohen,Neutralino_Wang,diCortona}).
The annihilation cross section for pure wino ${\tilde W}$  is determined by the vertex $\tilde W^0 \tilde W^\pm W^\mp$.
One has approximately \cite{welltempered}

\begin{equation}
\Omega_{\tilde W} h^2=0.13\left(\frac{M_2}{\mbox{2.5 TeV}}\right)^2 \,.
\end{equation}

\begin{figure}
	\centering \vspace{-0.0cm}
               \includegraphics[width=0.48\textwidth]{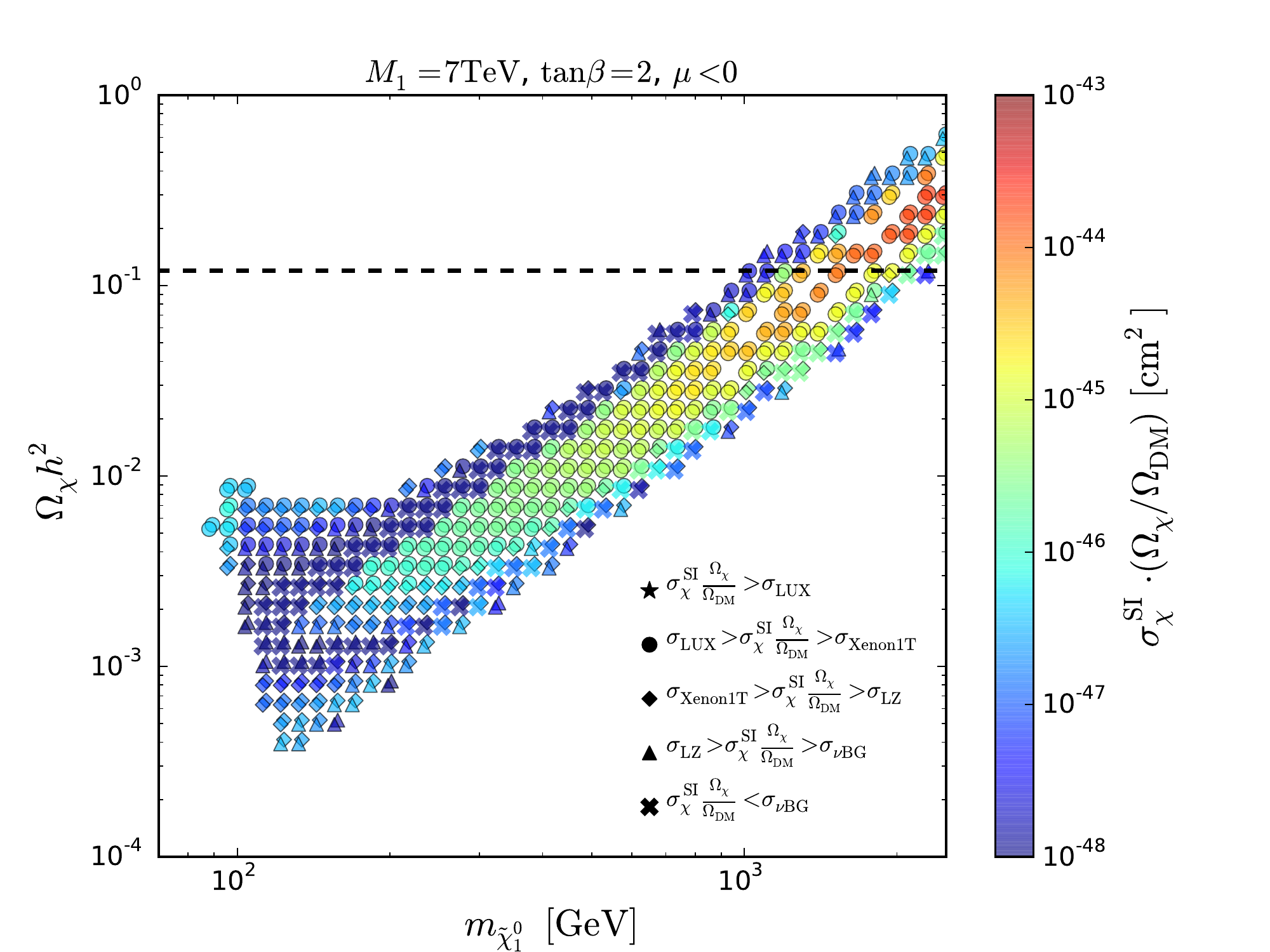}	
               \includegraphics[width=0.48\textwidth]{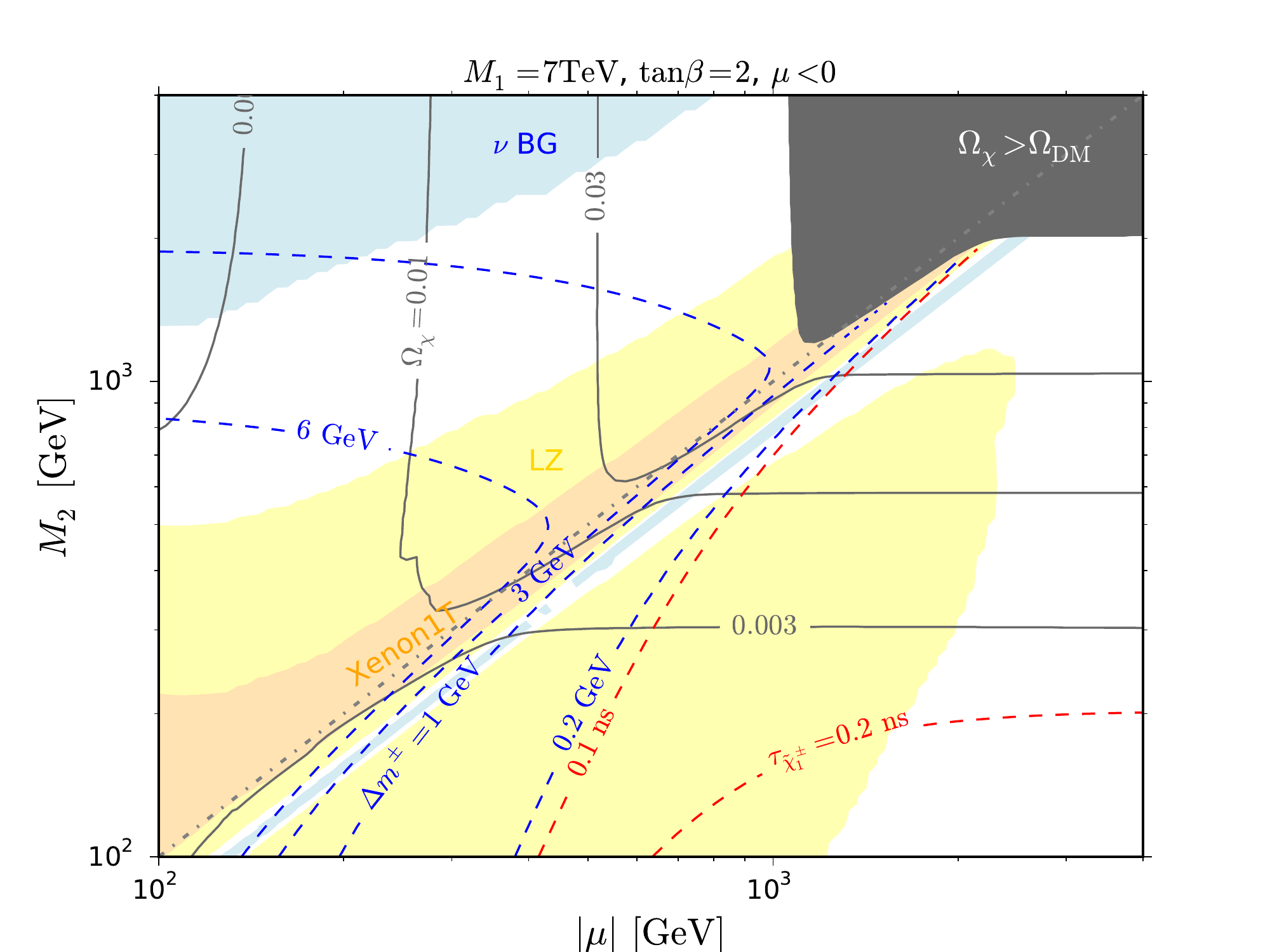}               
               \includegraphics[width=0.48\textwidth]{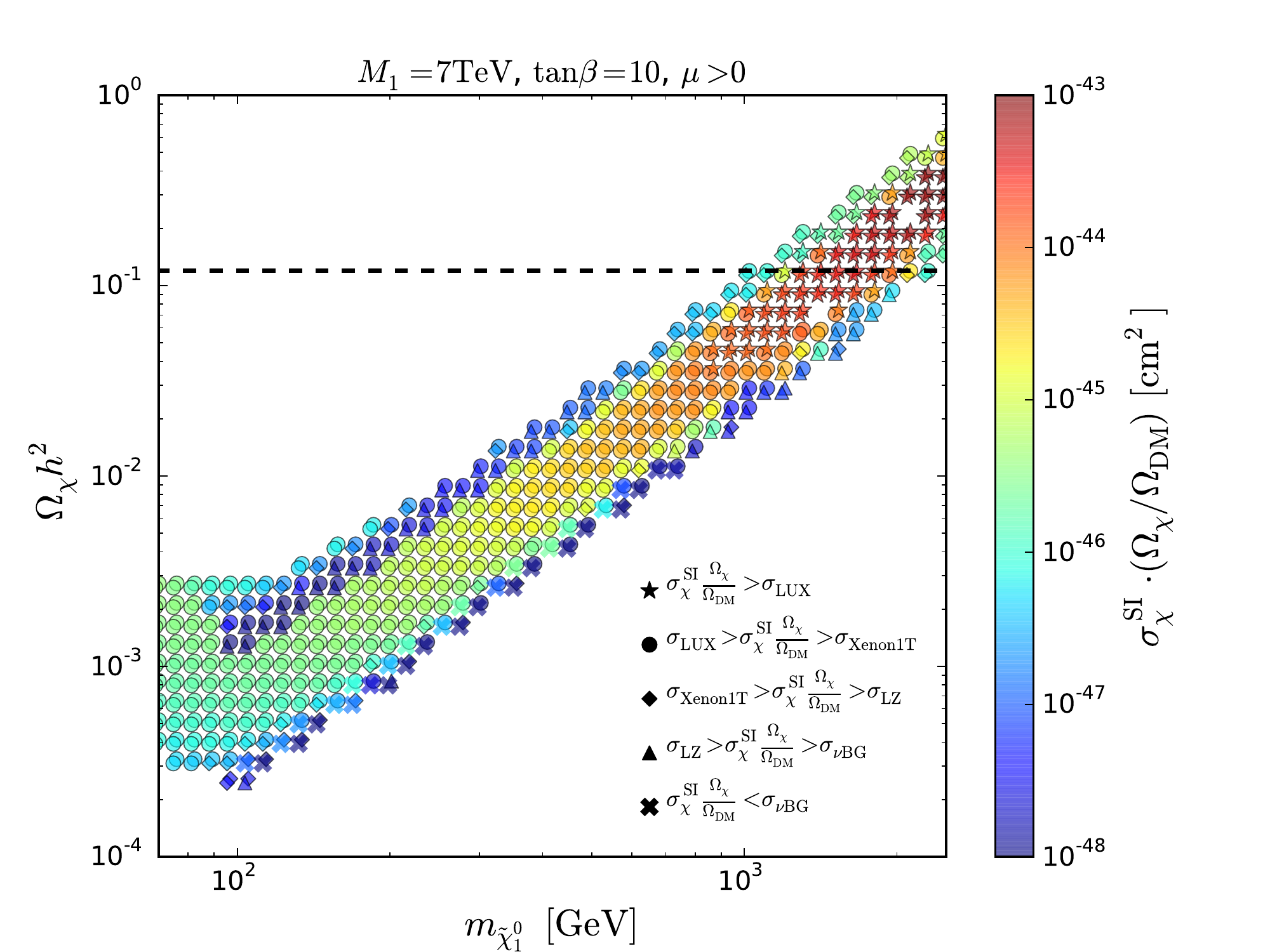}	
		\includegraphics[width=0.48\textwidth]{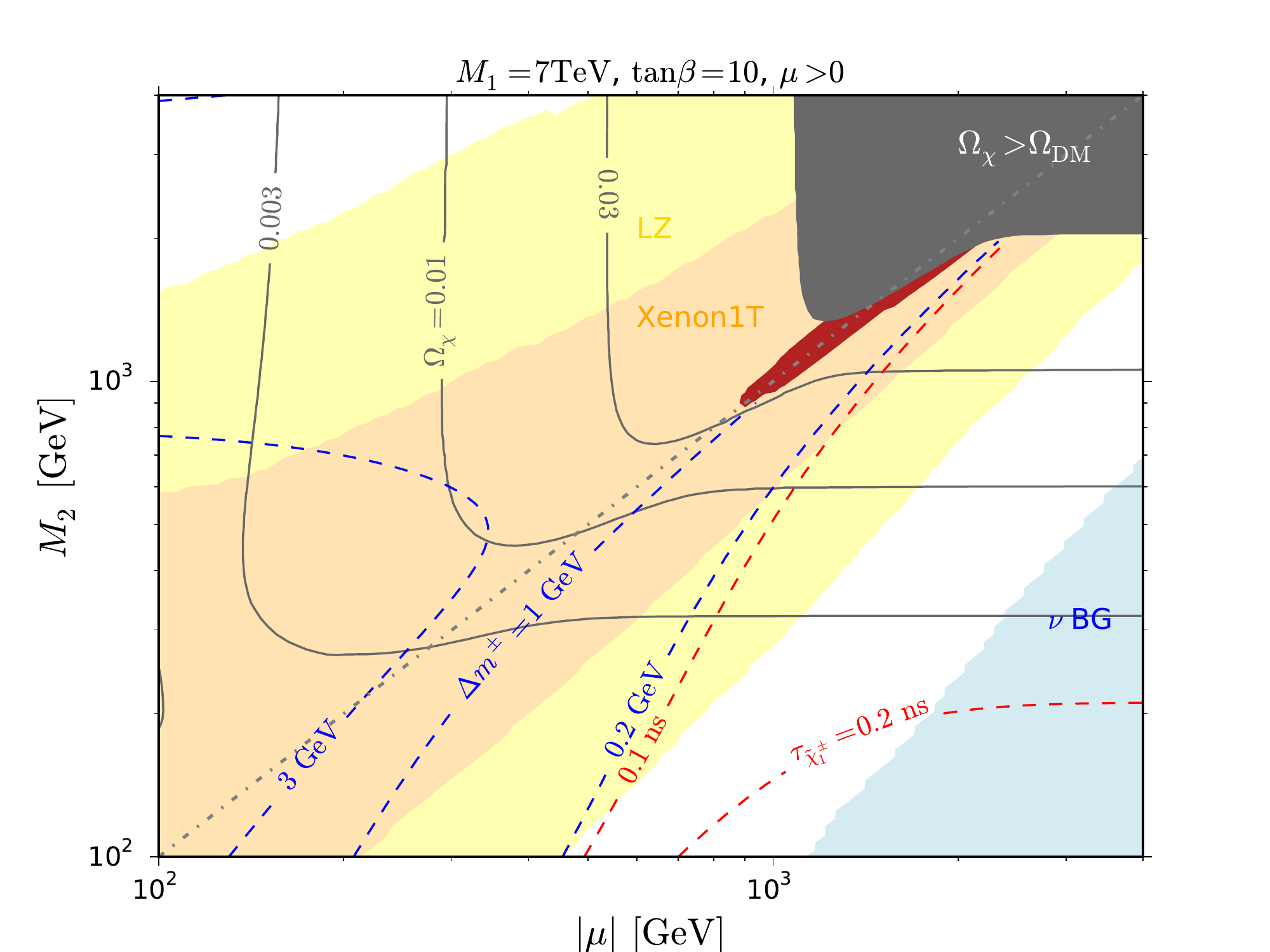}
\caption{
Same as in Fig.~\ref{fig:mu-m1_dec_1} but for the scan over $\mu$ and $M_2$, with decoupled binos.
\label{fig:m2-mu_dec_7}}
\end{figure}

This formula does not include the Sommerfeld enhancement for the wino component \cite{Sommerfeld}.   It is not included in the {\tt micrOMEGAs} code used by us, but this 
does not affect the main conclusions of the paper.\footnote{We have confirmed this by studying several representative points with the numerical package
\cite{Hryczuk} which is an extended version of {\tt DarkSUSY} \cite{DarkSUSY} that calculates Sommerfeld enhancement in MSSM. We thank A. Hryczuk for useful
correspondence about the usage of that package.}
The annihilation of winos is more efficient than that of higgsinos. The minimal value of
$\Omega_{\chi} h^2$ (for $M_2=100$ GeV) reaches  $2\cdot 10^{-4}$   and the value of 0.12 is reached  for $ M_2\approx 2.3$ GeV. This is seen in  the left plots of 
Fig.~\ref{fig:m2-mu_dec_7}, for $\tan\beta=2, \mu<0$  and $\tan\beta=10$ and  $\mu>0$.
An admixture of higgsinos increases the value of the neutralino relic abundance. 
The values of $\Omega_\chi h^2$, when shown as a function of the LSP mass, fall in the region between the two boundaries given by the pure  higgsino and wino states. 

The tree level spin independent scattering cross section  vanishes  for pure higgsino and wino states. 
The $h\chi\chi$ coupling  is approximately given  by  the  equations  \eqref{eq:c_mu>M1}, \eqref{eq:c_mu<M1} and \eqref{eq:diagonal}, with  the following replacements: $M_1 \rightarrow M_2, g_1\rightarrow g_2, \sin\theta_W\rightarrow \cos\theta_W$  and with the change of the sign of the first term in eq.~\eqref{eq:diagonal}.

Those equations explain the qualitative patterns  of the results of the scan  plotted in the ($|\mu|,\,M_2$) plane shown in the right column of Fig.~\ref{fig:m2-mu_dec_7}.\footnote{
We scan over   
($|\mu|,\,|M_2|$), both  in the range (0.1$-$5 TeV).} 
In general, the $\sigma_\chi^{\rm SI}$ decreases in the direction  perpendicular to the diagonal. There is an interesting dependence on  $\tan\beta$ and on the sign
of $\mu$.   One effect is again  the existence of blind spots for small $\tan\beta$ and negative $\mu$ and a strong enhancement of the spin  independent
scattering cross sections for positive $\mu$. For
$\tan\beta=2$ and $\mu>0$ (not shown in the Figure) a large part of the parameter space near the diagonal $|M_2|\approx|\mu|$ is already excluded by the LUX
experiment. For all combinations of $\tan\beta$ and signs of $\mu$ there can be seen certain asymmetries     in the values of the $\sigma_\chi^{\rm SI}$  with 
respect to the diagonal $|M_2|\approx|\mu|$. They can be understood  by looking at the  equations \eqref{eq:c_mu>M1},~\eqref{eq:c_mu<M1}
and~\eqref{eq:diagonal},  updated to the present case.  For instance, for large $\tan\beta$ the
$\sigma_\chi^{\rm SI}$ is larger for $|\mu|<|M_2|$
(dominantly higgsino)  than for $|M_2|<|\mu|$ (dominantly wino) because of the $\frac{1}{M_2}$ versus  $\frac{1}{\mu^2}$ suppression of the couplings in eqs.~\eqref{eq:c_mu<M1} and~\eqref{eq:c_mu>M1}, respectively.  There is also some residual dependence on the sign of $\mu$ coming from the numerator in eqs.~\eqref{eq:c_mu>M1} and \eqref{eq:c_mu<M1}.  For small $\tan\beta$ and negative $\mu$ the asymmetry with respect to the diagonal is reversed, as can be seen from the interplay of the numerators and denominators 
in eqs.~\eqref{eq:c_mu>M1} and \eqref{eq:c_mu<M1}.  
The important conclusion is that the vast parameter range is accessible in the DD experiments.  
However, it is clear that the dominantly wino or higgsino LSP regions will not be reachable in the DD experiments and 
these regions still take up sizeable portion of the ($\mu,\,M_2$) plane, particularly in the dominantly wino case.
\begin{figure}
	\centering \vspace{-0.0cm}
		\includegraphics[width=0.48\textwidth]{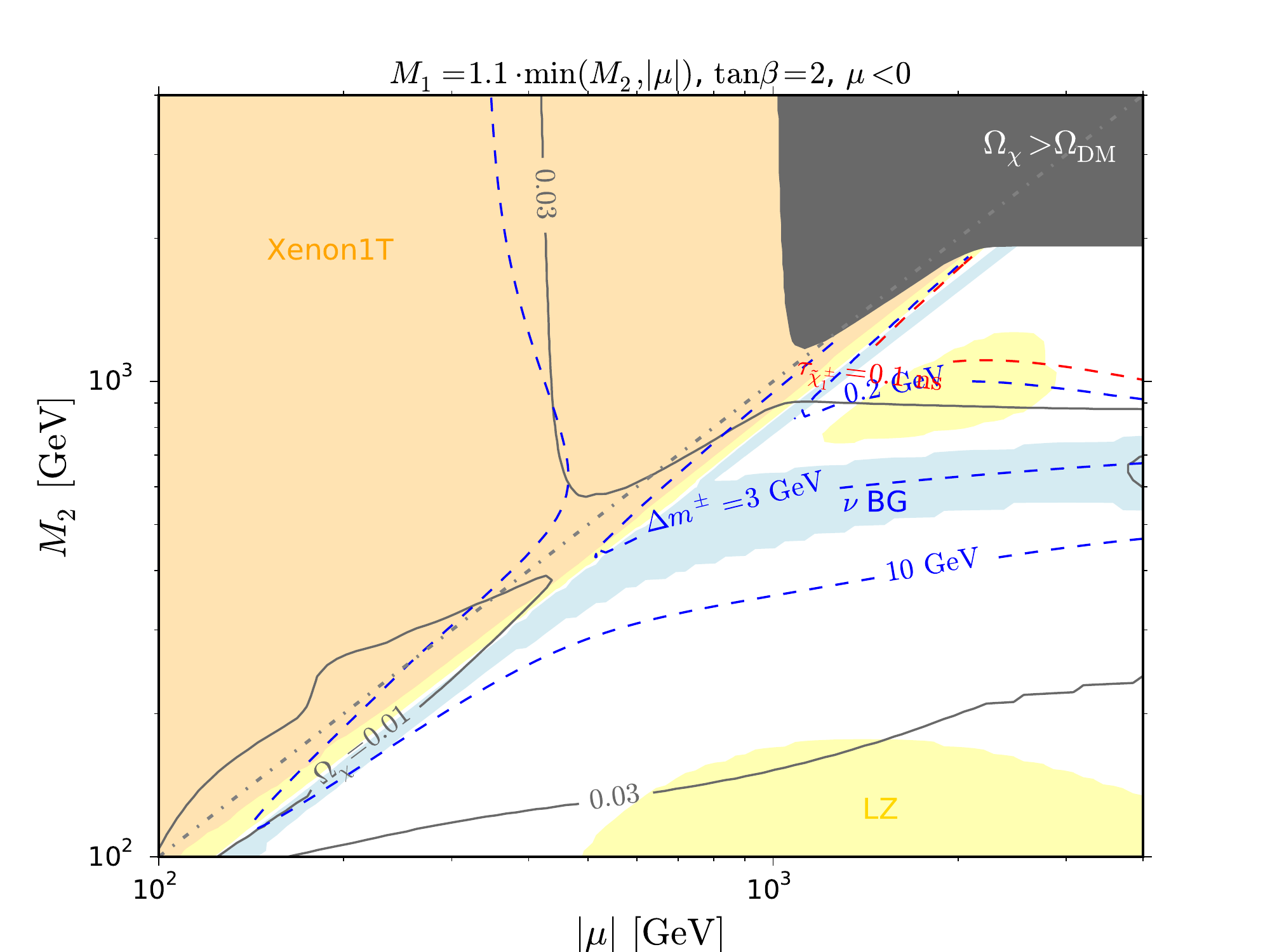}
		\includegraphics[width=0.48\textwidth]{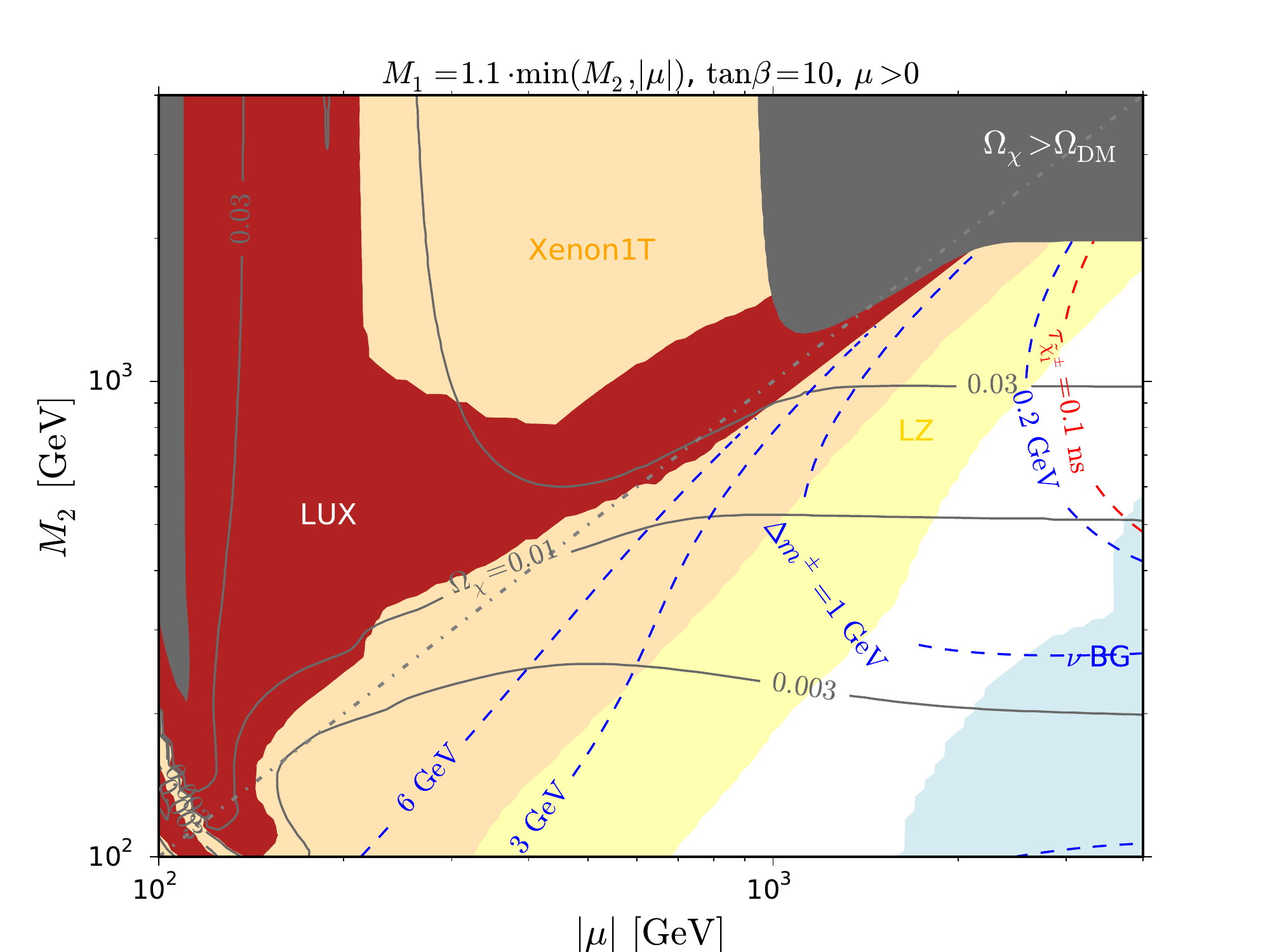}
\includegraphics[width=0.48\textwidth]{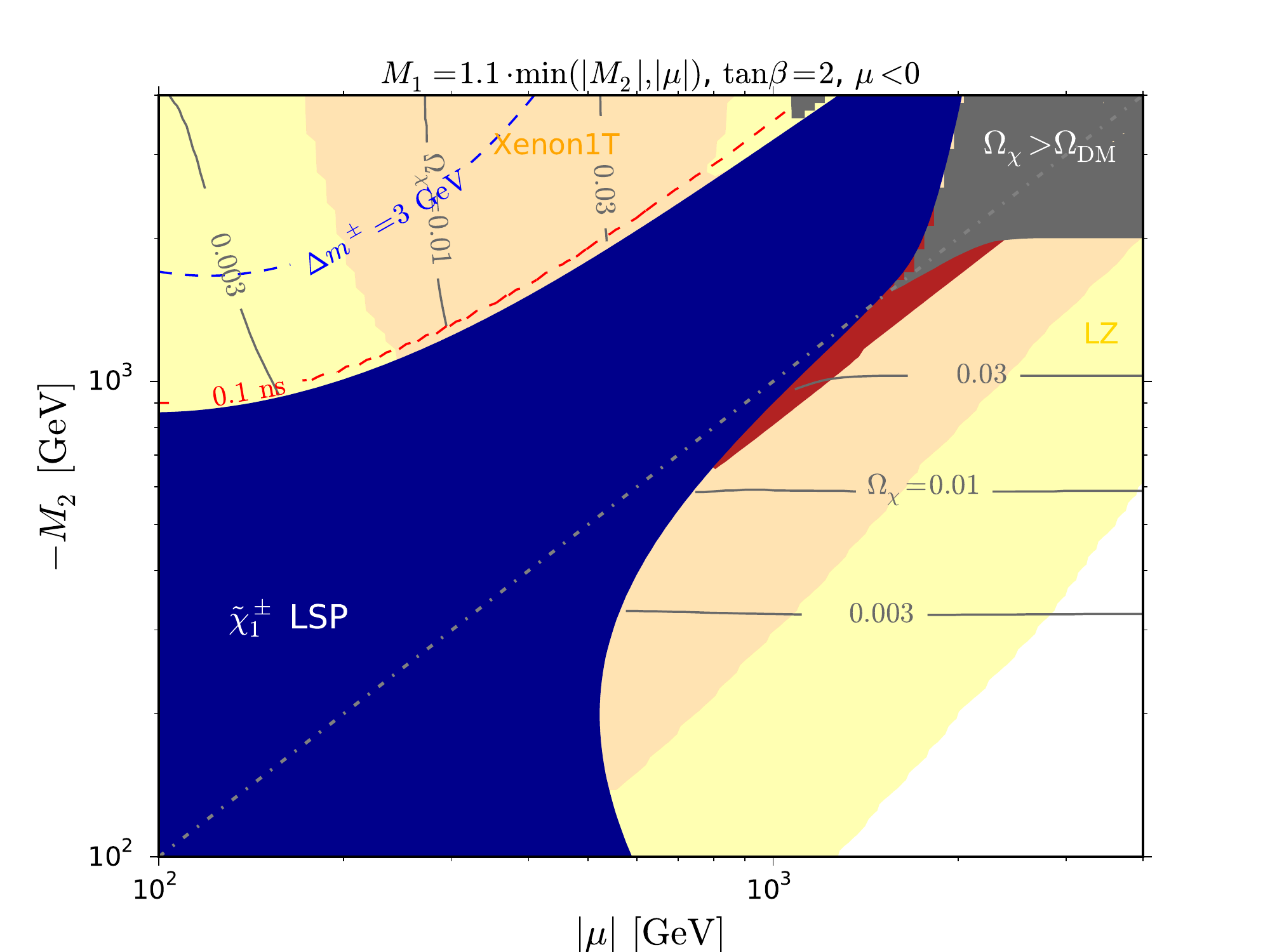}
		\includegraphics[width=0.48\textwidth]{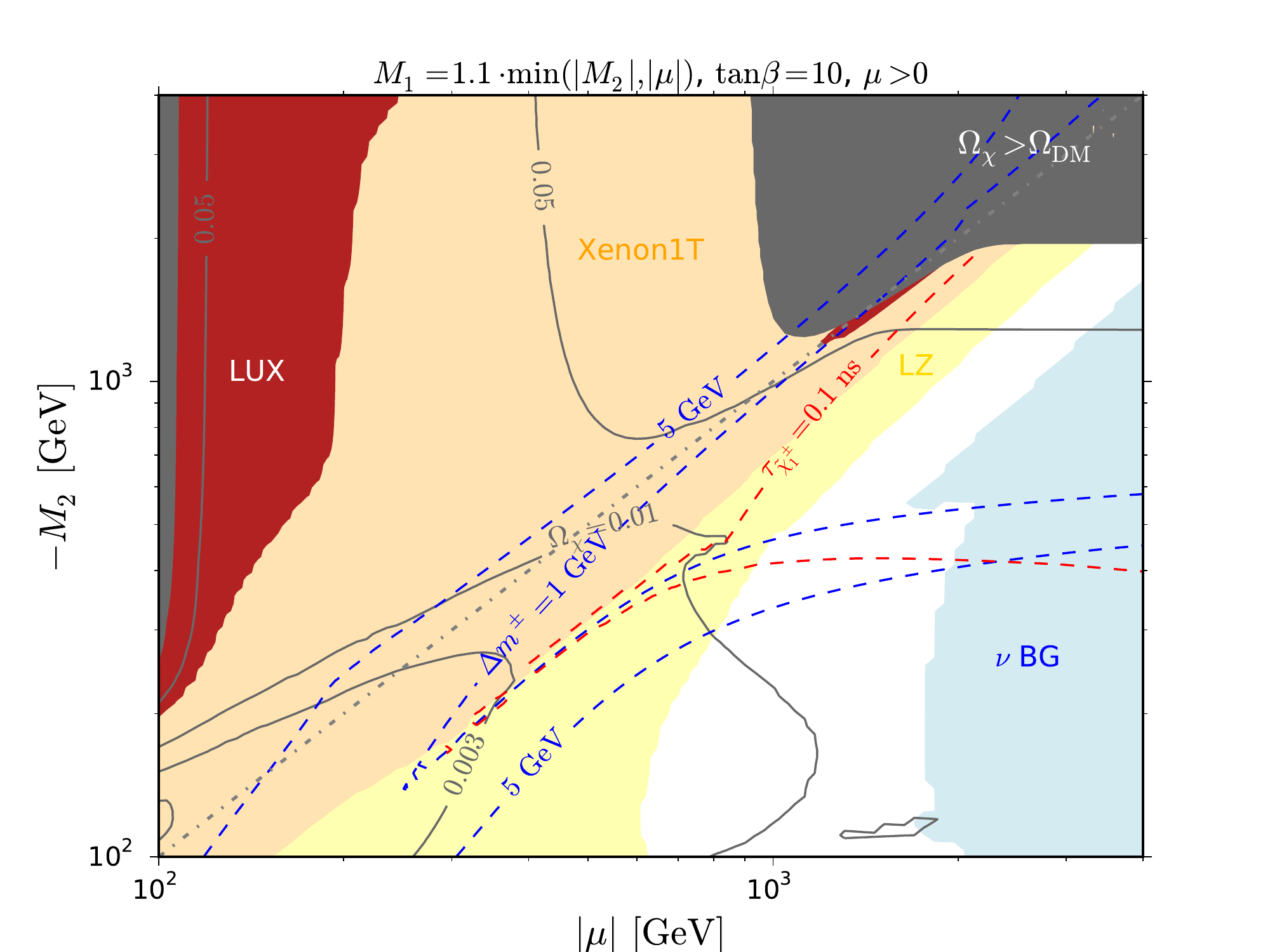}
\caption{
Same as Fig.~\ref{fig:m2-mu_dec_7} but for non-decoupled binos. The values of $M_1$ are marked on the plots.
\label{fig:m2-mu_plots_8}}
\end{figure}

The separation into  the two regions,  $|\mu|<|M_2|$ and  $|M_2|<|\mu|$,  is important  for understanding the pattern of mass differences and the NLSP lifetime. The higgsino-like LSP region ($|\mu|<|M_2|$) is
qualitatively similar to the bino-higgsino mixed LSP region discussed in the previous section.
We  observe that $\Delta m^{\pm}$ is also ${\mathcal O}(1-10)$ GeV but slightly larger than in the bino-higgsino case, because  of the $\cos\theta_W^2/M_2$
corrections versus $\sin\theta_W^2/M_1$  corrections in the previous case (see
eq.~\eqref{mu-mC1N1}). Also, as
before, $\Delta m^0 \approx 2\times\Delta m^{\pm}$.

In the wino-like LSP region ($|M_2|<|\mu|$) the tree  level $\Delta m^{\pm}$   vanishes up to very small corrections 
(see eq.\ (\ref{M2-mC1N1}) and (\ref{M2-mC1N1'}) in the Appendix) 
and $\Delta m^{\pm} \approx 160-170$ MeV is given by the loop effects 
\cite{wino_masssplitting}.
The lifetime becomes large, of the order of  ${\mathcal O}(0.1)$ ns.  
On the other hand,  there can be arbitrarily large mass difference $\Delta m^0 \approx |\mu|-|M_2|$ . 

The effects of non-decoupling  of $M_1$ in the scan over $M_2$ and $\mu$ are illustrated in Fig.~\ref{fig:m2-mu_plots_8}.
One can see, comparing Fig.~\ref{fig:m2-mu_plots_8} with Fig.~\ref{fig:m2-mu_dec_7}, that the main patterns are quite similar to the decoupled case. In
particular, this is true for the mass differences $\Delta m^0$ (not shown in the Figures) and $\Delta m^{\pm}$ but the $\sigma_\chi^{\rm SI}$  is generically larger, 
especially for $|M_2|>|\mu|$, 
making easier the detection of the LSP in the future DD experiments.
Adding the bino component makes the values
of $\Omega_{\chi} h^2$  for the same neutralino mass somewhat larger, going  above the higgsino bound seen in Fig.~\ref{fig:m2-mu_dec_7}.

%%%%%%%%%%%%%%%%%%%%%%%%%%%%%%%%%%%%%%%%%%%%%%%%%%%%%%%%%%%%%%%%%%%%%%
%%%%%%%%%%%%%%%%%%%%%%%%%%%%%%%%%%%%%%%%%%%%%%%%%%%%%%%%%%%%%%%%%%%%%%

\subsection{Collider}

\begin{figure}[b]
%\hspace{-0.7cm}
	 \vspace{-0.0cm}
		\includegraphics[width=1.0\textwidth]{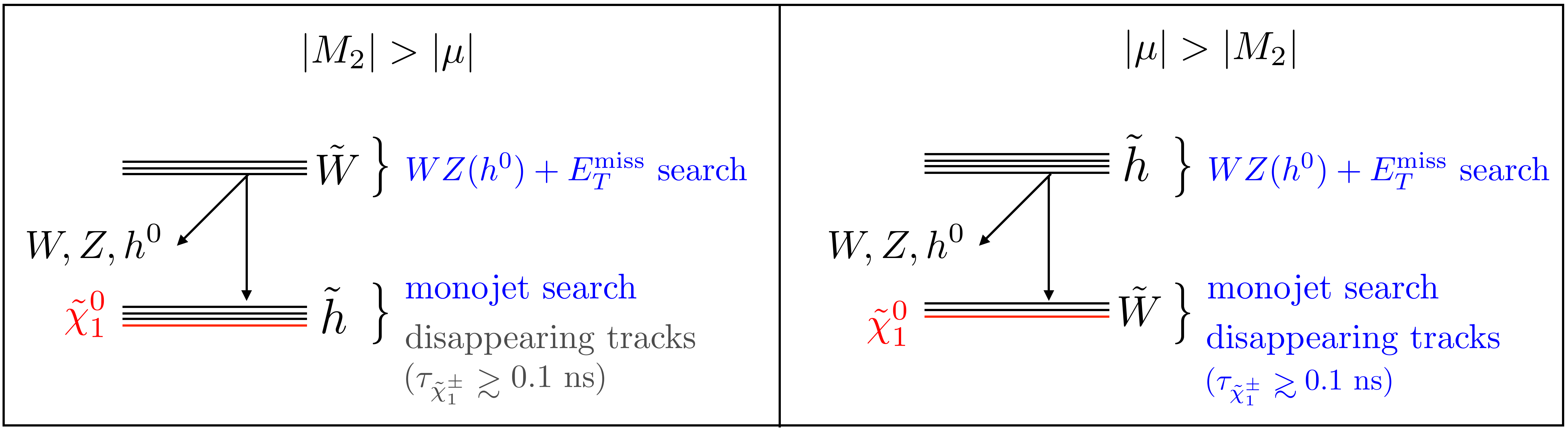}
\caption{Strategy for collider search.
\label{fig:m2-mu_schematic}}
\end{figure}

\begin{figure}
	\centering \vspace{-0.0cm}
		\includegraphics[width=0.48\textwidth]{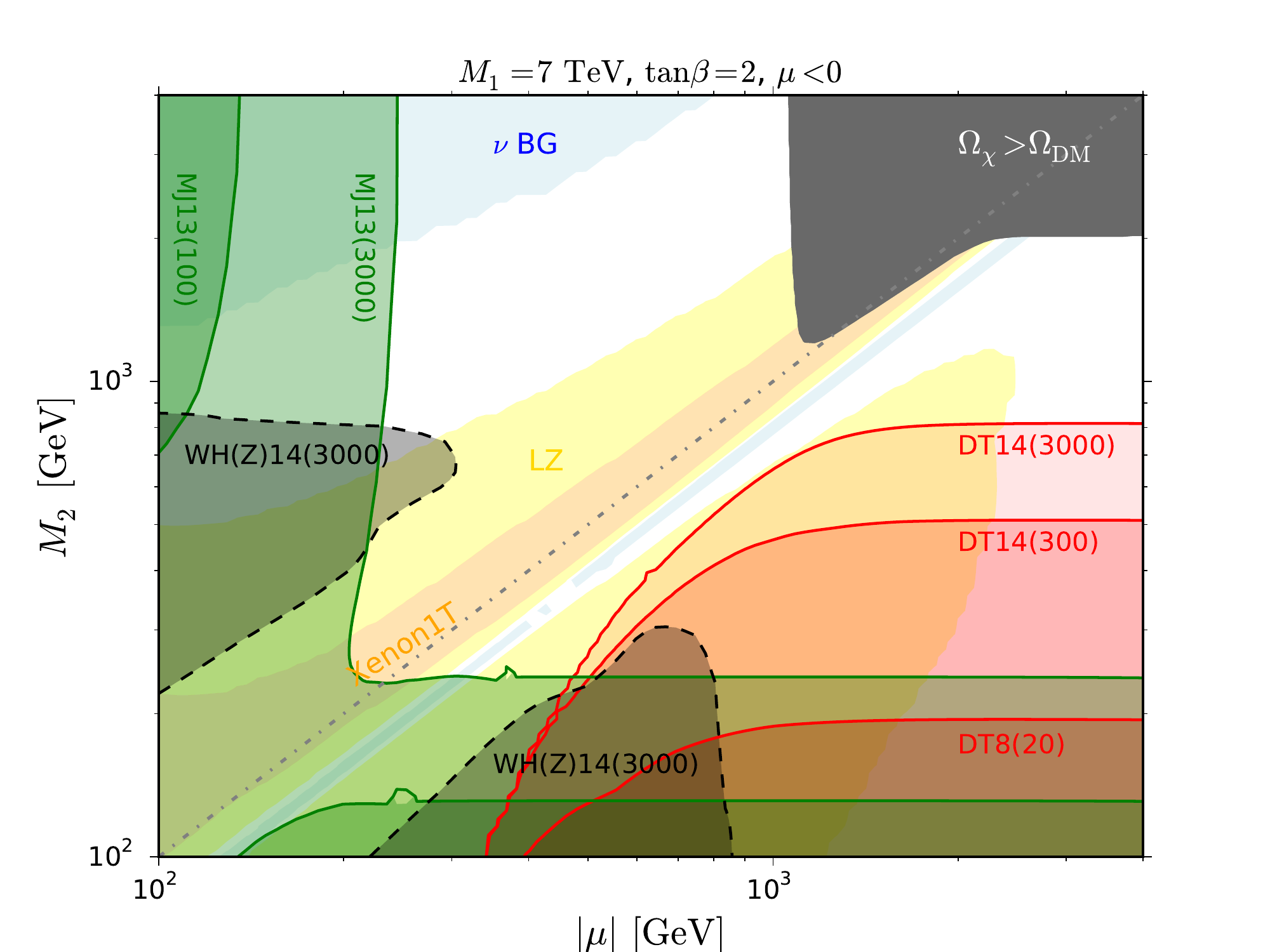}
		\includegraphics[width=0.48\textwidth]{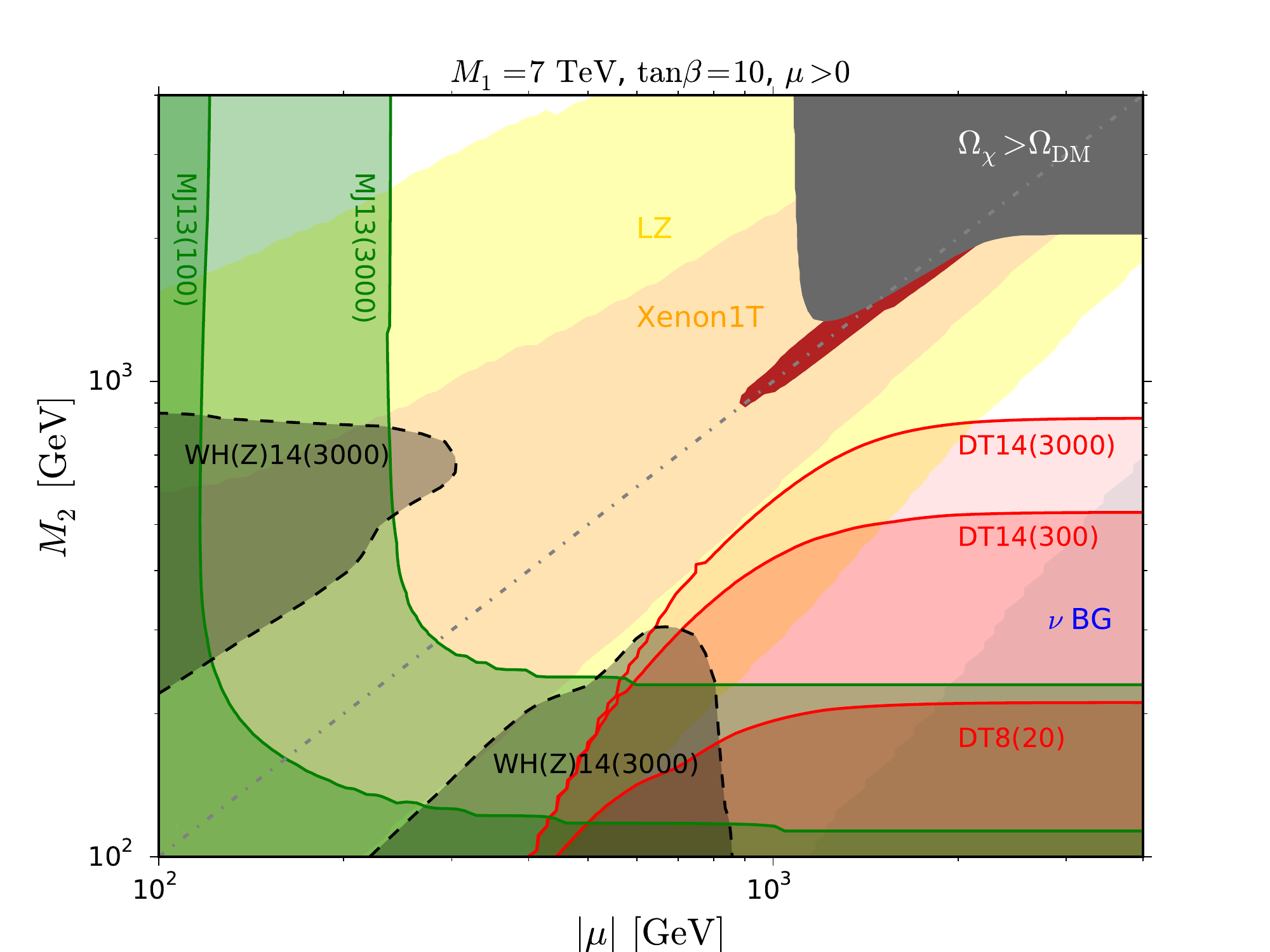}
\caption{
The LHC sensitivity regions for decoupled binos in the ($|\mu|, M_2$) plane:
the $WZ + E_T^{\rm miss}$ and $Wh + E_T^{\rm miss}$ combined search (grey),
the monojet search (green) and the disappearing track search (red). 
\label{fig:m2-mu_dec_9}}
\end{figure}

For the wino-higgsino scenario the strategy of collider search has new elements as compared to  the bino-higgsino case.
Both wino-like and higgsino-like states can have sizable production cross section.
For sufficiently large splitting in the values of $M_2$ and $\mu$, the
heavier of the wino-like or the higgsino-like states can decay to the higgsino-like or wino-like states, respectively,  accompanied by $W$, $Z$  or $W$, $h^0$ boson pairs
(see Fig.~\ref{fig:m2-mu_schematic}). An interesting signature for $|M_2| > |\mu|$ is then the decay
$\tilde W^\pm \tilde W^0 \to W^\pm Z(h) + E_T^{\rm miss}$, where $\tilde W^\pm$ and $\tilde W^0$ are the wino-like
$\tilde \chi_2^\pm$ and $\tilde \chi_3^0$, respectively and
$E_T^{\rm miss}$ includes higgsino-like $\tilde \chi_1^\pm$, $\tilde \chi_2^0$ and $\tilde \chi_1^0$, neglecting soft particles arising from the $\tilde \chi_1^\pm$ and $\tilde \chi_2^0$ decays into an almost  mass degenerate $\tilde \chi_1^0$. Similarly, for $|\mu| > |M_2|$,
we have  the decays
$\tilde h^\pm \tilde h^0 \to W^\pm Z(h) + E_T^{\rm miss}$ and
in this case, $\tilde h^0$ and $\tilde h^\pm$ are higgsino-like $\tilde \chi_2^0$, $\tilde \chi_3^0$ and $\tilde \chi_2^\pm$, respectively, and
$E_T^{\rm miss}$ includes wino-like $\tilde \chi_1^\pm$ and $\tilde \chi_1^0$ assuming the decay products of the $\tilde \chi_1^\pm$ decay does not contribute to the signal regions in the analysis.
Such  $W+Z(h) +E_T^{\rm miss}$ signatures depend on the  production cross sections of the heavy gaugino/higgsino  pairs and on their branching ratios into those final states.  Using the Goldstone equivalence theorem \cite{Jung:2014bda} one  can estimate \cite{Jung:2014bda,Goldstone}  that the branching ratio of each initial particle into
$WZ(h)$ is 50\%, with 25\% for $Z$ and $h$ each, so that  we have 50\% probability for the  $W+Z(h) +E_T^{\rm miss}$ signature and the remaining 50\%   goes into $WW$, $ZZ$ and $hh$ final states. There is no dedicated analysis of the LHC discovery potential for such signatures 
(such an analysis for 100 TeV colliders is presented in \cite{Goldstone}). 
CMS have studied the projected sensitivity in the 14 TeV LHC with 3000 fb$^{-1}$ for the $\tilde \chi_1^\pm \tilde \chi_2^0 \to (W^\pm \tilde \chi_1^0) (Z \tilde \chi_1^0)$ and the $\tilde \chi_1^\pm \tilde \chi_2^0  \to (W^\pm \tilde \chi_1^0) (h^0 \tilde \chi_1^0)$ processes \cite{CMS:2015vka}, assuming wino-like production cross sections for the initial states and 100\% 
$W+Z(h) +E_T^{\rm miss}$ signature. In this study, 5-$\sigma$ sensitivities for three different models, 
${\rm BR}(\tilde \chi_2^0 \to [Z \tilde\chi_1^0, h \tilde\chi_1^0]) = (100, 0), (0, 100), (50, 50)\,\%$, are presented and they are found to be similar to each other. We can use those results
for the purpose of a qualitative illustration of a ballpark sensitivity to our $(|\mu|, M_2)$ plane,
keeping in mind that the CMS assumptions about the production cross sections and the decay signature
are not strictly valid in our case, but just up to  a factor of 2-3.
We map the CMS 5-$\sigma$ sensitivity 
for the ${\rm BR}(\tilde \chi_2^0 \to [Z \tilde\chi_1^0, h \tilde\chi_1^0]) = (50, 50)\,\%$ case (in accordance with the Goldstone equivalence theorem)
by identifying 
$m_{\tilde \chi_1^\pm}^{\rm CMS} = M_2$, $m_{\tilde \chi_1^0}^{\rm CMS} = |\mu|$ for $M_2 > |\mu|$, and   
$m_{\tilde \chi_1^\pm}^{\rm CMS} = |\mu|$, $m_{\tilde \chi_1^0}^{\rm CMS} = M_2$ for $|\mu| > M_2$. 
The result is shown by the grey regions in Fig.~\ref{fig:m2-mu_dec_9} surrounded by the dashed curves.  One should stress again that this is only a qualitative  illustration of ``an order of magnitude'' for the expected sensitivity, because of the mentioned above simplified CMS assumptions about the production and decay rates, inadequate for our case.
\begin{figure}
	\centering \vspace{-0.0cm}
\includegraphics[width=0.48\textwidth]{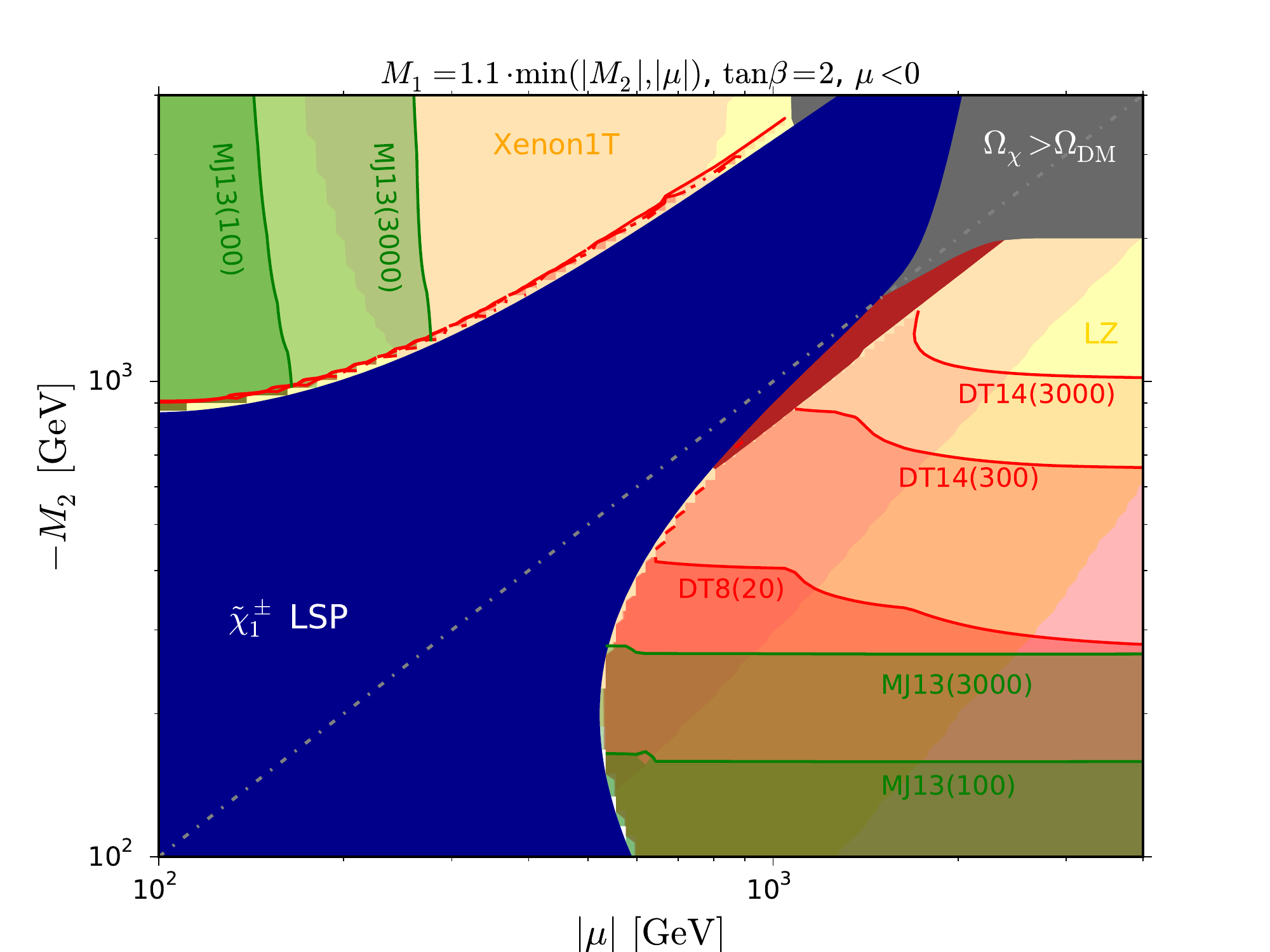}
        \includegraphics[width=0.48\textwidth]{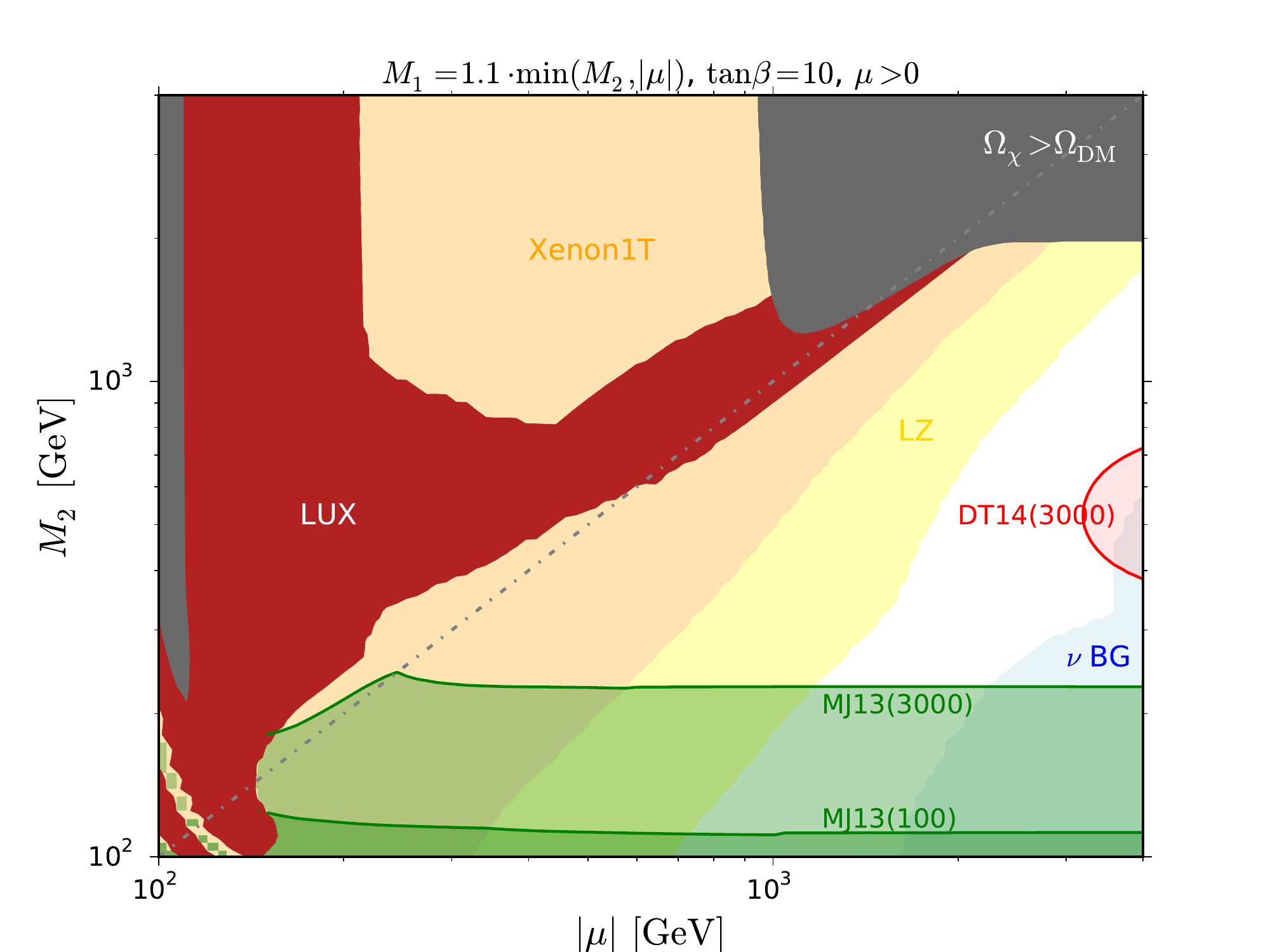}        
\caption{
The LHC sensitivity regions for non-decoupled binos in the ($|\mu|, -M_2$) (left) and ($|\mu|, M_2$) (right) planes:
the monojet search (green) and the disappearing track search (red). 
\label{fig:m2-mu_plots_10}}
\end{figure}
%
%

%%%%%%%%%%%%%%%%%%%%%%%%%%%%%%%

As we discussed in section~\ref{sec:Bino-Higgsino-DM}, the production of the lighter multiplet (lighter of the wino- or higgsino-like states) can be probed 
by the monojet search and the disappearing track search. 
In Fig.~\ref{fig:m2-mu_dec_9} and \ref{fig:m2-mu_plots_10} we show the 2-$\sigma$ sensitivity region found in \cite{Barducci}
using the same procedure as in Fig.~\ref{fig:mu-m1_dec_6}. 
Since the study of \cite{Barducci} assumes the higgsino-like LSP scenario, the green region in the wino-like LSP region should be regarded as conservative since the wino cross section is larger than the higgsinos at the same mass, although we believe it is not too conservative by the same reason mentioned above.

It is known that the lifetime of charged wino can be as long as the collider scale in the almost pure wino region.
In Fig.~\ref{fig:m2-mu_dec_7} the charged wino lifetime is longer than 0.1 ns to the  right  of the black dashed curves,
where the search exploiting the disappearing tracks can be sensitive.
ATLAS \cite{ATLAS_disapppearing} and CMS \cite{CMS_disapppearing} have looked for disappearing tracks in the 8 TeV proton-proton collision data and identified
the 
95\% CL excluded regions in the ($m_{\tilde \chi_1^\pm}$, $\Delta m^\pm$) plane.
Since we can compute $m_{\tilde \chi_1^\pm}$  and $\Delta m^\pm$ at each point in the ($|\mu|, M_2$) plane,
we can map these 95\% CL excluded regions onto our parameter plane.  
We translate the ATLAS excluded region (Fig.~6 of \cite{ATLAS_disapppearing}) in our ($|\mu|, M_2$) plane, which are shown by the darkest red regions.
We also project the 8 TeV excluded regions into the 14 TeV LHC using {\tt Collider Reach} program \cite{ColliderReach}
assuming the projected limit is obtained with the same number of signal events after the event selection
as observed at the 8 TeV exclusion boundary.\footnote{
  This is commonly used assumption. 
  For the different energy and luminosity, the cross section, the optimal event selection and its acceptance for signal and background change.  However, the signal yield at the exclusion boundary usually do not change much after optimising the selection cuts.  This is an empirical observation but good agreement is often found between the results from this approach and from the full simulation. 
  See e.g. \cite{Gori:2014oua, Buchmueller:2015uqa}.  
} 
The projected sensitivity for the 14 TeV LHC with 300 (3000) fb$^{-1}$ is shown by the darker (light) red region.

%%%%%%%%%%%%%%%%%%%%%%%%%%%%%%%%%%%%%%%%%%%%%%%%%%%%%%%%%%%%%%%%%%%%%%
%%%%%%%%%%%%%%%%%%%%%%%%%%%%%%%%%%%%%%%%%%%%%%%%%%%%%%%%%%%%%%%%%%%%%%

\section{Bino-Wino } 

%%%%%%%%%%%%%%%%%%%%%%%%%%%%%%%%%%%%%%%%%%%%%%%%%%%%%%%%%%%%%%%%%%%%%%
%%%%%%%%%%%%%%%%%%%%%%%%%%%%%%%%%%%%%%%%%%%%%%%%%%%%%%%%%%%%%%%%%%%%%%

In the limit of a  decoupled $\mu$, the picture is very simple because there is no mixing between the binos and winos
except $M_1 = M_2$.  
Without the bino-wino mixing, the heavier of the bino and wino cannot decay into lighter one if scalars are decoupled \cite{long-lived}.
The lifetime of the heavier EW gaugino exceeds $c \tau \gsim {\cal O}(1)$ cm
with $|\mu| \gsim {\cal O}(10^{2-6})$ TeV, depending on the bino-wino mass splitting.
 
\begin{figure}
	\centering \vspace{-0.0cm}
		\includegraphics[width=0.48\textwidth]{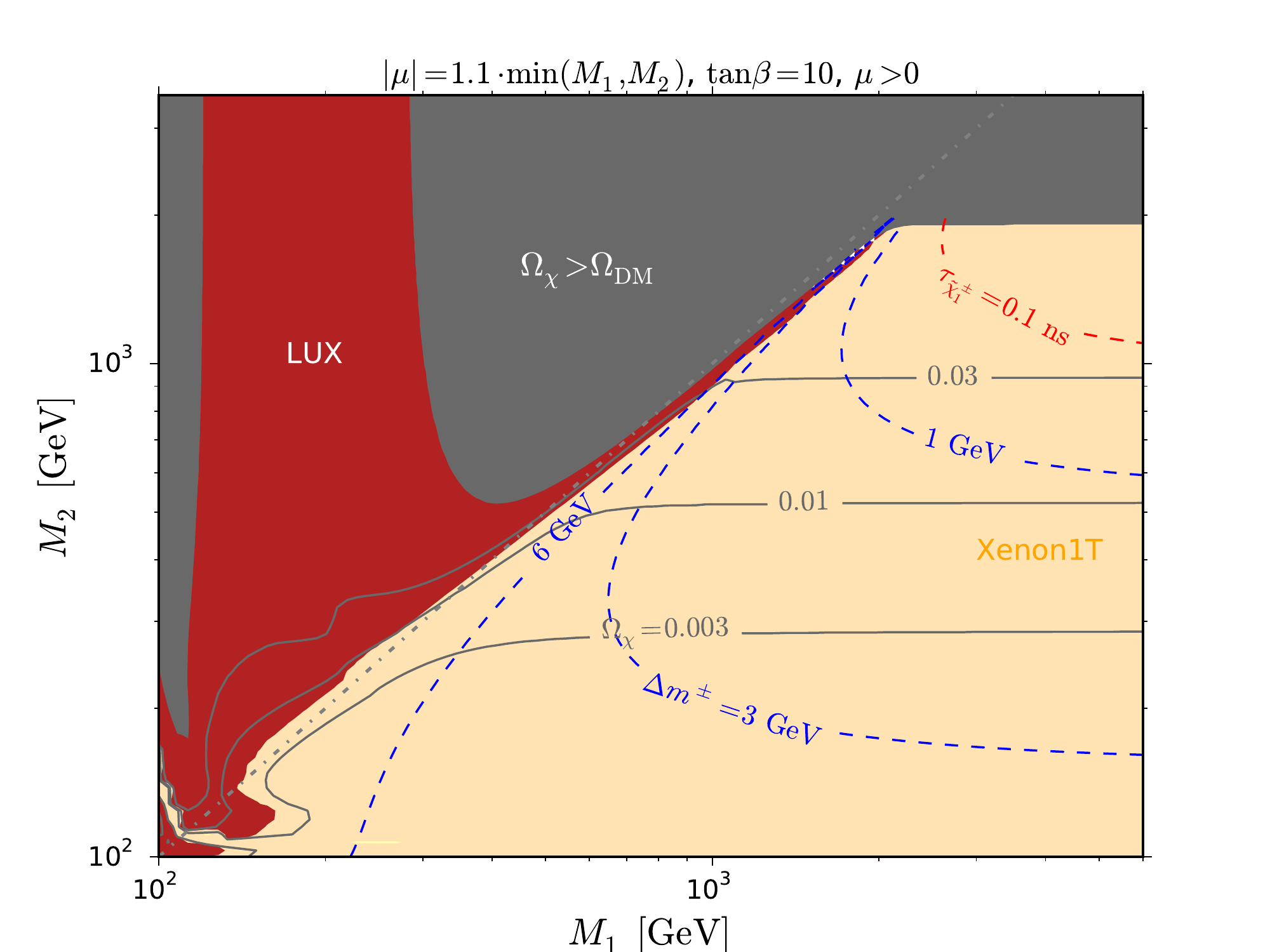}
               \includegraphics[width=0.48\textwidth]{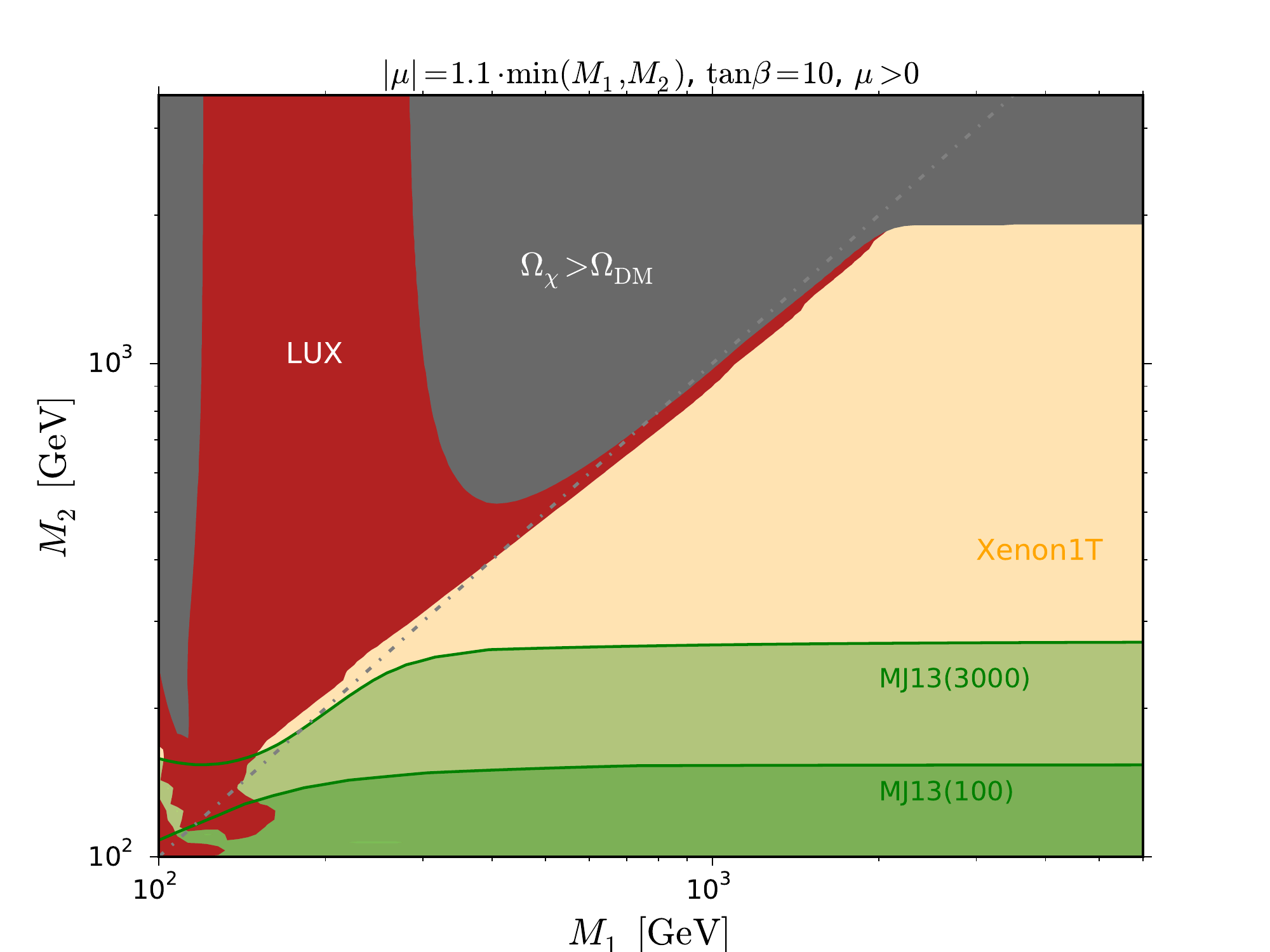}
\caption{
The results of the scan over $M_1$ and $M_2$ with $\mu$ fixed at the values marked on the plots.
In the  right plot, the estimated LHC discovery limit  is shown.
\label{fig:m1-m2_12}}
\end{figure}

\begin{figure}
	\centering \vspace{-0.0cm}
		\includegraphics[width=0.48\textwidth]{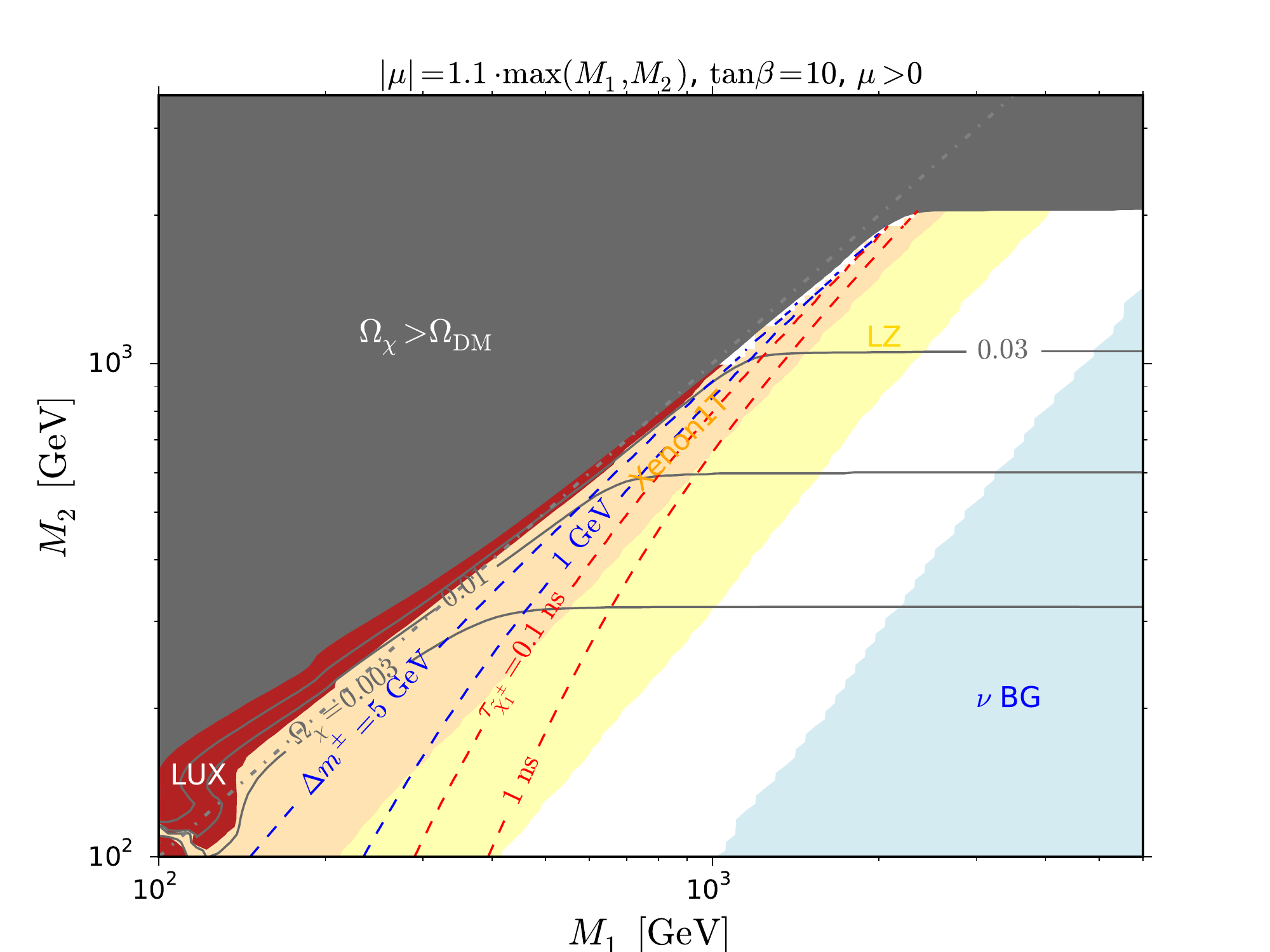}
                \includegraphics[width=0.48\textwidth]{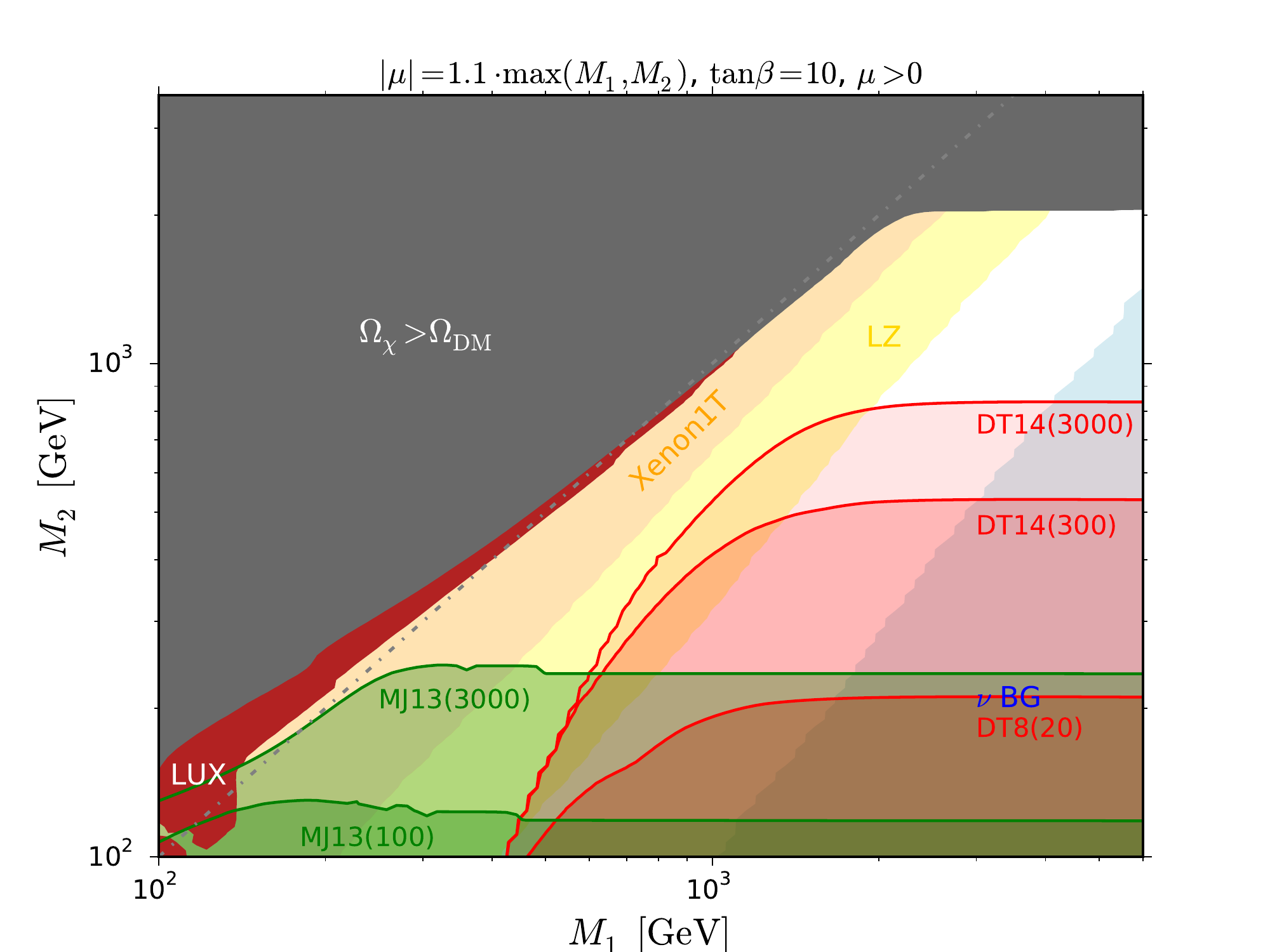}
\caption{
The results of the scan over $M_1$ and $M_2$ with $\mu$ fixed at the values marked on the plots.
In the  right plot, the estimated LHC discovery limit  is shown.
\label{fig:m1-m2_11}}
\end{figure}

For $|M_2|<M_1$ the LSP  and the NLSP are
neutral and charged winos, respectively.
The LSP relic abundance is that of a pure wino, already discussed in the previous section.  
The mass difference $\Delta m^{\pm}$ is about 160$-$170
MeV, given
by the loop effects and the NLSP decay 
has a collider scale lifetime, 
$c \tau_{\tilde \chi_1^\pm} \sim 5-10$ cm. The mass difference  $\Delta m^0$  is fixed by the difference
$M_1-|M_2|$.  The spin-independent scattering cross sections are very small, given by loop effects and below the neutrino background. The region $M_1<|M_2$| is excluded by the bound $\Omega_{\chi} h^2<0.12$.
The above simple picture is a good reference frame for discussing what happens when the values of
the $|\mu|$ parameter come closer to the values of $M_1$ and/or $|M_2|$ (see also \cite{SchwallerZurita}). In Fig.~\ref{fig:m1-m2_11} and in Fig.~\ref{fig:m1-m2_12}  we show the plots for $|\mu|=1.1 \max (M_1,|M_2|)$   and for  $|\mu|=1.1 \min(M_1,|M_2|)$, respectively, for $\tan\beta=10, \mu>0$ and positive $M_2$. For the other combinations of the values of $\tan\beta$  and the signs of $\mu$ and $M_2$ the results are very similar.

In the scan exemplified by Fig.~\ref{fig:m1-m2_11} the  higgsino is heavy, it mediates the bino-wino mixing but its admixture remains small. In Fig.~\ref{fig:m1-m2_12} the higgsino component in the LSP plays already important role and  the observed patterns can be understood by  comparing   with the plots for 10\% degeneracies  discussed in the previous section.

{\boldmath $|M_2|>M_1$}

We see in Fig.~\ref{fig:m1-m2_11}  that most of the bino-dominated region is still excluded by the bound $\Omega_{\chi} h^2\leq 0.12$. 
However, for $M_1\approx M_2$ an acceptable region opens up, although for $\mu >0$ mostly excluded by the LUX limits. 
For $\mu <0$ (not shown in the paper), in that well-tempered bino-wino region the spin-independent cross sections are often within the reach of the Xenon1T experiment.
The recent result of a general 10 parameter MSSM scan, including the $g-2$, the collider constraints as well as the DM constraints with $\Omega_{\chi} = \Omega_{DM}$, also prefers this parameter region \cite{deVries:2015hva}.

With $|\mu|=1.1M_1$ (see Fig.~\ref{fig:m1-m2_12}) the LSP becomes  a mixture of all three components, with large higgsino-bino mixing,  and only a part of the 
previously excluded  region is still  excluded by the bound $\Omega_\chi h^2 \leq 0.12$. This region resembles the regions with $M_1=1.1|\mu|$ in
Fig.~\ref{fig:m2-mu_plots_8}, that is the regions above the diagonal.  The regions excluded by the overproduction of the neutralinos are, however, larger because  of
the stronger bino component. The spin independent scattering  cross sections are large and the mass differences are of order ${\mathcal O}(10)$ GeV.

{\boldmath $|M_2|<M_1$}

For  $|\mu|=1.1 M_1$, the LSP remains dominantly wino-like but some admixture of bino makes the spin-independent cross sections partly accessible in the DD experiments. The mass differences are very small and the NLSP is long-lived, with strong implications for collider searches.

With  $|\mu|=1.1|M_2|$, the LSP is dominantly wino-higgsino. The patterns seen in Fig.~\ref{fig:m1-m2_12} resemble  those   for $M_2=1.1|\mu|$ in Fig.~\ref{fig:mu-m1_5} (i.e. above the diagonal), although the mass differences are typically smaller, closer to the pure wino case.
The scattering cross section is below the LUX bound  but above the neutrino bound.

In  the right plots of  Fig.~\ref{fig:m1-m2_11} and \ref{fig:m1-m2_12} we show the estimated LHC discovery limits. We plot again the 2-$\sigma$ sensitivity of the monojet search \cite{Barducci} in the $(M_1, M_2)$ plane  with the dark (light) green region corresponding to the 13 TeV LHC with 100 (3000) fb$^{-1}$ of the luminosity.  The projected sensitivity for the 14 TeV LHC with 300 (3000) fb$^{-1}$ to the disappearing track search, estimated as described in the previous section, is shown by the darker (light) red region.

Both, for $|\mu|=1.1 \max(M_1,|M_2|)$ and $|\mu|=1.1 \min(M_1,|M_2|)$,  the bino-like LSP region ($|M_2| > M_1$) is mostly excluded because of
the overproduction of the neutralinos.
In the other half of the parameter space ($M_1 > |M_2|$),  for $|\mu|=1.1 \max(M_1,|M_2|)$ the LSP is composed dominantly of wino, with some bino components. As seen in Fig.~\ref{fig:m1-m2_11} the DD and the collider searches are complementary to each other and cover a large part of that parameter space.

With  $|\mu|=1.1 \min(M_1,|M_2|)$, the lighter chargino cannot have long enough lifetime to contribute to the disappearing track signature and the only channel 
which can probe the wino production events would be the monojet search. 
As seen  in Fig.~\ref{fig:m1-m2_12}, for strong enough wino component the LHC 
can detect such  production events,
which will be an important cross check with the direct detection experiments such as XENON1T and LZ.

%%%%%%%%%%%%%%%%%%%%%%%%%%%%%%%%%%%%%%%%%%%%%%%%%%%%%%%%%%%%%%%%%%%%%%
%%%%%%%%%%%%%%%%%%%%%%%%%%%%%%%%%%%%%%%%%%%%%%%%%%%%%%%%%%%%%%%%%%%%%%

\section{Final Comments and Conclusions \label{sec:concl}}

%%%%%%%%%%%%%%%%%%%%%%%%%%%%%%%%%%%%%%%%%%%%%%%%%%%%%%%%%%%%%%%%%%%%%%
%%%%%%%%%%%%%%%%%%%%%%%%%%%%%%%%%%%%%%%%%%%%%%%%%%%%%%%%%%%%%%%%%%%%%%

\begin{figure}[t]
	\centering \vspace{-0.0cm}
                       \includegraphics[width=0.48\textwidth]{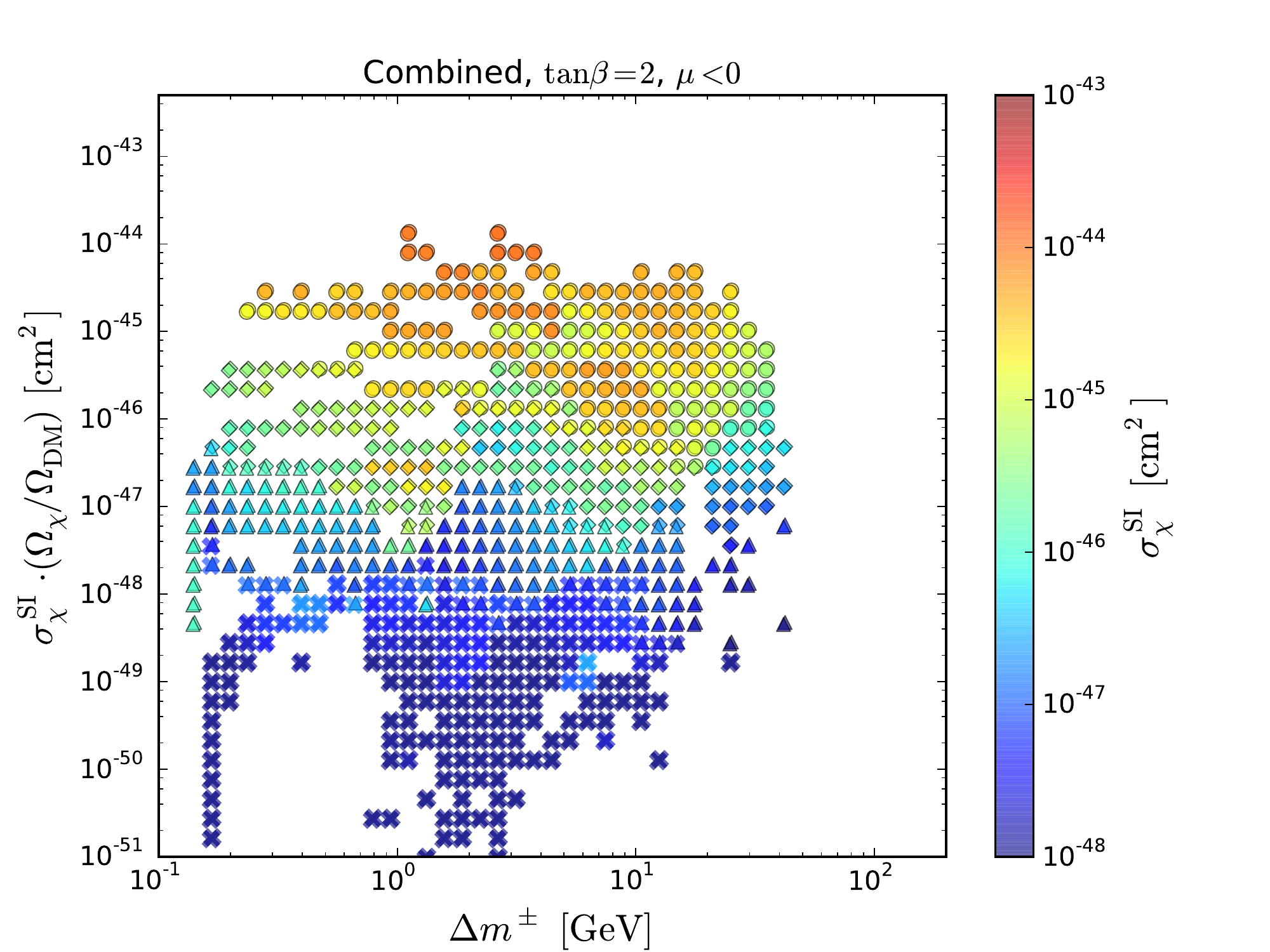} 
                       \includegraphics[width=0.48\textwidth]{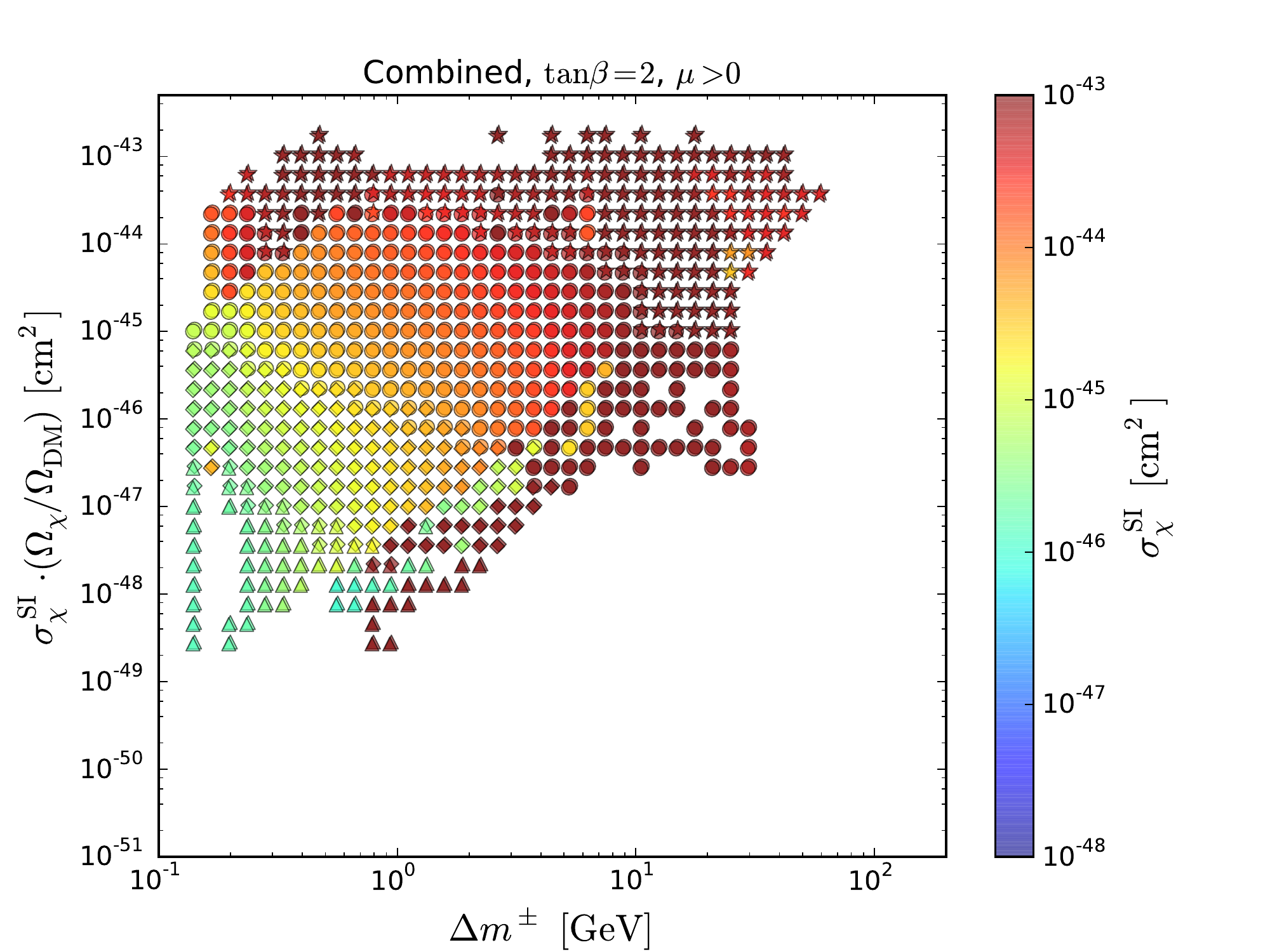}  
                       \includegraphics[width=0.48\textwidth]{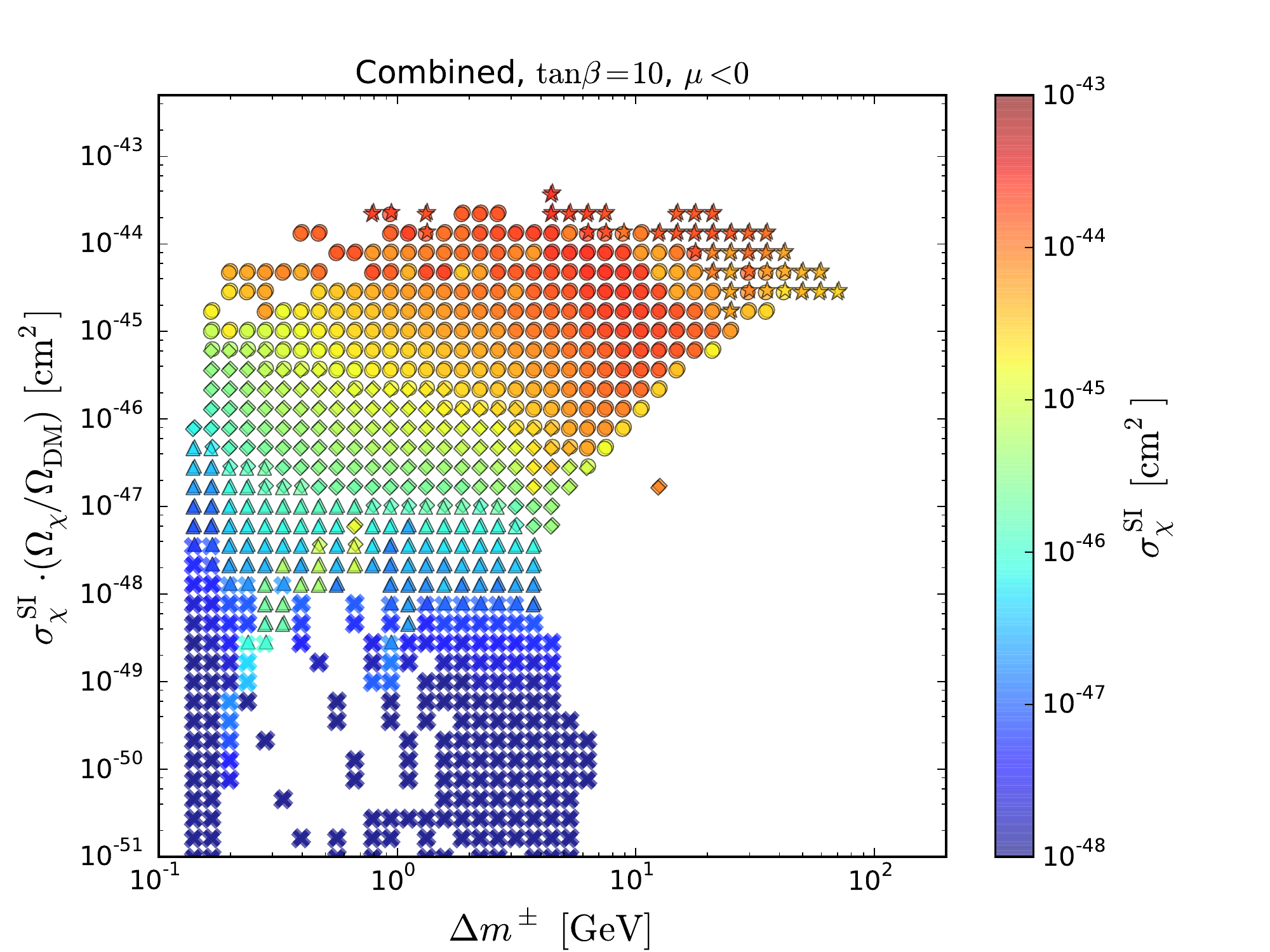}	
                       \includegraphics[width=0.48\textwidth]{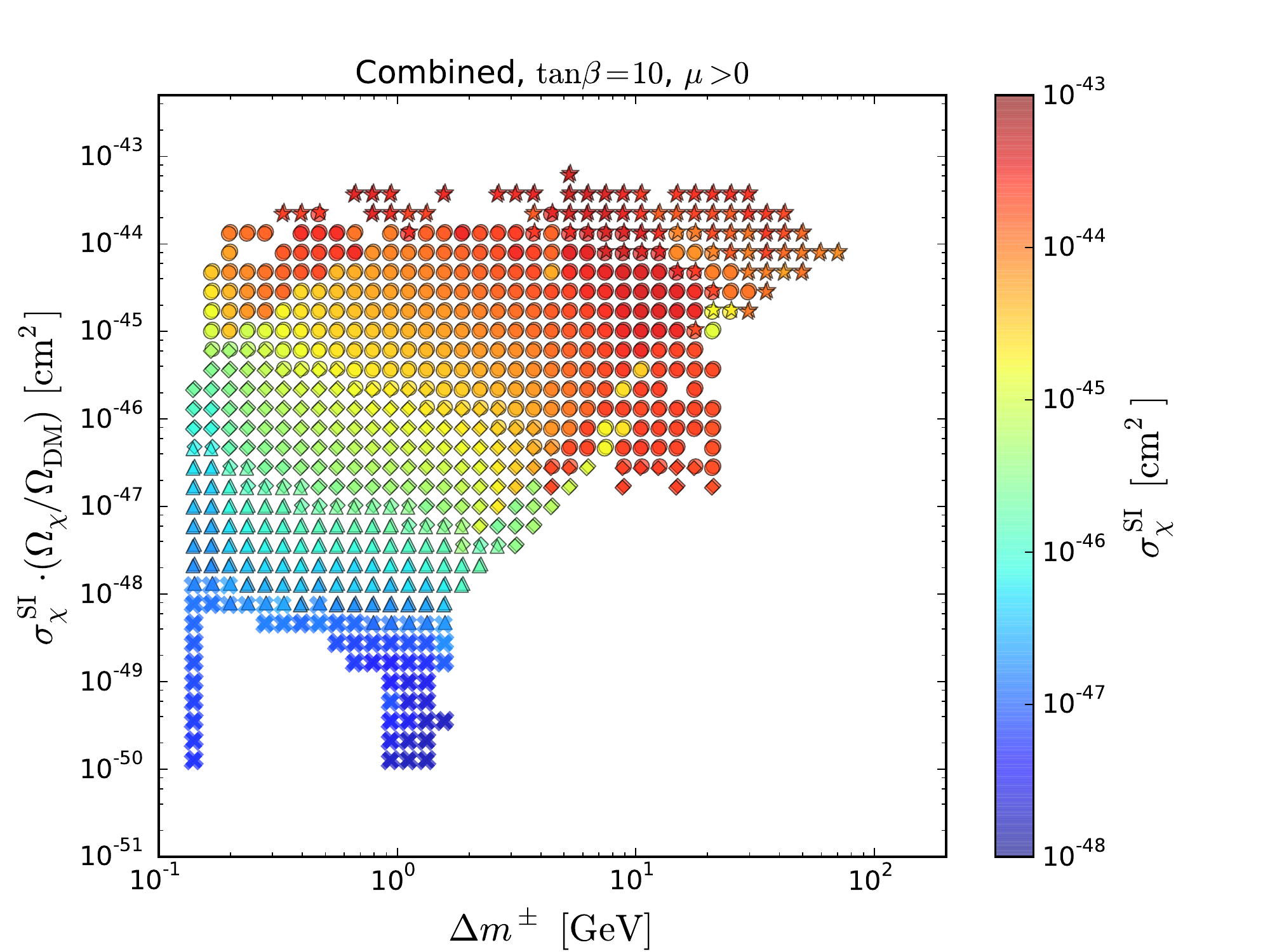}	
	
\caption{
The combined results of the scans over $M_1, M_2$  and $\mu$ for some examples of $\tan\beta$ values and signs of $\mu$.  The stars, circles, triangles, diamonds and crosses are defined in Fig.~\ref{fig:mu-m1_dec_1}
\label{fig:m2-mu_dec_13}}
\end{figure}

In the previous sections we have presented the results of several scans over certain ranges of parameters, which together  cover  essentially the whole relevant
parameter space of the MSSM-like
electroweak sector. Two conclusions have emerged from that study. One is that the two DD experiments, XENON1T and LZ, will be sensitive to neutralinos (LSPs) in
almost the entire range  of neutralino masses and compositions allowed by the bound $\Omega_{\chi} h^2\leq 0.12$. There are some exceptions, though. One to be mentioned are some regions with negative values  of 
$\mu$, particularly for small $\tan\beta$, with a dominant higgsino  component in  the LSP, where some ``blind spot" regions remain. This  is also true for the LSP with a dominant wino component, irrespectively of the sign of $\mu$. 
The second conclusion is that in the electroweakino spectrum, $\Delta m^\pm$ (defined as the mass difference between the lighter
chargino and the LSP) is smaller than 30 (10) GeV after imposing the LUX (expected from XENON1T) bounds on the neutralino spin independent scattering cross sections on nuclei. 
One should note however that, again, the small $\tan\beta$ and $\mu<0$ is the exception, as  the mass differences can remain up to 40 GeV even after the XENON1T and LZ results. Some  of the ``blind  spots" regions are accessible at  the LHC but not all of them. 
To make the generic character of those conclusions even more manifest, we show in Fig.~\ref{fig:m2-mu_dec_13} 
the combined results of the scans for  a given value of $\tan\beta$ and a given sign of $\mu$.

\begin{figure}
	\centering \vspace{-0.0cm}

\includegraphics[width=0.48\textwidth]{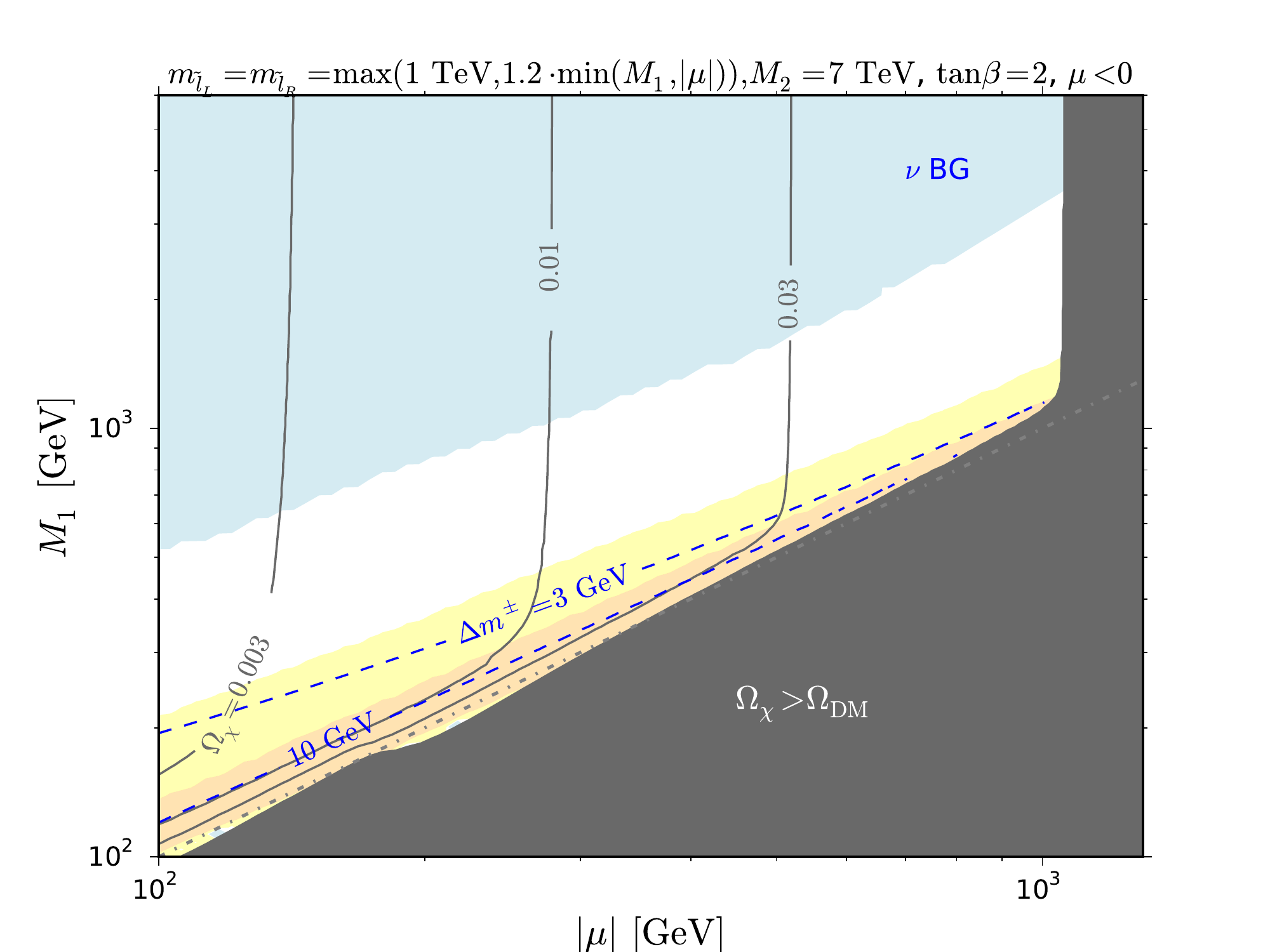}
\includegraphics[width=0.48\textwidth]{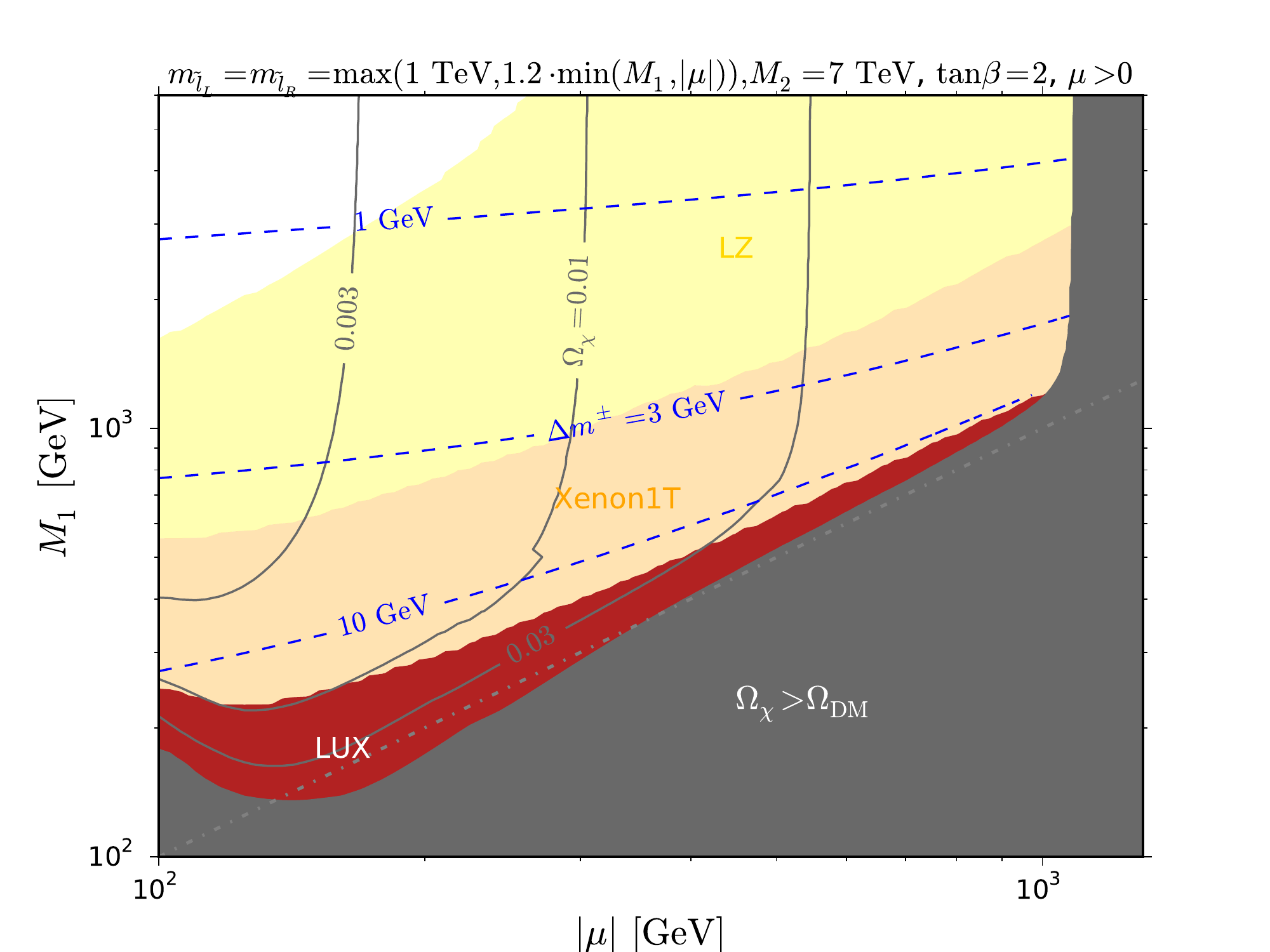}
                \includegraphics[width=0.48\textwidth]{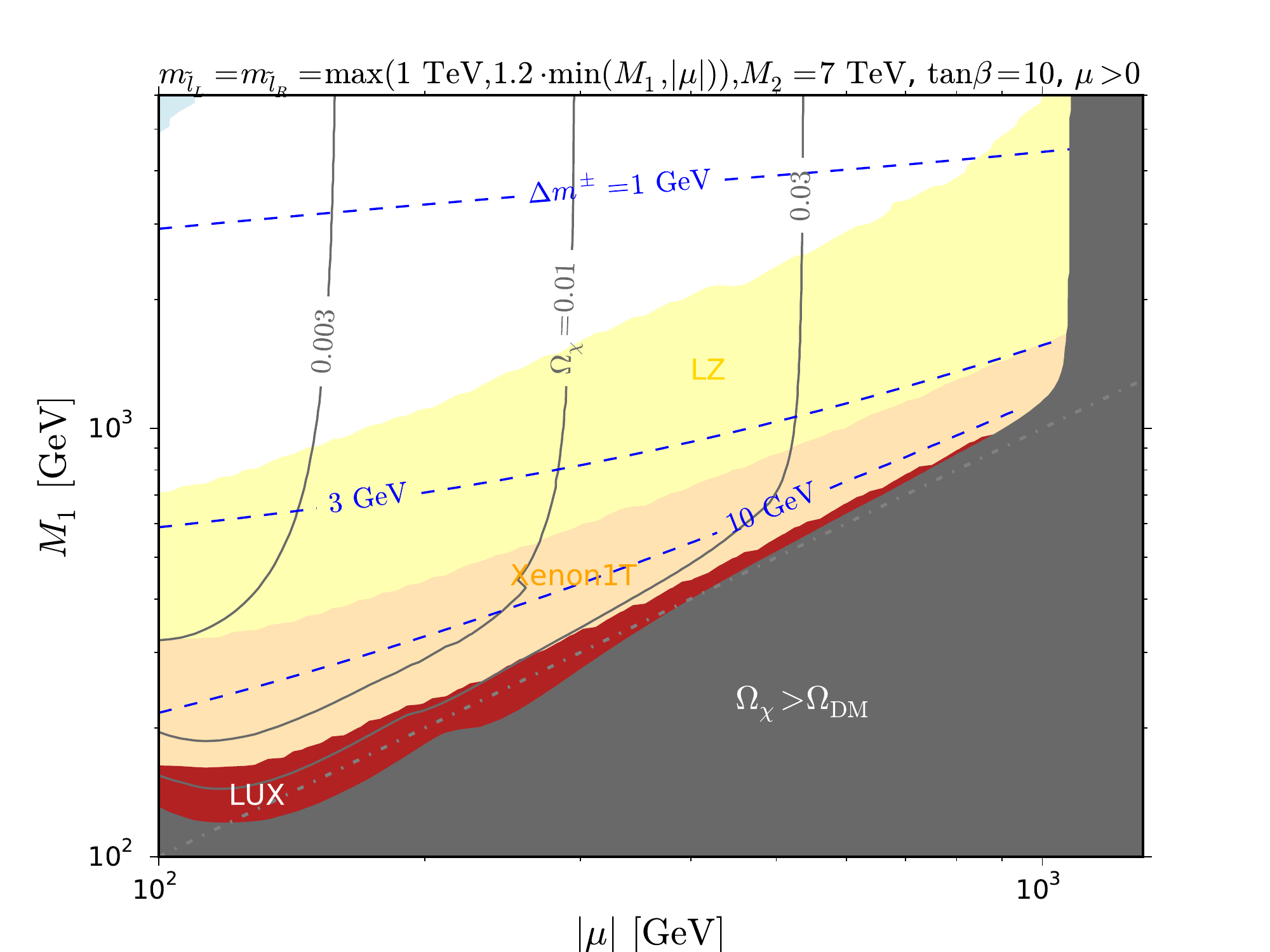}
               \includegraphics[width=0.48\textwidth]{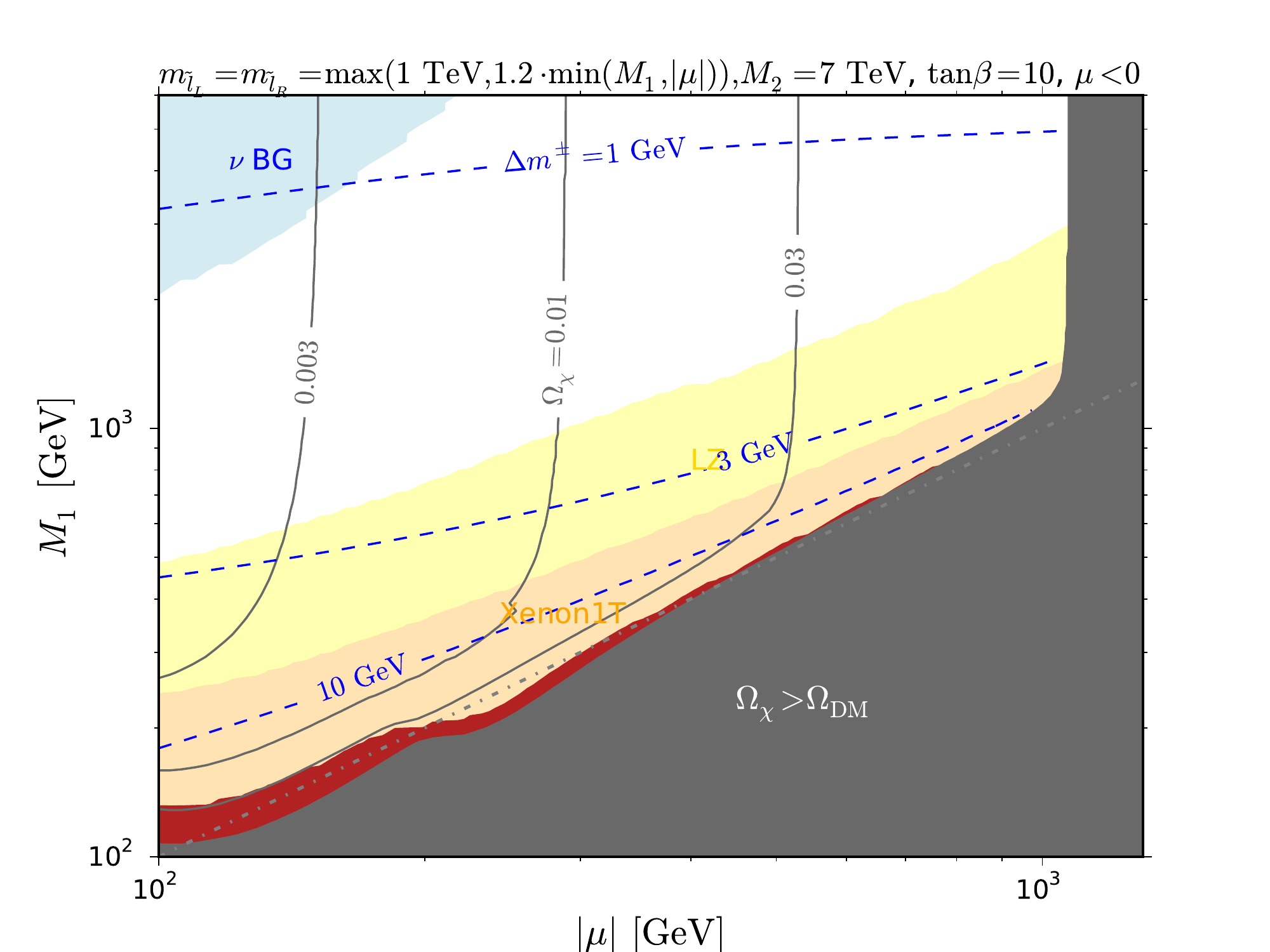}
\caption{
The results of the scan over $M_1, \mu$ with sleptons  lighter than squarks
\label{fig:leptons}}
\end{figure}

So far, we  have shown the results obtained with both slepton and squark  masses set to 7 TeV. It is interesting to check how much  our conclusions change when
the sfermions are lighter, particularly when the sleptons are lighter.  It turns out that they remain valid still when the sleptons (and of course the
squarks, too) are 20$\%$ heavier than the LSP (but not lighter than 1 TeV). This is illustrated in  Fig.~\ref{fig:leptons}, which shows the results of a scan over $|\mu|,\,M_1$  with
$m_{\tilde \ell}= \max(1 \,{\rm TeV}, 1.2 \min(M_1,|\mu|))$. Indeed, Fig.~\ref{fig:leptons} and Fig.~\ref{fig:mu-m1_dec_1}
are very similar.

\begin{figure}
	\centering \vspace{-0.0cm}		
                \includegraphics[width=0.48\textwidth]{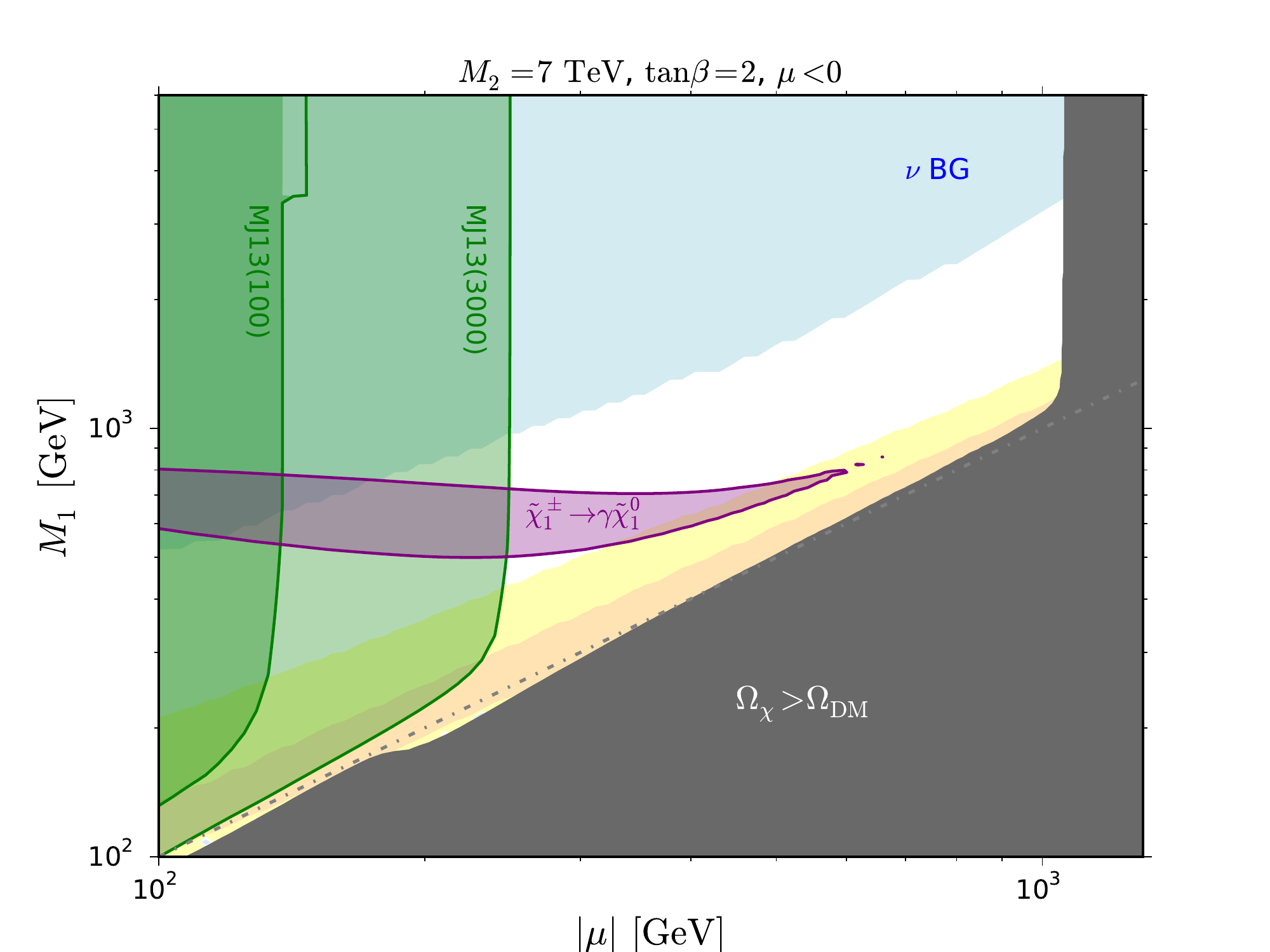}
                \includegraphics[width=0.48\textwidth]{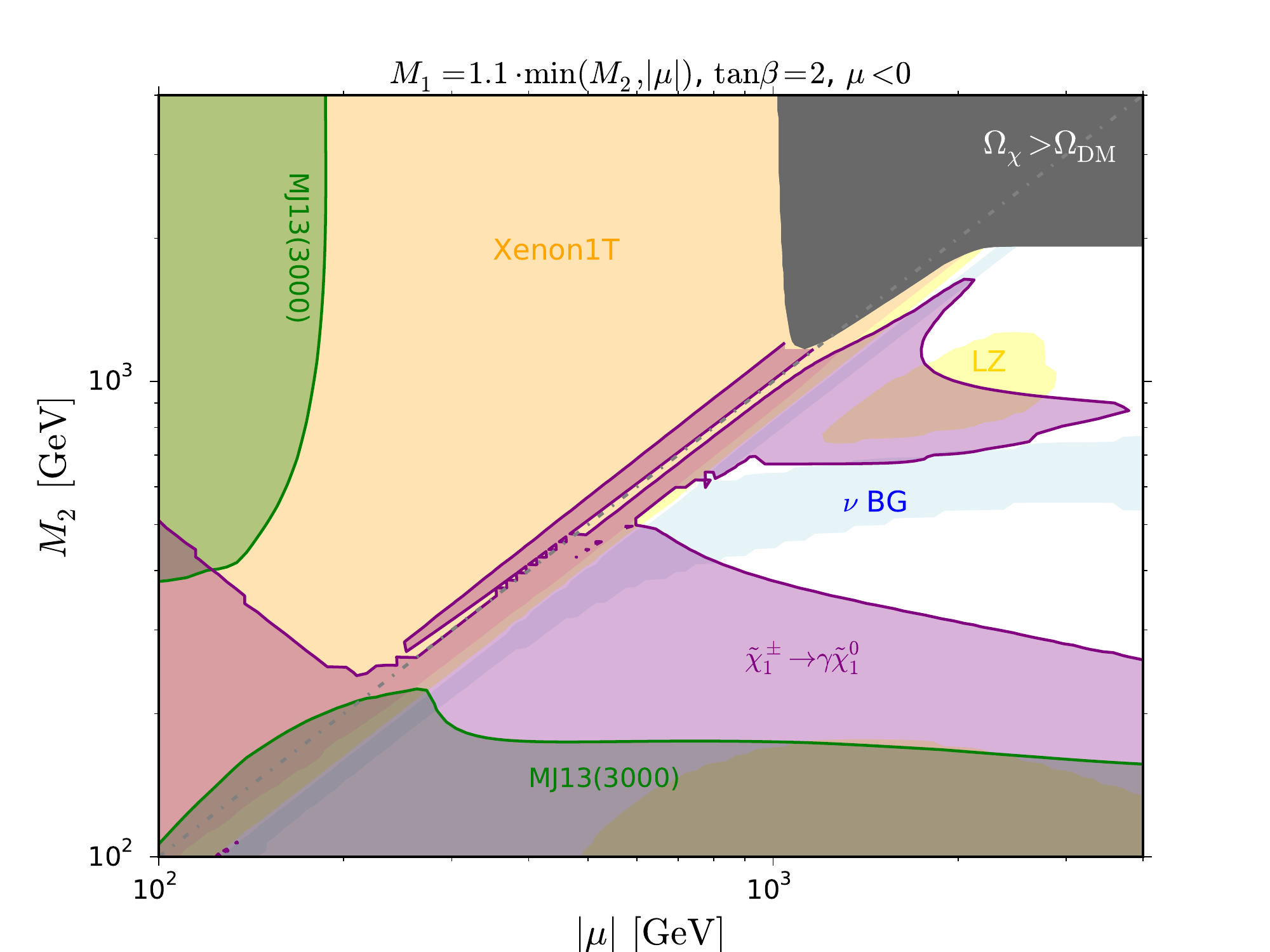}
\caption{ The region of potential sensitivity to the $\tilde \chi_2^0 \to \gamma \tilde \chi_1^0$ decay is shown by purple
\label{fig:mu-m1_dec_20}}
\end{figure}

Finally,  we comment on the collider searches.
We have presented rough estimates of the LHC sensitivity, based on the previous results of ref.~\cite{Barducci} for the monojet search,
ref.~\cite{CMS:2015vka} for $\tilde \chi_1^\pm \tilde \chi_2^0 \to (W^\pm \tilde \chi_1^0) (Z (h^0) \tilde \chi_1^0)$ search 
and ref.~\cite{ATLAS_disapppearing} for the disappearing track search. 
A more dedicated analysis focused on the parameter range of interest emerging from the present paper would certainly be worthwhile.
It should include a search for new interesting channels and techniques.  
For instance,
some studies at 13 TeV LHC \cite{Bramante_neutralino} and a 100 TeV \cite{NeutralinoSurface100} $pp$ collider have been done to attack the compressed spectra that appear in the higgsino-like LSP scenarios.
The idea is to look for a soft dilepton plus a soft photon originated from $\tilde \chi_1^\pm \to W^{*} \tilde \chi_1^0$ and $\tilde \chi_2^0 \to \gamma \tilde \chi_1^0$ decays, respectively.
Here we simply show in  Fig.~\ref{fig:mu-m1_dec_20} by purple an example of the
region with $\Delta m^0 > 3$ GeV and ${\rm Br}(\tilde \chi_2^0 \to \gamma \tilde \chi_1^0) > 0.1$.\footnote{
${\rm Br}(\tilde \chi_1^\pm \to W^* \tilde \chi_1^0) \sim 1$ across our ($|\mu|$, $M_1$) parameter plane. }
This region is potentially sensitive to more extensive analysis dedicated to this channel.

The overall conclusion of this paper is that the electroweakino sector,  being subject to important physical constraints, is a very promising territory in search for supersymmetry.  In some regions of the parameter space the DD  and the LHC experiments have comparable discovery potential and supplement each other and in some other regions they are complementary to each other. The collider discovery potential would significantly improve if new techniques were developed to search for the weakly produced states with  the mass differences between the NLSP and LSP of  order of 1 GeV.

%%%%%%%%%%%%%%%%%%%%%%%%%
\section*{Acknowledgments}
%%%%%%%%%%%%%%%%%%%%%%%%%

We thank Malcolm Fairbairn and Marc Kamionkowski for helpful discussion. 
This work was partially supported by Polish National Science Centre 
under research grants DEC-2012/05/B/ST2/02597, DEC-2014/15/B/ST2/02157
and DEC-2012/04/A/ST2/00099.
The work of AD was supported in part by the National Science Foundation 
under Grant No. PHY-1215979.
K.S. is supported in part by
the London Centre for Terauniverse Studies (LCTS), using funding from
the European Research Council 
via the Advanced Investigator Grant 267352.
MB was partially supported by the Foundation for Polish Science through its programme HOMING PLUS and the MNiSW grant IP2012 030272. 
%%%%%%%%%%%%%%%%%%%%%%%%%

%%%%%%%%%%%%%%%%%%%%%%%%%
\appendix
\section{Appendix}

In this Appendix we present approximate formulae for 
the masses of $\chi^0_1$, $\chi^0_2$ and $\chi^+_1$, 
the composition of LSP and NLSP and the LSP-Higgs 
coupling in some limits. The make the formulae more 
compact we use the notation: $s_{\theta_W}=\sin\theta_W$,
$c_{\theta_W}=\cos\theta_W$, $s_\beta=\sin\beta$, $s_{2\beta}=\sin(2\beta)$ 
etc.

\subsection{$|\mu|\ll|M_1|$, $|M_2|$}

The masses of the higgsino-dominated states up to terms of 
the second order in $1/M_{1,2}$ are given by:
\begin{align}
\label{mu-mN1}
m_{\chi^0_1}
\approx&
|\mu|
-\frac{M_Z^2}{2}
\left|
\left(\frac{c_{\theta_W}^2}{M_2}+\frac{s_{\theta_W}^2}{M_1}\right)
+\mu s_{2\beta}
\left(\frac{c_{\theta_W}^2}{M_2^2}+\frac{s_{\theta_W}^2}{M_1^2}\right)
\right|
\nn
&%\qquad
-\frac{M_Z^2}{2} \left[
\sgn(\mu)s_{2\beta}
\left(\frac{c_{\theta_W}^2}{M_2}+\frac{s_{\theta_W}^2}{M_1}\right)
+ |\mu|
\left(\frac{c_{\theta_W}^2}{M_2^2}+\frac{s_{\theta_W}^2}{M_1^2}\right)
\right]
%\nn
%&\qquad
+\frac{M_Z^4}{8|\mu|}
\left(\frac{c_{\theta_W}^2}{M_2}+\frac{s_{\theta_W}^2}{M_1}\right)^2
 c_{2\beta}^2
\,,\\
%\end{align}
%
%
%\begin{align}
\label{mu-mN2}
m_{\chi^0_2}
\approx&
|\mu|
+\frac{M_Z^2}{2}
\left|
\left(\frac{c_{\theta_W}^2}{M_2}+\frac{s_{\theta_W}^2}{M_1}\right)
+\mu s_{2\beta}
\left(\frac{c_{\theta_W}^2}{M_2^2}+\frac{s_{\theta_W}^2}{M_1^2}\right)
\right|
\nn
&%\qquad
-\frac{M_Z^2}{2} \left[
\sgn(\mu)s_{2\beta}
\left(\frac{c_{\theta_W}^2}{M_2}+\frac{s_{\theta_W}^2}{M_1}\right)
+ |\mu|
\left(\frac{c_{\theta_W}^2}{M_2^2}+\frac{s_{\theta_W}^2}{M_1^2}\right)
\right]
%\nn
%&\qquad
+\frac{M_Z^4}{8|\mu|}
\left(\frac{c_{\theta_W}^2}{M_2}+\frac{s_{\theta_W}^2}{M_1}\right)^2
 c_{2\beta}^2
\,,\\
%\end{align}
%
%
%\begin{align}
\label{mu-mC1}
m_{\chi^+_1}
\approx&
\sgn(\mu)\left[
\mu
-M_Z^2 \frac{c_{\theta_W}^2}{M_2}\left( s_{2\beta} + \frac{\mu}{M_2}\right)
\right]
\,.
\end{align}
In the same approximation the mass differences read:
\begin{align}
\label{mu-mN2N1}
m_{\chi^0_2} - m_{\chi^0_1}
\approx&
M_Z^2
\left|
\left(\frac{c_{\theta_W}^2}{M_2}+\frac{s_{\theta_W}^2}{M_1}\right)
+\mu s_{2\beta}
\left(\frac{c_{\theta_W}^2}{M_2^2}+\frac{s_{\theta_W}^2}{M_1^2}\right)
\right|
\,,\\
%\end{equation}
%
%
%\begin{align}
\label{mu-mC1N1}
m_{\chi^+_1} - m_{\chi^0_1}
\approx&
\frac{M_Z^2}{2}
\left|
\left(\frac{c_{\theta_W}^2}{M_2}+\frac{s_{\theta_W}^2}{M_1}\right)
+\mu s_{2\beta}
\left(\frac{c_{\theta_W}^2}{M_2^2}+\frac{s_{\theta_W}^2}{M_1^2}\right)
\right|
\nn
&
-\frac{M_Z^2}{2} \left[
\sgn(\mu)s_{2\beta}
\left(\frac{c_{\theta_W}^2}{M_2}-\frac{s_{\theta_W}^2}{M_1}\right)
+ |\mu|
\left(\frac{c_{\theta_W}^2}{M_2^2}-\frac{s_{\theta_W}^2}{M_1^2}\right)
\right]
%\nn&
-\frac{M_Z^4}{8|\mu|}
\left(\frac{c_{\theta_W}^2}{M_2}+\frac{s_{\theta_W}^2}{M_1}\right)^2 
 c_{2\beta}^2
\,.
\end{align}
The composition of the LSP (NLSP) is given by
\begin{align}
\label{mu-Ni1}
N_{i1}\approx
&-\frac{1}{\sqrt{2}}\,\frac{M_Z}{M_1}
\left(1-(-1)^i\eta \,\frac{|\mu|}{M_1}\right)
s_{\theta_W} \left( s_\beta-(-1)^i\eta \sgn(\mu) c_\beta\right)
\nn
&+\frac{1}{4\sqrt{2}}\,\frac{M_Z^3}{|\mu|M_1}
\left(\frac{c_{\theta_W}^2}{M_2}+\frac{s_{\theta_W}^2}{M_1}\right)
s_{\theta_W} c_{2\beta}
\left((-1)^i\eta \, s_\beta+\sgn(\mu) c_\beta\right)
\,,\\
\label{mu-Ni2}
N_{i2}\approx
&+\frac{1}{\sqrt{2}}\,\frac{M_Z}{M_2}
\left(1-(-1)^i\eta \,\frac{|\mu|}{M_2}\right)
c_{\theta_W} \left( s_\beta-(-1)^i\eta \,\sgn(\mu) c_\beta\right)
\nn
&-\frac{1}{4\sqrt{2}}\,\frac{M_Z^3}{|\mu|M_2}
\left(\frac{c_{\theta_W}^2}{M_2}+\frac{s_{\theta_W}^2}{M_1}\right)
c_{\theta_W} c_{2\beta}
\left((-1)^i\eta \, s_\beta+\sgn(\mu) c_\beta\right)
\,,\\
\label{mu-Ni3}
N_{i3}\approx
&\,\,\frac{\sgn(\mu)}{\sqrt{2}}
\left[(-1)^i\eta  
+\frac14\,\frac{M_Z^2}{|\mu|}
\left(\frac{c_{\theta_W}^2}{M_2}+\frac{s_{\theta_W}^2}{M_1}\right)
c_{2\beta}
%\right.
%\nn
%&\qquad
+\frac{M_Z^2}{4}
\left(\frac{c_{\theta_W}^2}{M_2^2}+\frac{s_{\theta_W}^2}{M_1^2}\right)
\left(\sgn(\mu)s_{2\beta}-2(-1)^i\eta  c^2_\beta\right)
\right.
\nn
&\qquad
\left.+\frac{1}{32}\,\frac{M_Z^4}{\mu^2}
\left(\frac{c_{\theta_W}^2}{M_2}+\frac{s_{\theta_W}^2}{M_1}\right)^2
c_{2\beta}\left(4\sgn(\mu)s_{2\beta}-(-1)^i\eta c_{2\beta}\right)
\right]
\,,\\
\label{mu-Ni4}
N_{i4}\approx
&\,\,\frac{1}{\sqrt{2}}
\left[1
-(-1)^i\eta \,\frac14\,\frac{M_Z^2}{|\mu|}
\left(\frac{c_{\theta_W}^2}{M_2}+\frac{s_{\theta_W}^2}{M_1}\right)
c_{2\beta}
%\right.
%\nn
%&\qquad
-\frac{M_Z^2}{4}
\left(\frac{c_{\theta_W}^2}{M_2^2}+\frac{s_{\theta_W}^2}{M_1^2}\right)
\left(-(-1)^i\eta \sgn(\mu)s_{2\beta}+2 s^2_\beta\right)
\right.
\nn
&\qquad
\left.-\frac{1}{32}\,\frac{M_Z^4}{\mu^2}
\left(\frac{c_{\theta_W}^2}{M_2}+\frac{s_{\theta_W}^2}{M_1}\right)^2
c_{2\beta}\left(4(-1)^i\eta \sgn(\mu)s_{2\beta}+c_{2\beta}\right)
\right]
\,,
\end{align}
where $i=1,2$ and
\begin{equation}
\label{mu-eta_i}
\eta
=
\sgn
\left(\frac{c_{\theta_W}^2}{M_2}+\frac{s_{\theta_W}^2}{M_1}\right)
\,.
\end{equation}
The coupling of the LSP to the Higgs scalar, 
in the decoupling limit, is given by
\begin{equation}
\label{c_hchichi}
c_{h\chi\chi}
=
\frac12\,\left(g_2 N_{12}-g_1 N_{11}\right)
\left(s_\beta N_{14} - c_\beta N_{13}\right)
\,.
\end{equation}
Using eqs.\ (\ref{mu-Ni1})--(\ref{mu-Ni4}) for $i=1$, 
and keeping terms up to the first subleading order we get
\begin{equation}
\label{mu-c}
c_{h\chi\chi}
\approx
\frac{M_Z}{4}\eta
\left[\left|\frac{g_2 c_{\theta_W}}{M_2}+\frac{g_1 s_{\theta_W}}{M_1}\right|
+
|\mu|\left(
\frac{g_2 c_{\theta_W}}{M_2^2}+\frac{g_1 s_{\theta_W}}{M_1^2}
\right)
\right]
\left(1+\eta\sgn(\mu)s_{2\beta}\right).
\end{equation}
For decoupled wino this simplifies to
\begin{equation}
\label{mu-c_noM2}
c_{h\chi\chi}
\approx
\frac{g_1M_Z s_{\theta_W}}{4M_1}\left(1+\frac{|\mu|}{|M_1|}\right)
(1+\sgn(\mu M_1)s_{2\beta})
\end{equation}
which, for positive $M_1$ and with one higher order term added, 
gives eq.\ (\ref{eq:c_mu<M1}).
Using eqs.\ (\ref{mu-Ni3}) and (\ref{mu-Ni4}) for $i=1$, 
one may calculate the coupling of the LSP to the $Z$ boson, 
$c_{Z\chi\chi}=\frac{g_2}{2c_{\theta_W}}(N_{13}^2-N_{14}^2)$. 
The result, up to the first subleading order in $M_{1,2}^{-1}$, reads:
\begin{equation}
\label{mu-cZ}
c_{Z\chi\chi}
\approx
-\frac{g_2 c_{2\beta}}{4 c_{\theta_W}}
%-\frac14\sqrt{g_1^2+g_2^2}\, c_{2\beta}
\left[
\frac{M_Z^2}{|\mu|}
\left|\frac{c_{\theta_W}^2}{M_2}+\frac{s_{\theta_W}^2}{M_1}\right|
+M_Z^2\left(\frac{c_{\theta_W}^2}{M_2^2}+\frac{s_{\theta_W}^2}{M_1^2}\right)
+\frac{\eta}{2}s_{2\beta}\frac{M_Z^4}{\mu|\mu|}
\left(\frac{c_{\theta_W}^2}{M_2}+\frac{s_{\theta_W}^2}{M_1}\right)^{\!\!2}
\right].
\end{equation}
%
%

%%%%%%%%%%%%%%%%%%%%%%%%%%%%%%%%%%%%%%%%%%%%%%%%%%%%%%%%%%%%%%%%%%%%%

\subsection{$|M_1|$, $|M_2|\ll|\mu|$}

Masses of light gaugino-dominated states:
\begin{align}
\label{M1M2-mNW}
m_{\chi^0_W}
&\approx
\left|M_2
-\frac{M_Z^2}{\mu} c_{\theta_W}^2 s_{2\beta}
-\frac{M_2M_Z^2}{\mu^2} c_{\theta_W}^2 
+\frac14\,\frac{M_Z^4}{\mu^2(M_2-M_1)}
\ s^2_{2\theta_W} s_{2\beta}^2\right|
\,,\\
%\end{equation}
%
%
%\begin{equation}
\label{M1M2-mNB}
m_{\chi^0_B}
&\approx
\left|M_1
-\frac{M_Z^2}{\mu} s_{\theta_W}^2 s_{2\beta}
-\frac{M_1M_Z^2}{\mu^2} s_{\theta_W}^2
-\frac14\,\frac{M_Z^4}{\mu^2(M_2-M_1)}
\ s^2_{2\theta_W} s_{2\beta}^2\right|
\,,\\
%\end{equation}
%
%
%\begin{equation}
\label{M1M2-mC1}
m_{\chi^+_1}
&\approx
\left|
M_2
-\frac{M_Z^2}{\mu}\,c_{\theta_W}^2 s_{2\beta}
-\frac{M_2 M_Z^2}{\mu^2}\,c_{\theta_W}^2
\right|
\,.
\end{align}
When $|M_1|<|M_2|$ the LSP is bino-dominated $\chi^0_B$ and 
\begin{equation}
\label{M1M2-mC1NB}
m_{\chi^+_1} - m_{\chi^0_1}
\approx
m_{\chi^0_2} - m_{\chi^0_1}
\approx
|M_2|-|M_1|
+\frac{M_Z^2}{|\mu|}\,s_{2\beta}
\left[\sgn(M_1)\,s_{\theta_W}^2-\sgn(M_2)\,c_{\theta_W}^2\right]
\end{equation}
while for $|M_2|<|M_1|$ the LSP is wino-dominated $\chi^0_W$ and 
\begin{equation}
\label{M1M2-mC1NW}
m_{\chi^+_1} - m_{\chi^0_1}
\approx
\frac14\,\sgn(M_1M_2)
\frac{M_Z^4}{\mu^2|M_1-M_2|}
\ s^2_{2\theta_W} s_{2\beta}^2
\,.
\end{equation}
So, this mass difference is positive (negative) when $M_1$ and $M_2$ 
have the same (oppositve) sign. 
The LSP and NLSP composition read:
\begin{align}
\label{M1M2-NB1}
N_{B1}\approx&\,\,
1-\frac12\,\frac{M_Z^2}{\mu^2}\,s_{\theta_W}^2
-\frac18\,\frac{M_Z^4}{\mu^2(M_2-M_1)^2}\,\ s^2_{2\theta_W}
 s_{2\beta}^2
\,,\\
\label{M1M2-NB2}
N_{B2}\approx&
-\frac12\,\frac{M_Z^2}{\mu(M_2-M_1)}\, s_{2\theta_W} s_{2\beta}
%\nn&
-\frac12\,\frac{M_1M_Z^2}{\mu^2(M_2-M_1)}\, s_{2\theta_W} 
-\frac14\,\frac{M_Z^4}{\mu^2(M_2-M_1)^2}\, s_{4\theta_W}  s_{2\beta}^2
\,\\
\label{M1M2-NB3}
N_{B3}\approx&
+\frac{M_Z}{\mu}\,s_{\theta_W}  s_\beta
+\frac{M_1M_Z}{\mu^2}\,s_{\theta_W}  c_\beta
+\frac{M_Z^3}{\mu^2(M_2-M_1)}\,s_{\theta_W} c_{\theta_W}^2  s_\beta s_{2\beta}
\,,\\
\label{M1M2-NB4}
N_{B4}\approx&
-\frac{M_Z}{\mu}\,s_{\theta_W}  c_\beta
-\frac{M_1M_Z}{\mu^2}\,s_{\theta_W}  s_\beta
-\frac{M_Z^3}{\mu^2(M_2-M_1)}\,s_{\theta_W} c_{\theta_W}^2  c_\beta s_{2\beta}
\end{align}
\begin{align}
\label{M1M2-NW1}
N_{W1}\approx&
+\frac12\,\frac{M_Z^2}{\mu(M_2-M_1)}\, s_{2\theta_W} s_{2\beta}
%\nn&
+\frac12\,\frac{M_2M_Z^2}{\mu^2(M_2-M_1)}\, s_{2\theta_W} 
+\frac14\,\frac{M_Z^4}{\mu^2(M_2-M_1)^2}\, s_{4\theta_W}  s_{2\beta}^2
\,,\\
\label{M1M2-NW2}
N_{W2}\approx&\,\,
1-\frac12\,\frac{M_Z^2}{\mu^2}\,c_{\theta_W}^2
-\frac18\,\frac{M_Z^4}{\mu^2(M_2-M_1)^2}\,\ s^2_{2\theta_W}
 s_{2\beta}^2
\,,\\
\label{M1M2-NW3}
N_{W3}\approx&
-\frac{M_Z}{\mu}\,c_{\theta_W}  s_\beta
-\frac{M_2M_Z}{\mu^2}\,c_{\theta_W}  c_\beta
+\frac{M_Z^3}{\mu^2(M_2-M_1)}\,s_{\theta_W}^2 c_{\theta_W}  s_\beta s_{2\beta}
\,,\\
\label{M1M2-NW4}
N_{W4}\approx&
+\frac{M_Z}{\mu}\,c_{\theta_W}  c_\beta
+\frac{M_2M_Z}{\mu^2}\,c_{\theta_W}  s_\beta
-\frac{M_Z^3}{\mu^2(M_2-M_1)}\,s_{\theta_W}^2 c_{\theta_W}  c_\beta s_{2\beta}
\,.
\end{align}
The approximation used for $|M_1|$, $|M_2|\ll|\mu|$
brakes down for very small values of the difference $|M_2-M_1|$.

The coupling to the Higgs boson is approximately given by
\begin{equation}
\label{M1M2-cB}
c_{h\chi\chi}
\approx
\frac{g_1}{2}
\,\frac{M_Z s_{\theta_W}}{\mu}\left(s_{2\beta}
+\frac{M_1}{\mu}\right)
\left[
1+\frac{M_Z^2 c_{\theta_W}^2}{\mu(M_2-M_1)}\, s_{2\beta}
\right]
\end{equation}
for the bino-dominated LSP and 
\begin{equation}
\label{M1M2-cW}
c_{h\chi\chi}
\approx
\frac{g_2}{2}
\,\frac{M_Z c_{\theta_W}}{\mu}\left(s_{2\beta}
+\frac{M_2}{\mu}\right)
\left[
1+\frac{M_Z^2 s_{\theta_W}^2}{\mu(M_1-M_2)}\, s_{2\beta}
\right]
\end{equation}
for the wino-dominated one.
The second term in the square bracket of eq.\ (\ref{M1M2-cB}) 
formally gives contributions of the sub-sub-leading order. 
However, for small values of the difference 
$(M_2-M_1)$ it may lead to important numerical effects. For small 
$\tan\beta$ and negative (positive) product $\mu(M_2-M_1)$ this term 
gives quite substantial decrease (increase) of $c_{h\chi\chi}$ of the 
bino-dominated LSP. Analogous term in eq.\ (\ref{M1M2-cW}) 
is somewhat less important for the case of the wino-dominated LSP 
because $s_{\theta_W}^2$ is more than 3 times smaller than $c_{\theta_W}^2$.

The coupling to the $Z$ boson is approximately equal
\begin{equation}
\label{M1M2-cZB}
c_{Z\chi\chi}
\approx
-\frac{g_1 s_{\theta_W}}{2}\,c_{2\beta}\,\frac{M_Z^2}{\mu^2}
\left[
1-\frac{M_1^2}{\mu^2}+\frac{2M_Z^2 c_{\theta_W}^2}{\mu(M_2-M_1)}\, s_{2\beta}
\right]
\end{equation}
for the  bino-dominated LSP and 
\begin{equation}
\label{M1M2-cZW}
c_{Z\chi\chi}
\approx
-\frac{g_2 c_{\theta_W}}{2}\,c_{2\beta}\,\frac{M_Z^2}{\mu^2}
\left[
1-\frac{M_2^2}{\mu^2}+\frac{2M_Z^2 s_{\theta_W}^2}{\mu(M_1-M_2)}\, s_{2\beta}
\right]
\end{equation}
for the wino-dominated one.

%%%%%%%%%%%%%%%%%%%%%%%%%%%%%%%%%%%%%%%%%%%%%%%%%%%%%%%%%%%%%%%%%%%%%%%%

\subsection{$|M_2|\ll|\mu|,|M_1|$}

The masses of the wino-dominated states are approximated by:
\begin{align}
\label{M2-mN1}
m_{\chi^0_1}
\approx&
\left|M_2
-\frac{M_Z^2}{\mu} c_{\theta_W}^2 s_{2\beta}
-\frac{M_2M_Z^2}{\mu^2} c_{\theta_W}^2 
%\right.\nn&\left.
+\frac{M_Z^4}{\mu^3} c^4_{\theta_W}s_{2\beta}
-\frac{M_2^2M_Z^2}{\mu^3}c_{\theta_W}^2s_{2\beta}
-\frac14\,\frac{M_Z^4}{\mu^2M_1}
\ s^2_{2\theta_W} s_{2\beta}^2
\right|
\,,\\
%\end{align}
%
%
%
%\begin{align}
\label{M2-mC1}
m_{\chi^+_1}
\approx&
\left|M_2
-\frac{M_Z^2}{\mu} c_{\theta_W}^2 s_{2\beta}
%\right.
-\frac{M_2M_Z^2}{\mu^2} c_{\theta_W}^2 
%\nn&\left.
+\frac{M_Z^4}{\mu^3} c^4_{\theta_W}s_{2\beta}
-\frac{M_2^2M_Z^2}{\mu^3}c_{\theta_W}^2s_{2\beta}
\right|
\,.
\end{align}
We keep the third order terms in the above expansions 
because the leading contribution to the NLSP(chargino)--LSP 
mass difference is of this order:
\begin{equation}
\label{M2-mC1N1}
m_{\chi^+_1} - m_{\chi^0_1}
\approx
\frac14\,\sgn(M_2)
\frac{M_Z^4}{\mu^2 M_1 }
\ s^2_{2\theta_W} s_{2\beta}^2
\,.
\end{equation}
As in the previous case this mass difference is positive 
(negative) when $M_1$ and $M_2$ have the same (oppositve) 
sign. But its absolute 
value is small so $m_{\chi^+_1} - m_{\chi^0_1}$ is dominated by the loop 
contribution.
For totally decoupled bino the tree level mass difference 
of the NLSP and LSP starts at the 4-th order: 
\begin{equation}
\label{M2-mC1N1'}
m_{\chi^+_1} - m_{\chi^0_1}
\approx
\frac12  |M_2|  c^4_{\theta_W}   c_{2\beta}^2  \frac{M_Z^4}{\mu^4}
\,.
\end{equation}
The LSP is much lighter than all other nautralinos:
\begin{equation}
\label{M2-mN2N1}
m_{\chi^0_2} - m_{\chi^0_1}
\approx
\min\left(|\mu|,|M_1|\right)-|M_2|
\,.
\end{equation}
The composition of the LSP is given by:
\begin{align}
\label{M2-N11}
N_{11}\approx&\,\,
-\frac12\,\frac{M_Z^2}{\mu M_1}\, s_{2\theta_W} s_{2\beta}
\,,\\
\label{M2-N12}
N_{12}\approx&\,\,
1-\frac12\,\frac{M_Z^2}{\mu^2}\,c_{\theta_W}^2
\,,\\
\label{M2-N13}
N_{13}\approx&
-\frac{M_Z}{\mu}\,c_{\theta_W}  s_\beta
-\frac{M_2M_Z}{\mu^2}\,c_{\theta_W}  c_\beta
\,,\\
\label{M2-N14}
N_{14}\approx&
+\frac{M_Z}{\mu}\,c_{\theta_W}  c_\beta
+\frac{M_2M_Z}{\mu^2}\,c_{\theta_W}  s_\beta
\,.
\end{align}

Substituting the above components into eq.\ (\ref{c_hchichi})
one can find the following leading terms of the LSP-Higgs 
coupling
\begin{equation}
\label{M2-c}
c_{h\chi\chi}
\approx
\frac{g_2}{2} c_{\theta_W} M_Z\,\frac{M_2+\mu s_{2\beta}}{\mu^2}
\,.
\end{equation}
Taking into account more subleading terms, not given in eqs.\ 
(\ref{M2-N11})-(\ref{M2-N14}), one can complete the denominator 
in the above equation to the form $\mu^2-M_2^2$. 
The coupling $c_{Z\chi\chi}$ is the same as in eq.~(\ref{M1M2-cZW}) 
with the last term in the square bracket less important because of 
a bigger value of $M_1$.

%%%%%%%%%%%%%%%%%%%%%%%%%%%%%%%%%%%%%%%%%%%%%%%%%%%%%%%%%%%%%%%%%%%%

\subsection{$|M_1|\ll|\mu|$, $|M_2|$}

The mass splittings are big and dominated by: 
\begin{equation}
\label{M1-masses}
m_{\chi^+_1} - m_{\chi^0_1}
\approx
m_{\chi^0_2} - m_{\chi^0_1}
\approx
\min\left(|\mu|,|M_2|\right)-|M_1|
\,.
\end{equation}
The LSP is bino-dominated with the composition is given by:
\begin{align}
\label{M1-N11}
N_{11}\approx&\,\,
1-\frac12\,\frac{M_Z^2}{\mu^2}\,s_{\theta_W}^2
\,,\\
\label{M1-N12}
N_{12}\approx&\,\,
-\frac12\,\frac{M_Z^2}{\mu M_2}\, s_{2\theta_W} s_{2\beta}
\,,\\
\label{M1-N13}
N_{13}\approx&
+\frac{M_Z}{\mu}\,s_{\theta_W}  s_\beta
+\frac{M_1M_Z}{\mu^2}\,s_{\theta_W}  c_\beta
\,,\\
\label{M1-N14}
N_{14}\approx&
-\frac{M_Z}{\mu}\,s_{\theta_W}  c_\beta
-\frac{M_1M_Z}{\mu^2}\,s_{\theta_W}  s_\beta
\,.\end{align}
The LSP-Higgs coupling can be calculated in the same 
way as in the previous subsection and reads
\begin{equation}
\label{M1-c}
c_{h\chi\chi}
\approx
\frac{g_1}{2} s_{\theta_W} M_Z\,\frac{M_1+\mu s_{2\beta}}{\mu^2}
\,.
\end{equation}
Also here the denominator can be completed to the form 
$\mu^2-M_1^2$ presented in eq.\ (\ref{eq:c_mu>M1}).
The coupling $c_{Z\chi\chi}$ is the same as in eq.~(\ref{M1M2-cZB}) 
with the last term in the square bracket less important because of 
a bigger value of $M_2$.

%%%%%%%%%%%%%%%%%%%%%%%%%%%%%%%%%%%%%%%%%%%

\subsection{$|M_i|\sim|\mu|$ with $|M_j|\to\infty$}

In the previous subsections of the Appendix we used the approximation 
of small mixings. The LSP was dominated by one of the gauginos
or a mixture of higgsinos (with coefficients close to $\pm\frac{1}{\sqrt{2}}$).
Now we turn to the approximation of the higgsino-gaugino mixing 
close to the maximal one in which the LSP component is close to 
 $\pm\frac{1}{\sqrt{2}}$ for one of the gauginos and $\pm\frac12$ for 
the higgsinos. This corresponds to the situation when the off-diagonal 
entries of the neutralino mass matrix are much smaller than the 
diagonal ones but (much) bigger then the difference between a pair 
of the diagonal terms. Let us start with the case of the bino-higgsino LSP. 
Using the approximation 
$|M_1|+|\mu| \gg M_Z\sin\theta_W \gg ||M_1|-|\mu||$ 
one finds:\footnote
{
In addition one has to assume that 
$|M_1|\lesssim|\mu|+\sqrt{2}M_Z s_{\theta_W}\sqrt{1+\zeta_1 s_{2\beta}}
+M_Z^2 s_{\theta_W}^2 (1-\zeta_1 s_{2\beta})/[2(|M_1|+|\mu|)]$. 
Otherwise the LSP is higgsino-dominated with the mass given 
by (\ref{M1=mu-mN2}).
}

\begin{align}
\label{M1=mu-mN1}
m_{\chi^0_1}
\approx&\,\,
\frac{|M_1|+|\mu|}{2}
-\frac{\sqrt{2}}{2} M_Z s_{\theta_W} \sqrt{1+\zeta_1 s_{2\beta}}
+\frac14 \frac{M_Z^2 s_{\theta_W}^2}{|M_1|+|\mu|} \left(1-\zeta_1 s_{2\beta}\right)
-\frac{\sqrt{2}}{8}\,
\frac{(|M_1|-|\mu|)^2}{M_Z s_{\theta_W}\sqrt{1+\zeta_1 s_{2\beta}}}
\,,\\
\label{M1=mu-mN2}
m_{\chi^0_2}
\approx&\,\,
|\mu|+\frac12 \frac{M_Z^2 s_{\theta_W}^2}{|M_1|+|\mu|} \left(1-\zeta_1 s_{2\beta}\right)
\,,\\
\label{M1=mu-mC1}
m_{\chi_1^+} 
\approx&\,\, 
|\mu|
\,,
\end{align}
where, in order to make the expressions in this subsection more compact, 
we introduced the notation $\zeta_i=\sgn(\mu M_i)$. 
The chargino-neutralino mass difference, $m_{\chi_1^+}-m_{\chi_1^0}$ calculated 
from the above formulae may be negative or positive. This sign  
is important because it determines the electric charge of the LSP
so we include the leading corrections from the finite value of $M_2$. 
With such corrections we get
\begin{align}
\label{M1=M2-mC1N1}
m_{\chi_1^+}-m_{\chi_1^0}
\approx&\,
\frac{|\mu|-|M_1|}{2} 
+\frac{\sqrt{2}}{2} M_Z\, s_{\theta_W} \sqrt{1+\zeta_1 s_{2\beta}}
-\frac14 \frac{M_Z^2 \,s_{\theta_W}^2}{|M_1|+|\mu|} \left(1-\zeta_1 s_{2\beta}\right)
\nn
&-\sgn(M_2\mu)\frac{M_Z^2}{|M_2|}\,s_{2\beta}
+\frac14\sgn(M_1M_2)\frac{M_Z^4\,s^2_{2\theta_W}}{\mu^2|M_2|}s^2_{2\beta}
\end{align} 
The composition of the lightest neutralino reads:
\begin{align}
\label{M1=mu-N11}
N_{11}\approx
\sgn(M_1)&\left[-\frac{\sqrt{2}}{2}
+\frac18 \frac{M_Z s_{\theta_W}}{|M_1|+|\mu|} 
\frac{1-\zeta_1 s_{2\beta}}{\sqrt{1+\zeta_1 s_{2\beta}}}
+\frac14 \frac{|M_1|-|\mu|}{M_Z s_{\theta_W}} \frac{1}{\sqrt{1+\zeta_1 s_{2\beta}}}
\right.\nn[4pt]
&+\frac{\sqrt{2}}{32}\, \frac{|M_1|-|\mu|}{|M_1|+|\mu|}\,
\frac{1-\zeta_1 s_{2\beta}}{1+\zeta_1 s_{2\beta}}
+\frac{\sqrt{2}}{32} \,\frac{(|M_1|-|\mu|)^2}{M_Z^2 s_{\theta_W}^2}\,
\frac{1}{1+\zeta_1 s_{2\beta}}
\nn[4pt]
&\left.+\frac{\sqrt{2}}{128}\,  \frac{M_Z^2 s_{\theta_W}^2}{(|M_1|+|\mu|)^2}
(1-\zeta_1 s_{2\beta})\left(5+\frac{6}{1+\zeta_1 s_{2\beta}}
\right)\right]
\,,\end{align}
\begin{align}
\label{M1=mu-N13}
N_{13}
\approx\sgn(M_1\mu)
&\left[
-\frac12
-\frac{\sqrt{2}}{16} \frac{M_Z s_{\theta_W}}{|M_1|+|\mu|} 
\frac{1-\zeta_1 s_{2\beta}-4c_{2\beta}}{\sqrt{1+\zeta_1 s_{2\beta}}}
-\frac{\sqrt{2}}{8} \frac{|M_1|-|\mu|}{M_Z s_{\theta_W}} 
\frac{1}{\sqrt{1+\zeta_1 s_{2\beta}}}
\right.\nn[4pt]
&+\frac{1}{32}\, \frac{(|M_1|-|\mu|)^2}{M_Z^2 s_{\theta_W}^2} 
\frac{1}{1+\zeta_1 s_{2\beta}}
+\frac{1}{32}\, \frac{|M_1|-|\mu|}{|M_1|+|\mu|}\,
\frac{1-\zeta_1 s_{2\beta}-4c_{2\beta}}{1+\zeta_1 s_{2\beta}}
\nn[4pt]
&\left.-\frac{1}{128}\,  \frac{M_Z^2 s_{\theta_W}^2}{(|M_1|+|\mu|)^2}
\left(
2\,\frac{1-\zeta_1 s_{2\beta}+8c_{2\beta}}{1+\zeta_1 s_{2\beta}}
-9(1-\zeta_1 s_{2\beta})-40c_{2\beta}
\right)\right]
\,,\\
\label{M1=mu-N14}
N_{14}\approx&
\,\,
\frac12
+\frac{\sqrt{2}}{16}\, \frac{M_Z s_{\theta_W}}{|M_1|+|\mu|} \,
\frac{1-\zeta_1 s_{2\beta}+4c_{2\beta}}{\sqrt{1+\zeta_1 s_{2\beta}}}
+\frac{\sqrt{2}}{8}\, \frac{|M_1|-|\mu|}{M_Z s_{\theta_W}} \,
\frac{1}{\sqrt{1+\zeta_1 s_{2\beta}}}
\nn[4pt]
&-\frac{1}{32}\, \frac{(|M_1|-|\mu|)^2}{M_Z^2 s_{\theta_W}^2} \,
\frac{1}{1+\zeta_1 s_{2\beta}}
-\frac{1}{32}\, \frac{|M_1|-|\mu|}{|M_1|+|\mu|}\,
\frac{1-\zeta_1 s_{2\beta}+4c_{2\beta}}{1+\zeta_1 s_{2\beta}}
\nn[4pt]
&+\frac{1}{128}  \frac{M_Z^2 s_{\theta_W}^2}{(|M_1|+|\mu|)^2}
\left(
2\frac{1-\zeta_1 s_{2\beta}-8c_{2\beta}}{1+\zeta_1 s_{2\beta}}
-9(1-\zeta_1 s_{2\beta})+40c_{2\beta}
\right)
\,.
\end{align}
Using the above equations for the $N_{1i}$ one can find the 
LSP-Higgs coupling. Up to the first subleading terms one finds
\begin{equation}
\label{M1=mu-c}
 c_{h\chi\chi} \approx \frac{\sqrt{2}\,g_1}{8}\,\sgn(M_1)
\left[
\sqrt{1+\zeta_1 s_{2\beta}}
-\frac{1}{\sqrt{2}}\,\frac{M_Z s_{\theta_W}}{|M_1|+|\mu|} (1-\zeta_1 s_{2\beta}) \right].
\end{equation}

The formulae for the LSP composition, (\ref{M1=mu-N11})-(\ref{M1=mu-N14}), 
can not be used for values of $\tan\beta$ close to 1 and 
opposite sign $M_1$ and $\mu$ because the term 
$(1+\zeta_1 s_{2\beta})\equiv(1+\sgn(\mu M_1)s_{2\beta})$ 
present in the denominators of many terms becomes very small. 
The problem is caused by the fact that in such limit one of the 
off-diagonal terms in the neutralino mass matrix (after diagonalization 
of the higgsino sub-matrix) becomes very small and one of the assumptions 
discussed before eq.\ (\ref{M1=mu-mN1}) is not fulfilled. However, 
the first subleading contributons to the $c_{h\chi\chi}$ with 
$\sqrt{1+\zeta_1 s_{2\beta}}$ in the denominator cancell out and do
not contribute to (\ref{M1=mu-c}).

The LSP coupling to the $Z$ boson is approximately given by
\begin{equation}
\label{M1=mu-cZ} 
c_{Z\chi\chi} 
\approx 
-\frac{g_1}{2\sqrt{2}}\,c_{2\beta}\,\frac{M_Z}{|M_1|+|\mu|}\,
\frac{1}{\sqrt{1+\zeta_1 s_{2\beta}}}
=
\frac{g_1}{2\sqrt{2}}\,\frac{M_Z}{|M_1|+|\mu|}\,
\sqrt{1-\zeta_1 s_{2\beta}}
\,.
\end{equation}
In the second equality we used the relation 
$c_{2\beta}=-\sqrt{1+s_{2\beta}}\sqrt{1-s_{2\beta}}$ 
to show that $c_{Z\chi\chi}$ has no singularity for $\zeta_1=-1$ 
and $s_{2\beta}\to1$.

The case of large wino-higgsino mixing with decoupled bino 
is quite similar. One has to 
make the following changes: $M_1\to M_2$, $g_1\to g_2$, 
$s_{\theta_W}\to c_{\theta_W}$, $\zeta_1\to\zeta_2$ 
and to mulitply $c_{h\chi\chi}$ by $-1$.
Only the chargino mass is different and reads
\begin{equation}
\label{M2=mu-mC1}
m_{\chi_1^+}
\approx
\frac{|M_2|+|\mu|}{2}
-\frac{M_Z c_{\theta_W}}{\sqrt{2}}\sqrt{1+\zeta_2 s_{2\beta}}
%\nn&
+\frac12\,\frac{M_Z^2 c^2_{\theta_W}}{|M_2|+|\mu|}\left(1-\zeta_2 s_{2\beta}\right)
-\frac{\sqrt{2}}{8}\,
\frac{(|M_2|-|\mu|)^2}{M_Z c_{\theta_W}\sqrt{1+\zeta_2 s_{2\beta}}}
\,.
\end{equation}
The chargino-neutralino mass difference, with the leading corrections
from finite $M_1$, is given by
\begin{equation}
\label{M2=mu-mC1N1}
m_{\chi_1^+}-m_{\chi_1^0}
\approx
\frac14\frac{M_Z^2\,c^2_{\theta_W}}{|M_2|+|\mu|}\left(1-\zeta_2 s_{2\beta}\right)
+\frac14\sgn(M_1M_2)\frac{M_Z^4\,c^2_{2\theta_W}}{\mu^2|M_1|}s^2_{2\beta}
\,.
\end{equation}

%%%%%%%%%%%%%%%%%%%%%%%%%%%%%%%%%%%%%%%%%%%%%%%%%%%%%%%%%%%%%%%%%%%%%%%%%%

%%%%%%%%%%%%%%%%%%%%%%%%%%%%%%%%%%%%%%%%%%%%%%%%%%%%%%%%%%%%%%%
%%%%%%%%%%%%%%%%%%%%%%%%%%%%%%%%%%%%%%%%%%%%%%%%%%%%%%%%%%%%%%%
%

\end{document}